\documentclass[12pt]{article}
\usepackage{geometry}
\usepackage{amsmath}
\usepackage{amssymb,epsfig,subfigure}
\usepackage{graphicx}
\numberwithin{equation}{section}

\newcommand{\be}{\begin{equation}}
\newcommand{\ee}{\end{equation}}
\newcommand{\ba}{\begin{aligned}}
\newcommand{\ea}{\end{aligned}}

\def\m1{\left(-1\right)^{F_i}}

\makeatletter
\def\sla@#1#2#3#4#5{{%
  \setbox\z@\hbox{$\m@th#4#5$}%
  \setbox\tw@\hbox{$\m@th#4#1$}%
  \dimen4\wd\ifdim\wd\z@<\wd\tw@\tw@\else\z@\fi
  \dimen@\ht\tw@
  \advance\dimen@-\dp\tw@
  \advance\dimen@-\ht\z@
  \advance\dimen@\dp\z@
  \divide\dimen@\tw@
  \advance\dimen@-#3\ht\tw@
  \advance\dimen@-#3\dp\tw@
  \dimen@ii#2\wd\z@  \raise-\dimen@\hbox to\dimen4{%
    \hss\kern\dimen@ii\box\tw@\kern-\dimen@ii\hss}%
  \llap{\hbox to\dimen4{\hss\box\z@\hss}}}}
\def\slashed#1{%
  \expandafter\ifx\csname sla@\string#1\endcsname\relax
    {\mathpalette{\sla@/00}{#1}}%
  \else
    \csname sla@\string#1\endcsname
  \fi}
\makeatother

\begin{document}


\thispagestyle{empty}
\begin{center}\footnotesize
\texttt{CALT-68-2765} $\qquad $ \texttt{EFI-09-35} $\qquad $
\texttt{NSF-KITP-09-209} $\qquad $ \texttt{PI-STRINGS-168}
\vspace{1.2cm}
\end{center}

\renewcommand{\thefootnote}{\fnsymbol{footnote}}
\setcounter{footnote}{0}

\begin{center}
{\Large\textbf{\mathversion{bold}
Compact F-theory GUTs with $U(1)_{PQ}$}\par}
\vspace{1.6cm}

\textrm{Joseph Marsano$^{1,4}$, Natalia Saulina$^{2,4}$ and Sakura Sch\"afer-Nameki$^{3,4}$}

\vspace{.5cm}

\textit{$^1$ Enrico Fermi Institute, University of Chicago\\ 5640 S Ellis Ave, Chicago, IL 60637, USA\\[1ex]
$^2$ Perimeter Institute for Theoretical Physics\\ 31 Caroline St N., Waterloo, Ontario N2L 2Y5, Canada \\[1ex]
$^3$ Kavli Institute for Theoretical Physics \\ University of California, Santa Barbara, CA 93106, USA\\[1ex]
$^4$ California Institute of Technology\\ 1200 E California Blvd, Pasadena, CA 91125, USA\\[1ex]}

\texttt{marsano,  saulina, ss299  theory.caltech.edu}

\bigskip


\par\vspace{0.5cm}

\textbf{Abstract}\vspace{5mm}
\end{center}

\noindent

We construct semi-local and global realizations of $SU(5)$ GUTs in F-theory that utilize a $U(1)_{PQ}$ symmetry to protect against dimension four proton decay.  Symmetries of this type, which assign charges to $H_u$ and $H_d$ that forbid a tree level $\mu$ term, play an important role in scenarios for neutrino physics and gauge mediation that have been proposed in  local F-theory model building.  As demonstrated in \cite{Marsano:2009gv}, the presence of such a symmetry implies the existence of non-GUT exotics in the spectrum, when hypercharge flux is used to break the GUT group and to give rise to doublet-triplet splitting.  These exotics are of precisely the right type to solve the unification problem in such F-theory models and might also comprise a non-standard messenger sector for gauge mediation.  We present a detailed description of models with $U(1)_{PQ}$ in the semi-local regime, which does not depend on details of any specific Calabi-Yau four-fold, and then specialize to the geometry of \cite{Marsano:2009ym} to construct three-generation examples with the minimal allowed number of non-GUT exotics.  Among these, we find a handful of models in which the D3-tadpole constraint can be satisfied without requiring the introduction of anti-D3-branes.  Finally, because $SU(5)$ singlets that carry $U(1)_{PQ}$ charge may serve as candidate right-handed neutrinos or can be used to lift the exotics, we study their origin in compact models and motivate a conjecture for how to count their zero modes in a semi-local setting.

\vspace*{\fill}

\setcounter{page}{1}
\renewcommand{\thefootnote}{\arabic{footnote}}
\setcounter{footnote}{0}

 \newpage

\tableofcontents
\newpage

\section{Introduction}

 It has become evident during the past two years that $F$-theory is a promising framework for building realistic supersymmetric $SU(5)$ GUTs in string theory.  Local models \cite{Donagi:2008ca,Beasley:2008dc} have shed light on mechanisms for breaking the GUT gauge group \cite{Beasley:2008kw,Donagi:2008kj}, achieving doublet-triplet splitting \cite{Beasley:2008kw,Donagi:2008kj}, supersymmetry breaking \cite{Marsano:2008jq, Heckman:2008qt},
 generating favorable flavor hierarchies \cite{Font:2008id,Heckman:2008qa,Hayashi:2009ge,Bouchard:2009bu,Heckman:2009mn,Heckman:2009de,Font:2009gq,Conlon:2009qq, Cecotti:2009zf, Hayashi:2009bt, Marchesano:2009rz}, and obtaining naturally small neutrino masses \cite{Tatar:2009jk,Heckman:2009mn}.  Trying to embed these local models into honest string compactifications \cite{Andreas:2009uf,Donagi:2009ra,Marsano:2009ym,Marsano:2009gv}, however, has proven to be quite constraining \cite{Marsano:2009gv}.  While there is a topological constraint that must be satisfied globally in order to prevent the $U(1)_Y$ gauge boson from becoming massive \cite{Buican:2006sn,Beasley:2008kw,Donagi:2008kj}, the most severe constraints at the moment seem to arise from the study of semi-local models \cite{Donagi:2009ra,Marsano:2009gv}, which focus on the geometry near the 7-branes where the $SU(5)$ GUT degrees of freedom are localized.  
 
 The approach using semi-local models is particularly powerful because it is independent of the specific details of the compactification, so that any constraints that arise apply to general $F$-theory compactifications.
Phenomenologically, this is very welcome, as embedding into a semi-local model highly constrains the viable low-energy GUT theories that are consistent with a UV-completion into F-theory.


\subsection{Symmetries and Proton Decay Operators}

One of the most urgent problems when constructing $F$-theory models is the suppression of dangerous dimension 4 proton decay operators.  After the analysis of \cite{Tatar:2009jk}, it seems apparent that these operators must be expressly forbidden by a symmetry of the low energy theory in order to avoid conflict with current bounds on the proton lifetime.  The possibility of using a discrete symmetry for this purpose has been investigated in \cite{Tatar:2009jk} but seems very delicate.
Such a symmetry must not only extend beyond $S_{\rm GUT}$ to the full geometry, it must also act in the right way on all of the zero mode wave functions from which MSSM matter fields originate.

The other option is to utilize a continuous global symmetry.  The MSSM superpotential is invariant under a two-parameter family of $U(1)$ symmetries that commute with the action of $SU(5)_{\rm GUT}$.  One particular member of this family, which we refer to as $U(1)_X$, is a linear combination of hypercharge and $U(1)_{B-L}$ under which MSSM fields carry charges
\begin{equation}\begin{array}{c|c}\text{Field} & U(1)_X \\ \hline
\mathbf{10}_M & -1\\
\mathbf{\overline{5}}_M & 3\\
\mathbf{5}_H & 2\\
\mathbf{\overline{5}}_H & - 2\end{array}\end{equation}
Such a symmetry is easy to implement in $F$-theory compactifications that engineer precisely the matter content of the MSSM \cite{Marsano:2009gv}, but using this to solve the dimension 4 proton decay problem has some drawbacks.  First, such a symmetry still allows for a nonzero $\mu$-term,
\begin{equation}W_{\mu}\sim \mu H_u H_d\,,\end{equation}
so another mechanism must be introduced in order to address this issue.
 It is known in principle how this can be done without introducing any more $U(1)$ symmetries \cite{Marsano:2009gv} but no explicit realization of the mechanism that we proposed there has been provided thus far.

Another drawback is that the $U(1)_X$ symmetry makes it impossible to realize any of the scenarios for neutrino physics described in \cite{Bouchard:2009bu}.  These scenarios are particularly nice because they do not require the generation of any new scales intrinsic to the neutrino sector.  In the Majorana scenario, the small hierarchy between the GUT scale and the right-handed neutrino mass arises from the replacement of a single right-handed neutrino (per generation) by an entire tower of suitable Kaluza-Klein modes\footnote{It is important to point out that the GUT-scale and KK-scale can differ by an order of magnitude \cite{Conlon:2009qa}.}.  A $U(1)_X$ symmetry would forbid Majorana masses for the relevant KK modes, rendering such a mechanism impossible.  The Dirac scenario of \cite{Bouchard:2009bu}, on the other hand, is based on electroweak-scale neutrino masses that are further suppressed by $\mu/M_{GUT}$, allowing one to obtain sufficiently light neutrinos once the $\mu$ problem is adequately solved.  This neutrino mass, however, is the first order correction to a leading unsuppressed contribution that must be otherwise forbidden.  While it is easy to construct global symmetries that can do this, $U(1)_X$ is not one of them.

These problems can all be avoided by utilizing a different $U(1)$ symmetry to forbid dimension 4 proton decay operators.  All choices other than $U(1)_X$ that commute with $SU(5)$ while preserving the MSSM superpotential share a single distinguishing feature, namely that $H_u$ and $H_d$ do not carry exactly opposite charges.  We will refer to any symmetry of this type as a $PQ$ symmetry
\begin{equation}PQ(H_u)+PQ(H_d)\ne 0 \,.
\end{equation}
A nice feature of $PQ$ symmetries is that, in addition to forbidding dimension 4 proton decay operators, they also forbid the $\mu$-term, which then can be generated naturally of the right size, when $PQ$ is broken.


\subsection{Models with $PQ$ Symmetries}

Unfortuantely, semi-local $F$-theory models equipped with a $PQ$ symmetry come with a seemingly unwelcome consequence: all such models that use internal $U(1)_Y$ flux to break $SU(5)_{\rm GUT}$ and solve the doublet-triplet splitting problem are guaranteed to have charged exotics beyond the standard MSSM matter content that \emph{do not} comprise complete GUT multiplets \cite{Marsano:2009gv}.
The reason for this stems from the lack of a common origin of the zero modes associated to the Higgs doublets, $H_u$ and $H_d$.  Through several topological relations, this affects the way that $U(1)_Y$ flux is distributed among the matter curves.  Keeping track of $U(1)_Y$ flux is important because its presence on a matter curve alters the counting of zero modes in a way that does not respect $SU(5)_{\rm GUT}$ invariance.
While $U(1)_Y$ flux is desirable on Higgs matter curves, where it can be used to lift Higgs triplets while retaining the doublets $H_u$ and $H_d$, it is problematic on $\mathbf{10}_M$ and $\mathbf{\overline{5}}_M$ matter curves because it guarantees a zero mode spectrum that is not comprised of complete GUT multiplets{\footnote{We take all $\mathbf{10}_M$ ($\mathbf{\overline{5}}_M$) fields to originate on a single $\mathbf{10}$ ($\mathbf{\overline{5}}$) matter curve in the sense of \cite{Tatar:2009jk} in order to prevent various components of a given multiplet from carrying different $U(1)$ charges.  This is also a necessary condition for the flavor models of \cite{Heckman:2008qa,Bouchard:2009bu,Heckman:2009mn} but addressing detailed flavor structure in a global context is beyond the scope of the present paper.}}.  Unfortunately, any semi-local $F$-theory model that realizes a $PQ$ symmetry will have nonzero $U(1)_Y$ flux on at least one matter curve other than those associated to $H_u$ and $H_d$ when this flux is used to solve the standard problems of GUT-breaking \cite{Marsano:2009gv}{\footnote{While this may be avoided if one can realize a different mechanism for breaking $SU(5)_{\rm GUT}$, such as the Wilson line approach proposed in \cite{Tatar:2008zj}, we will focus on models with $U(1)_Y$ flux in the remainder of this paper.}}.

One might be concerned that exotic fields that come in incomplete GUT multiplets will have a severe impact on gauge coupling unification.  This is certainly true even if a mechanism is introduced to lift them unless their mass is  very close to $M_{\rm GUT}$.  In light of this, it is interesting to recall that generic $F$-theory GUT models with $U(1)_Y$ flux do not exhibit gauge coupling unification at the scale "$M_{\rm GUT}$", but rather the weaker "F-unification" relation \cite{Blumenhagen:2008aw}.  The culprit that disrupts complete unification is the Chern-Simons coupling on the 7-brane worldvolume
\begin{equation}W_{CS}\sim \int_{\mathbb{R}^{3,1}\times S_{\rm GUT}}\, C_0\,\,\text{tr}(F^4) \,.
\label{CScoup}\end{equation}
In the presence of a nontrivial $U(1)_Y$ flux, $F_Y$, this term gives distinct contributions to the gauge kinetic functions.  This creates serious problems for $F$-theory models with only the MSSM fields because, in that case, low energy measurements indicate that the couplings should truly unify at some scale.  In principle, it is possible to cure this problem with a suitable "twist" of the bundle $F_Y$ through addition of a nonzero $B$-field on the 7-brane worldvolume \cite{Blumenhagen:2008aw}.  An alternative to this, however, is the inclusion of some extra exotic fields of just the right type that they can account for this splitting.  Quite nicely, the incomplete GUT multiplets that are forced on us when we require a $PQ$ symmetry are always of this type.

\subsubsection{Lifting the Exotics}

Even so, these extra exotic fields, which we denote hereafter by $f$ and $\overline{f}${\footnote{As this notation implies, the charged exotics always come in vector-like pairs with respect to $SU(5)$.}}, should not be massless so a mechanism must be introduced to lift them to a sufficiently high scale.  One way to do this is to couple them to a charged singlet field that picks up a nonzero expectation value
\begin{equation}W\sim \phi f\overline{f}\,,\qquad \langle \phi\rangle \ne 0\,.
 \label{OGM}\end{equation}
Unfortunately, the nonzero value of $\langle \phi\rangle$ represents a new scale that we have to generate.  This is particularly unwelcome because our primary motivation for studying models with $PQ$ symmetries was to reduce the number of independent scales that must be introduced by implementing one of the neutrino scenarios of \cite{Bouchard:2009bu}.  If we have to generate a new scale to lift charged exotics, we would not have really accomplished anything at all; rather, we would seem to be trading one tuning for another.

An interesting possibility arises, however, if we note that \eqref{OGM} is precisely the superpotential of ordinary gauge mediation with our exotic fields, $f$ and $\overline{f}$, playing the role of messengers.  If we take this idea seriously and suppose that $\phi$ picks up bosonic and $F$-component expectation values, then we don't really have any new tunings beyond those that we would already need for supersymmetry-breaking and its mediation.  This is the context in which we feel that models with $PQ$ symmetries are best motivated.

\subsubsection{Importance of Global Models}

Implementing such a scenario requires detailed knowledge of several features of the global compactification.  First, to obtain gauge mediation we must understand how supersymmetry breaking occurs and, in particular, how to obtain a gravitino mass that is sufficiently light.  Further, we must determine how supersymmetry breaking is communicated to the singlet field $\phi$ and, in addition, understand the dynamics responsible for providing $\phi$ with a bosonic expectation value.  Ideally, a simple D3-instanton model of the sort proposed in \cite{Heckman:2008es, Marsano:2008py} could suffice for everything but unfortunately the structure of singlet fields in global $F$-theory compactifications seems much more complicated than the simple model discussed there.

Beyond this, it is easy to see that the construction of global models is important for studying $SU(5)$ singlets in general because they do not localize on the same complex surface as the charged degrees of freedom.  Their wave functions are free to explore parts of the geometry far away from the $SU(5)$ surface so they can be nontrivially affected by physics not captured by semi-local models.

\subsection{Summary and Outline}

In this paper, we initiate the study of semi-local and global
$F$-theory models that exhibit $PQ$ symmetries.  After reviewing the
semi-local approach in section 2, we start in section 3 by
constructing a class of semi-local geometries capable of realizing a
$U(1)_{PQ}$ symmetry.   We then discuss how these semi-local
geometries can be completed to fully compact elliptic fibrations
over the base manifold $B_3$ constructed in
\cite{Marsano:2009ym}{\footnote{In \cite{Marsano:2009ym}, we use the
notation $\tilde{X}$ for the base manifold rather than $B_3$.}}.
This step requires a bit of care (see Appendix \ref{app:ss})
because the base manifold does not admit any holomorphic sections
capable of individually distinguishing between curves where $H_u$
and $H_d$ fields can be engineered{\footnote{This is connected to
the fact that $B_3$ is constructed to satisfy the "hypercharge
contraint" \cite{Buican:2006sn,Beasley:2008kw,Donagi:2008kj}
required for consistent implementation of GUT-breaking via $U(1)_Y$
flux.}}.  In section 4 we turn to the construction of $G$-fluxes,
which must be introduced in order to control the zero mode
structure.  In the semi-local context, we describe universal fluxes
that are always present as well as a class of non-universal fluxes
that can arise if we slightly refine the geometry.  Further, these
fluxes are constructed in such a way that they extend to the global
models based on $B_3$.  Using them, we are able to obtain entire
families of 3-generation models with $U(1)_{PQ}$ and the smallest
possible number of extra exotics of the type described above.  We
then verify that in several explicit examples, the D3-brane tadpole
induced by the combination of our base manifold, $B_3$, and the
$G$-fluxes is negative so that it can be cancelled through the
addition of D3-branes, as opposed to anti-D3-branes.  The result, then, is a set 
supersymmetric, 3-generation models with the smallest
number of exotics required to realize a $U(1)_{PQ}$ symmetry{\footnote{It should be emphasized that while some complex structure moduli are stabilized on our models, many remain unstabilized.  Addressing this will require us to turn on additional $G$-fluxes, which will contribute to the D3-brane tadpole.  For this reason, our study of the D3-tadpole is preliminary.}}.  We
make some preliminary comments about the phenomenology of these
models in section 5.  In section 6, we discuss the $F$-theoretic
origin of singlet fields and present a conjecture for how to count
them in a semi-local framework.  Finally, we make some concluding
remarks in section 7.



\section{Semi-local Model}
\label{sec:semilocal}


\subsection{Definitions and General Properties}
\label{subsec:Defs}

\begin{figure}
\begin{center}
\epsfig{file=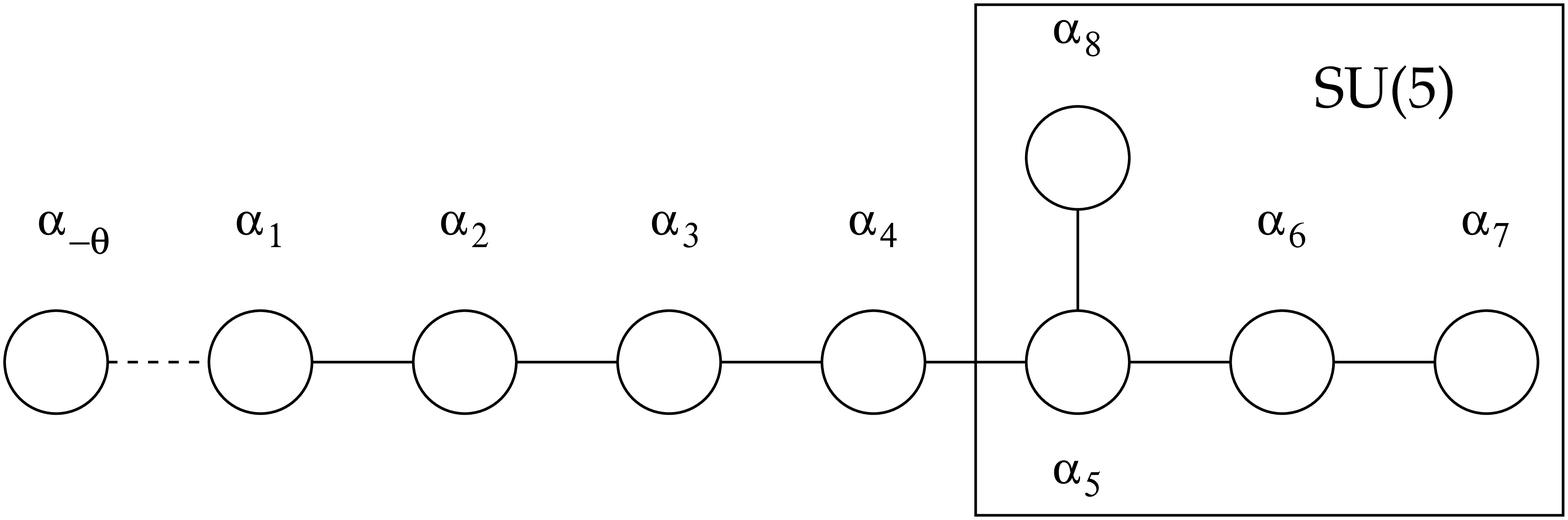, width=8cm}
\caption{Affine $E_8$ Dynkin diagram}
\label{fig:E8}
\end{center}
\end{figure}

The class of viable low energy effective theories that emerge from $F$-theory can be highly constrained by studying "semi-local" models, which describe the physics of  an ADE singularity fibered over an honestly compact 4-cycle{\footnote{This is to be contrasted with so-called local models, which are based on compactifications with an ADE singularity fibered over a copy of $\mathbb{C}^2$, which is intended to represent a single coordinate patch of a compact 4-cycle.}}.  In this section, we briefly review the properties of semi-local models with emphasis on nontrivial monodromies and how they can be succinctly encoded using the spectral cover \cite{Donagi:2009ra,Marsano:2009gv}.

Semi-local models, which purport to describe only "open string" degrees of freedom, depend on the leading behavior of an elliptically fibered Calabi-Yau four-fold $Y_4$ in the vicinity of a surface, $S_{\rm GUT}$, on which the $SU(5)_{\rm GUT}$ gauge degrees of freedom localize.  To describe the geometry in this neighborhood, we start with a Weierstrass equation for the elliptic fiber whose coefficients vary over the base, $B_3$, of the fibration.  If we require an $SU(5)$ degeneration of the fiber along $S_{\rm GUT}$ and include the minimal number of terms needed to ensure that $Y_4$ does not become too singular anywhere{\footnote{By this, we mean that only singularities of Kodaira type are allowed on $S_{\rm GUT}$.}}, the result is a deformed $E_8$ singularity
\be\label{E8Sing} y^2 = x^3 + b_5 x y + b_4 x^2 z + b_3 y z^2 + b_2
x z^3 + b_0 z^5 \,, \ee where $S_{\rm GUT}$ corresponds to $z=0$.  Because $Y_4$ is Calabi-Yau, the deformation parameters  $b_m$ must be
sections of line bundles whose first Chern classes are given by $\eta
- m c_1$, where \be\label{EtaDef} \eta = 6 c_1 - t \,,\qquad t= -
c_1 \left(N_{S_{\rm GUT}}\right) \,,\qquad c_1 = c_1 (S_{\rm GUT})
\, ,\ee
and $N_{S_{\rm GUT}}$ denotes the normal bundle of the 4-cycle $S_{\rm GUT}$ inside the base $B_3$ of the fibration.  We see, then, that the semi-local description depends on a very limited set of data, namely the first Chern class $c_1$ of $S_{\rm GUT}$ and the normal bundle, $-t$, which describes how $S_{\rm GUT}$ is embedded in $B_3$.  Deriving constraints from the semi-local model however does
not require us to specify these, and it is this property that makes
this analysis very powerful and general.

Generically (\ref{E8Sing}) exhibits an $SU(5)$ singularity along $S_{\rm GUT}$.  The singularity type can be enhanced in rank along curves or isolated points in $S_{\rm GUT}$ over which certain combinations of the $b_m$ vanish.  Charged matter localizes on curves where the rank enhances by at least one while Yukawa couplings receive their dominant contributions from isolated points where the rank enhances by two or more.  The matter fields and couplings that arise from various enhancements are easy to determine from the Kodaira classification and we present them in the following table
\begin{center}
\begin{tabular}{c|c|c}
Local Enhancement Type & GUT interpretation & Locus \cr
\hline
$SU(5)$ & Gauge dof's & $b_m\not=0$ for all $m$ \cr
$SU(6)$ &  ${\bf 5}$, ${\bf \bar{5}}$  &  $P= b_0 b_5^2 - b_2 b_3 b_5 + b_3^2 b_4 =0$   \cr
$SO(10)$ &  ${\bf 10}$ &  $b_5 =0$ \cr
$SO(12)$  & $\lambda_b \, {\bf \bar{5}}_H \times {\bf \bar{5}}_M \times {\bf 10}_M$ &  $b_5= b_3=0$ \cr
$E_6$ & $\lambda_t \, {\bf {5}}_H \times  {\bf 10}_M\times  {\bf 10}_M$ &  $ b_5 =b_4 =0$\cr
$E_8$ & $E_8$ "Yukawa"  &  $b_m =0$ for all $m\not=0$
\end{tabular}
\end{center}

Locally $E_8$ gauge group is broken to $SU(5)_{\rm GUT}\times
U(1)^4$ where $U(1)^4$ is the maximal torus of $SU(5)_{\perp}$-- the
commutant of $SU(5)_{\rm GUT},$ and one may think that these four
$U(1)$'s control superpotential couplings. However, this is not the
case in the semi-local model due to monodromies, which we now
review.

First, we introduce some notation: let $\alpha_i$ for $i=1, \cdots,
8$ be the simple roots of $E_8$ and $\alpha_5,\cdots, \alpha_8$ be
the simple roots of $SU(5)_{\rm GUT}$. We consider the decomposition
\be\label{E8SU5SU5} \ba E_8  & \quad \rightarrow SU(5)_{\perp}
\times SU(5)_{\rm GUT} \cr 248  & \quad \rightarrow ({\bf 24}, {\bf
1} ) + ({\bf 1},  {\bf 24 })  + ({\bf \overline{5} },  {\bf
\overline{10} }) + \boxed{({\bf \overline{10}},  {\bf 5})+({\bf  5},
{\bf  10 }) + ({\bf  10}, {\bf \overline{5}})} \ea \ee
 Note that the
boxed fields precisely give rise to the matter and Higgs fields in
the ${\bf 5},$ ${\bf 10}$, and ${\bf \bar{5}}$ representations.
Denote  by $\lambda_i  = \sum_{k =1}^i \alpha_{5-k}$ for $i=1,
\cdots, 5$ the weights of $SU(5)_\perp$, where \be \alpha_0 =
\alpha_{-\theta} = - 2 \alpha_1 - 3 \alpha_2 - 4 \alpha_3 - 5
\alpha_4 -  6 \alpha_5 - 4 \alpha_6 - 2 \alpha_7 - 3 \alpha_8 \,.
\ee Note that $\sum_{i=1}^5 \lambda_i =0$. In terms of the weights,
which we will implicitly assume to also denote the holomorphic
volumes\footnote{By holomorphic volumes we mean integrals of the
$(2,0)$ form in the ALE-fiber over $\mathbb{P}^1$'s which arise
due to deformation of $E_8$ singularity down to $SU(5)$
singularity.} of $\mathbb{P}^1$'s of the deformed $E_8$ singularity,
we can identify the parametrization of the matter and Yukawa
couplings as follows: From (\ref{E8SU5SU5}) we read off that the
${\bf 10}$ GUT multiplets are labeled in terms of the fundamental
weights of $SU(5)_\perp$, $\lambda_i$. Likewise, the ${\bf \bar{5}}$
GUT multiplets arise from the ${\bf 10}$ of $SU(5)_\perp$ and are
thus labeled by $\lambda_i + \lambda_j$. In summary we  can augment
our table as follows:

\begin{center}
\begin{tabular}{c|c|c|c}
Enhancement & GUT interpretation & Locus & Locus in terms of $\lambda_i$\cr
\hline
$SU(5)$ & Gauge dof's & $b_m\not=0$ for all $m$ &  $\lambda_i \not= 0$ for all $i$\cr
$SU(6)$ &  ${\bf 5}$, ${\bf \bar{5}}$  &  $P=0$   & $\pm (\lambda_i + \lambda_j)=0$ for some $i, j$ \cr
$SO(10)$ &  ${\bf 10}$ &  $b_5 =0$  &  $\lambda_i =0$ for some $i$  \cr
$SO(12)$  & $\lambda_b \, {\bf \bar{5}}_H \times {\bf \bar{5}}_M \times {\bf 10}_M$ &  $b_5= b_3=0$ &$ (\lambda_i + \lambda_j ) + (\lambda_k + \lambda_l) + (\lambda_m) =0$, \cr
    &&&  $\epsilon_{ijklm}\not= 0$ \cr
$E_6$ & $\lambda_t \, {\bf {5}}_H \times  {\bf 10}_M\times  {\bf 10}_M$ &  $ b_5 =b_4 =0$ & $(\lambda_i) + (\lambda_j) + (- \lambda_k- \lambda_l) =0 $ \cr
&&& $(ij) =  (kl)$ or $(lk)$ \cr
$E_8$ & $E_8$ "Yukawa"  &  $b_m =0$ for all $m\not=0$ &  $\lambda_i =0$ for all $i$
\end{tabular}
\end{center}

The deformation parameters $b_m$ of the $E_8$ singularity and the
holomorphic volume $\lambda_i$ of the i-th  $\mathbb{P}^1$, are
related by \be\ba b_5 & \sim b_0  \prod_{i=1}^5 \lambda_i  \cr b_4 &
\sim b_0 \sum_{i<j<k<l} \lambda_i \lambda_j\lambda_k \lambda_l  \cr
b_3 &\sim b_0  \sum_{i<j<k} \lambda_i \lambda_j\lambda_k \cr b_2
&\sim b_0 \sum_{i<j}\lambda_i \lambda_j\cr b_1 & \sim b_0   \sum_{i}
\lambda_i =0 \,. \ea \ee These relations show that the deformation
parameters of the geometry are not linearly related to the
$\lambda_i$. Put differently, inverting these algebraic relations
will yield $\lambda_i (b_m)$ to be functions with branch cuts, and
thus despite the apparent local linear independence of four
$U(1)$'s, labeled by the associated weights $\lambda_i$, globally
these may be identified under the action of a monodromy group.

\begin{figure}
\begin{center}
\epsfig{file=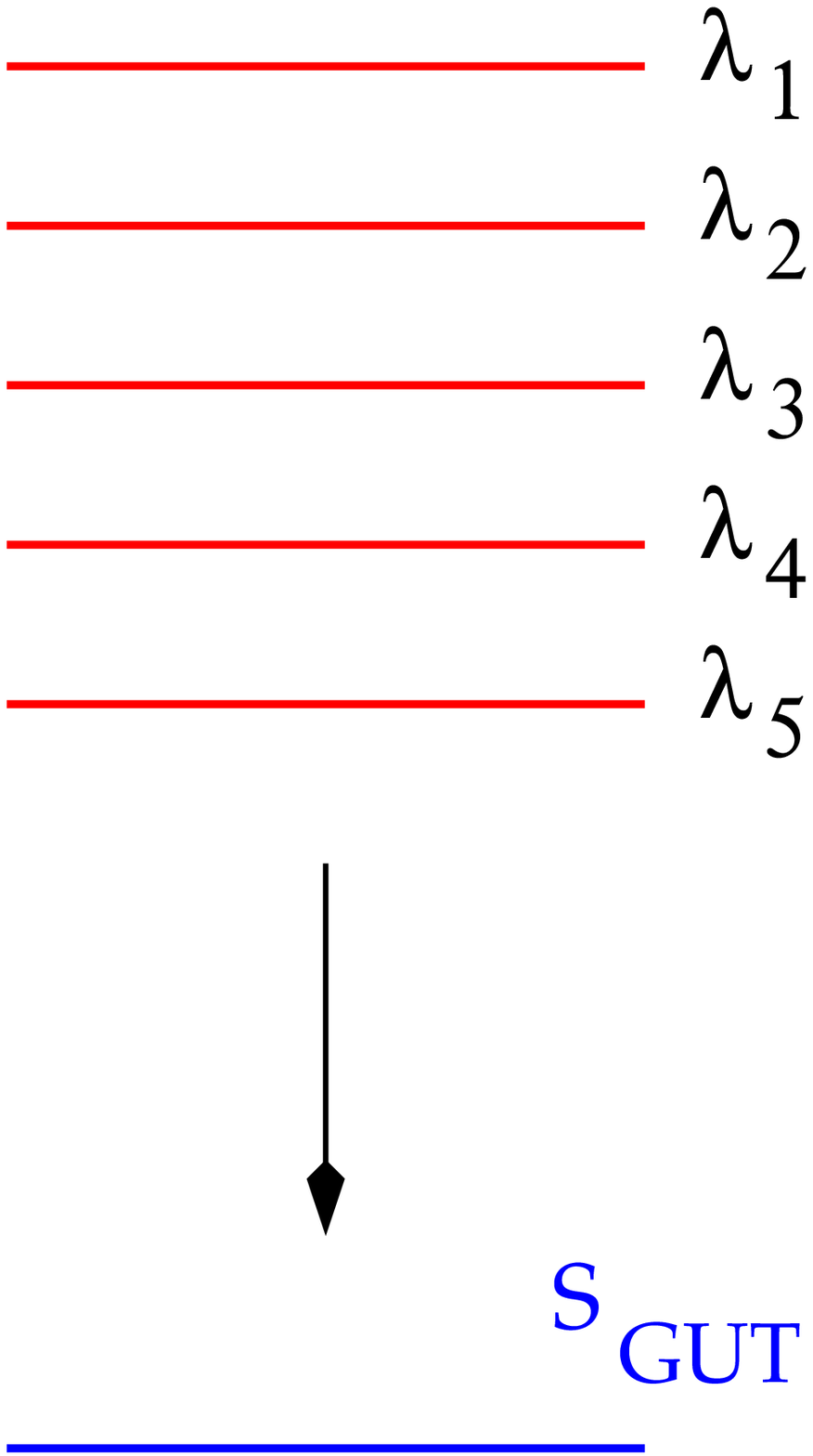, width=3.5cm}
$\qquad \qquad $
\epsfig{file=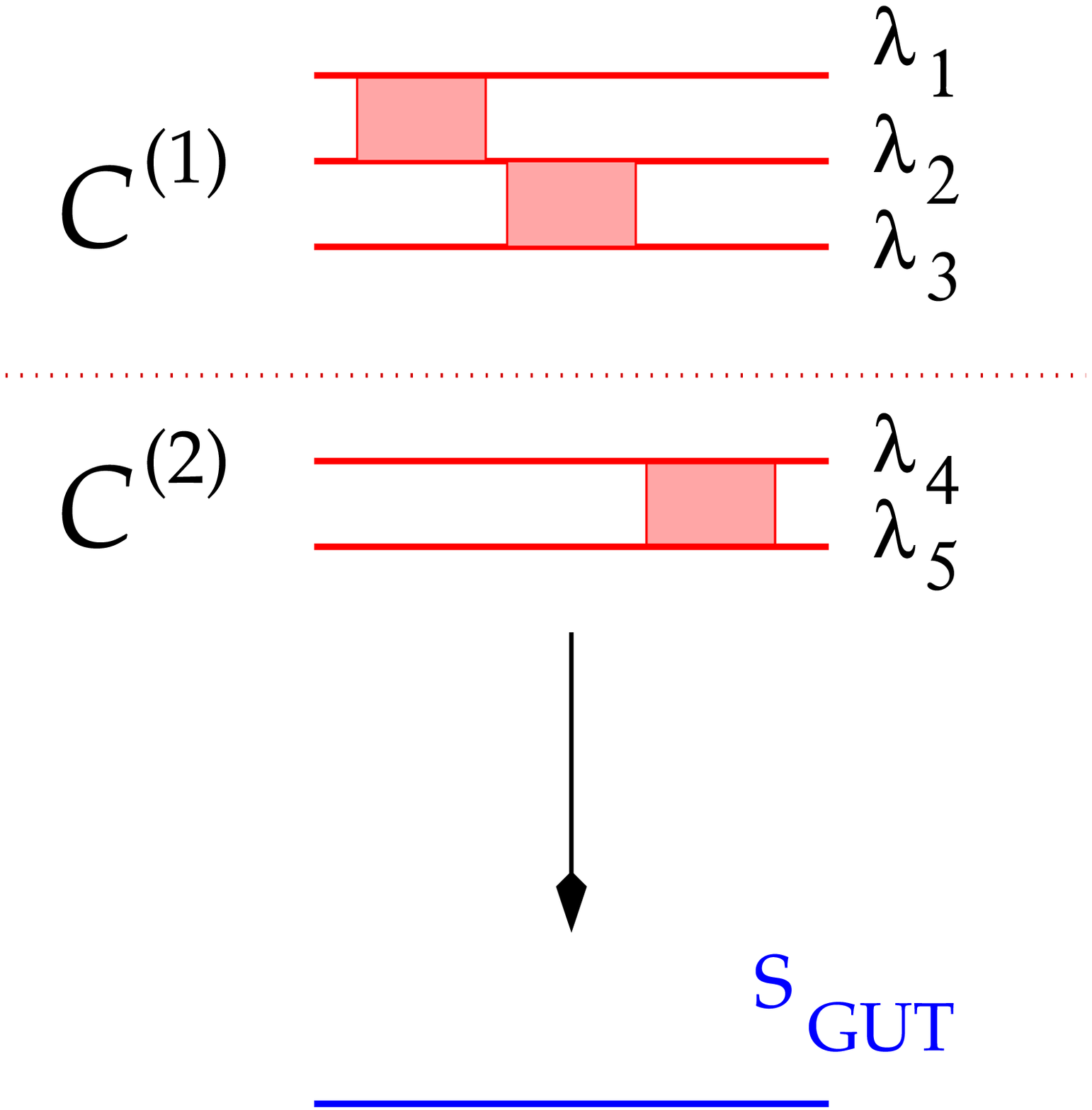, width=5.5cm}
\caption{Five-sheeted spectral cover labeled by $\lambda_i$. Generically these sheets will be connected by branch-cuts, leading to a factorization of the spectral cover into the connected pieces. The right hand picture shows a $3+2$ factored spectral cover over $S_{\rm GUT}$.}
\label{fig:SpecCovs}
\end{center}
\end{figure}

While all of the data of the monodromy group is contained in the
$b_m$, it is useful to introduce an auxiliary object, the
fundamental spectral surface ${\cal{C}}$, which very efficiently
encodes all the monodromy information \cite{Donagi:2009ra}.  A convenient realization of
this surface was given in \cite{Donagi:2009ra} as a submanifold of
the projective three-fold
\begin{equation}
X=\mathbb{P}({\cal{O}}_{S_{\rm GUT}}\oplus K_{S_{\rm GUT}})\,,
\label{Xdef}\end{equation}
defined by the equation
\begin{equation}{\cal{F}}_{ {\cal{C}}}=b_0 U^5 + b_2V^2U^3+b_3V^3U^2+b_4V^4U+b_5V^5=0\,.
\label{fund}\end{equation}
Here ${\cal{O}}_{S_{\rm GUT}}$ and $K_{S_{\rm GUT}}$ denote the trivial and canonical bundle on $S_{\rm GUT}$, respectively.  The homogeneous coordiantes $[U,V]$ on the $\mathbb{P}^1$ fiber are sections of ${\cal{O}}(1)\otimes K_{S_{\rm GUT}}$ and ${\cal{O}}(1)$, respectively, where ${\cal{O}}(1)$ is the line bundle of degree 1 on $\mathbb{P}^1$.

It will also be convenient to define the projection
\begin{equation}\pi:X\rightarrow S_{\rm GUT}\end{equation}
along with the map, $p_{\cal{C}}$, that it induces
\begin{equation}p_{{\cal C}}:{\cal{C}}\rightarrow S_{\rm GUT}\,.\label{pCdef}\end{equation}
The object ${\cal{F}}_{{\cal{C}}}$ is a projectivization of the
equation
\begin{equation}0=b_0s^5 + b_2s^3+b_3s^2+b_4s+b_5\sim b_0\prod_{i=1}^5 (s+\lambda_i)\,,
\end{equation}
with $s$ replaced by $U/V$ and whose roots, as indicated, are essentially the $\lambda_i$.  In any
local patch, the sheets of ${\cal{C}}$ provide a solution
$\lambda_i(b_n)$ while the monodromy group, $G$, is encoded by the
topology of the full surface. In particular the various sheets of the cover may be connected by branch-cuts. We depict this in figure \ref{fig:SpecCovs}.

An important consequence of nontrivial monodromies is that they can project out some of the $U(1)$'s that remain when $E_8$ is broken to $SU(5)_{\rm GUT}$.  In general, the number of $U(1)$'s that survive is determined by the number of components into which ${\cal{C}}$ factors as
\be \# \hbox{ independent } U(1)
= \left(\# \hbox{ factors of } \mathcal{C}\right) -1 \,. \ee
In particular, this means that, in the generic situation where ${\cal{C}}$ is an irreducible surface, all $U(1)$'s are projected out and only $SU(5)_{\rm GUT}$ remains to constrain the superpotential.  In figure
\ref{fig:SpecCovs} we depict an example spectral cover that
factors into three sheets and two sheets connected by branch
cuts, in which case a single $U(1)$ symmetry remains.


\subsection{Constraints from Phenomenology and F-unification}
\label{subsec:Consts}

In \cite{Marsano:2009gv} we demonstrated that by imposing minimal phenomenological requirements, one can highly constrain the class of viable semi-local models.
The phenomenological constraints that we initially impose are:
\begin{itemize}
\item $SU(5)$ GUT with 3 families of ${\bf \bar{5}}_M$ (${\bf 10}_M$) matter localized on a single curve{\footnote{We require that MSSM $\mathbf{\overline{5}}$ and $\mathbf{10}$ fields live on a single matter curve in order to avoid a situation in which components of different families are distinguished by $U(1)$ charges, which would give rise to apparently unwanted Yukawa textures.  It may be possible to utilize such $U(1)$ charges to induce good flavor structure \cite{Font:2008id} but we do not investigate that possibility here.}}
\item ${\bf 5}_H$ and ${\bf \bar{5}}_H$ Higgs multiplets
\item GUT-breaking through $U(1)_Y$ flux that also lifts the Higgs triplets
\item Absence of Exotics
\item Absence of dim 4 proton decay operators\footnote{Dimension 5 proton decay
operators are absent by putting the Higgs fields ${\bf 5}_H$ and $\bar{\bf{5}}_H$ onto different
matter curves in the sense of \cite{Tatar:2009jk}, which realizes the well-known `missing partner'
mechanism.}
\item Absence of tree-level $\mu$-term
\end{itemize}
Note that the absence of exotics is essentially a condition that the hypercharge flux, $F_Y$, restricts trivially on all curves where MSSM matter fields localize
\be\label{FY10}
F_{Y} \cdot \Sigma_{{\bf 10}_M} =0 \,,\qquad F_Y\cdot \Sigma_{\mathbf{\overline{5}}_M}=0\,.
\ee
The only allowed models under these assumptions are:
\begin{center}
\underline{{\bf 4+1} factored models:}  $\qquad  \mathcal{C} =
\mathcal{C}^{(1)}\mathcal{C}^{(4)}$  with monodromy $G =
\mathbb{Z}_4, D_4, \hbox{Klein}_4$.
\end{center}
Models with this factorization and the assignments of matter to the spectral cover components as in \cite{Marsano:2009gv} 
\be
\ba
{\bf 10}_M: &\qquad \mathcal{C}^{(4)} \cr 
{\bf 5}_H : & \qquad \mathcal{C}^{(4)} \cap \tau \mathcal{C}^{(4)} \cr
{\bf \bar{5}}_H :&\qquad \mathcal{C}^{(4)}\cap \tau \mathcal{C}^{(4)} \cr
{\bf \bar{5}}_M: & \qquad  \mathcal{C}^{(4)}\cap \tau \mathcal{C}^{(1)}\,.
\ea
\ee
give rise to minimal $SU(5)$ GUTs with all Yukawa couplings required. 
These models are distinguished by the fact that, in each of them, dimension 4 proton decay operators are forbidden by a $U(1)_X$ symmetry with respect to which MSSM fields carry charges
\begin{equation}\begin{array}{c|c|c|c|c}
\text{Field} & \mathbf{10}_M & \mathbf{\overline{5}}_M & \mathbf{5}_H & \mathbf{\overline{5}}_H \\ \hline
U(1)_X & -1 & 3 & 2 & -2
\end{array}\end{equation}
This is a linear combination of $U(1)_Y$ and $U(1)_{B-L}$ so these can be viewed as "gauged $B-L$" models.  Note, however, that $U(1)_{X}$ is typically anomalous if we do not engineer exactly 3 right-handed neutrino fields.  In this case, the corresponding vector multiplet will typically acquire a string scale mass, as usual.

One highly worrying feature of these models is that their failure to exhibit a $U(1)_{PQ}$ symmetry makes it impossible to realize any of the neutrino scenarios of \cite{Bouchard:2009bu}.  To implement these scenarios, then, we must rethink of our original set of constraints.  As we pointed out in  \cite{Marsano:2009gv}, it is very natural to relax the condition that no exotic fields appear because the presence of exotics can be motivated from a different perspective altogether.  This is because the use of $F_Y$ to break the gauge group, while providing a nice mechanism for solving the doublet-triplet splitting problem and removing unwanted $SU(5)_{\rm GUT}$ adjoint degrees of freedom, comes at the price of spoiling gauge coupling unification \cite{Blumenhagen:2008aw}.   The resulting gauge couplings satisfy the F-unification relation
\be\label{Fun}
\alpha_1^{-1} (M_{GUT}) - {3\over 5} \alpha_2^{-1} (M_{GUT}) - {2\over 5}\alpha_3^{-1} (M_{GUT}) =0 \,
\ee
at the fiducial GUT scale, $M_{\rm GUT}$, but they are not equivalent to one another in general.  As shown in \cite{Marsano:2009gv}, it is possible to reconcile this with the apparent gauge coupling unification that we infer from low energy data by adding new exotic fields of precisely the type that are needed to build models with a $U(1)_{PQ}$ symmetry.  More specifically, to realize a $U(1)_{PQ}$ symmetry we are forced to introduce fields that do not comprise a complete GUT multiplet but that nevertheless induce shifts, $\delta b_i$, of the MSSM $\beta$ function coefficients, $b_i$, that always satisfy
 \be\label{Deltab}
\delta b_1 - {3\over 5} \delta b_2 - {2\over 5} \delta b_3 =0 \,.
\ee
Said differently, GUT-breaking by hypercharge flux and the resulting spoilt unification essentially require us to add additional, non-GUT multiplet exotic fields.

In the rest of this paper, we  construct and study semi-local and global models that exhibit a $U(1)_{PQ}$ symmetry.  The starting point of our constructions is to consider models in which the spectral surface ${\cal{C}}$ factors into cubic and quadratic pieces:
\begin{center}
\underline{${\bf 3+2}$ factored models}: $\qquad \mathcal{C} =
\mathcal{C}^{(3)} \mathcal{C}^{(2)}$
\end{center}
One can also consider ${\bf 4+1}$ models with exotics, which have $U(1)_{PQ}$ symmetry, and an assignment of matter as follows
\be
\ba
{\bf 10}_M: &\qquad \mathcal{C}^{(4)} \cr 
{\bf 5}_H : & \qquad \mathcal{C}^{(4)} \cap \tau \mathcal{C}^{(4)} \cr
{\bf \bar{5}}_H :&\qquad \mathcal{C}^{(4)}\cap \tau \mathcal{C}^{(1)} \cr
{\bf \bar{5}}_M: & \qquad  \mathcal{C}^{(4)}\cap \tau \mathcal{C}^{(4)}\,.
\ea
\ee
However, this requires to further fix complex structure moduli  (i.e. to tune the coefficients $b_n$ futher) in order to realize a refined monodromy group that puts ${\bf 5}_H$ and ${\bf \bar{5}}_M$ on different curves. No such tuning is required in the ${\bf 3+2}$ models, which is why we prefer to study these in the present paper. 

One could also study models in which ${\cal{C}}$ splits into 3 or more components, leading to multiple $U(1)$ symmetries.  In doing so, however, one must be careful that the particular solution to the constraint $b_1=0$ does not introduce any non-Kodaira type singularities on $S_{\rm GUT}$.  This seems a bit tricky to avoid so we typically stick to models with only one $U(1)$ symmetry.


\section{Semi-local and Global Models with $U(1)_{PQ}$}
\label{sec:model}

We turn now to the construction of semi-local and global models that exhibit $U(1)_{PQ}$ symmetries.  By this, we mean that a suitable identification of matter curves with MSSM multiplets makes possible the realization of the MSSM superpotential along with a $U(1)_{PQ}$ symmetry.  We will not worry about engineering 3 generations of MSSM matter on the corresponding matter curves until the next section, where we discuss $G$-fluxes.



\subsection{${\bf 3 +2}$ factored spectral cover}

In this paper, we consider models based on a spectral surface ${\cal{C}}$
\be
\mathcal{C}\,: \qquad   b_0 U^5 + b_2 V^2 U^3 + b_3 V^3 U^2 + b_4 V^4 U + b_5 V^5 =0 \,,
\ee
that factors into cubic and quadratic pieces according to
\be
\mathcal{C}= \mathcal{C}^{(1)} \mathcal{C}^{(2)}\,:\qquad (a_0 U^3 + a_1 U^2 V + a_2 U V^2 + a_3 V^3 ) (e_0 U^2 + e_1 U V + e_2 V^2) =0 \,.
\label{32fact}\ee
To achieve this, we require the $b_m$ take the form
\be
\ba
b_5 &= a_3 e_2 \cr
b_4 &= a_3 e_1+a_2 e_2 \cr
b_3 &= a_3 e_0+a_2 e_1+a_1 e_2\cr
b_2 &=a_2 e_0+a_1 e_1+a_0 e_2 \cr
b_0 &= a_0 e_0 \,.
\label{bmae}\ea
\ee
Furthermore, $b_1=0$ implies
\be\label{b1Cond}
a_1 e_0+ a_0 e_1 =0\,.
\ee
In order to solve this constraint, we will make the following ansatz:
\be\label{b1Sol}
\ba
a_0 = \alpha e_0 \,,\qquad a_1 = - \alpha e_1 \,.
\ea
\ee

The classes of the various coefficients can be determined as follows: recall that the complete spectral cover $\mathcal{C}$ is in the class $\eta + 5 \sigma$ in the auxiliary space $X$ \eqref{Xdef}, where $\eta$ is defined as in (\ref{EtaDef}).
This constrains the remaining classes as follows:
\be
\begin{tabular}{c|c}
\hbox{Section} & \hbox{Divisor Class in }X \cr
\hline
$U$ & $\sigma$ \cr
$V$ & $\sigma + \pi^*c_1 $\cr
$b_m$ & $\pi^*(\eta - m c_1)$  \cr
$e_2$ & $\pi^*\xi $\cr
$e_1$ & $\pi^* (c_1 + \xi) $\cr
$e_0$ & $\pi^* (2c_1 + \xi) $ \cr
$a_m$ & $\pi^*(\eta - (m+2)  c_1 - \xi )$ \cr
$\alpha$ & $\pi^* (\eta - 4 c_1 - 2 \xi)$
\end{tabular}
\ee
Here, $\xi$ is an element of $H_2(S_{\rm GUT},\mathbb{Z})$ that we are free to choose, as long as all sections remain holomorphic.
The various components of the spectral cover are then in the following classes in $X$
\be
\ba
\mathcal{C}^{(1)}: &\quad  3\sigma + \pi^*(\eta - 2 c_1  -\xi )   \cr
\mathcal{C}^{(2)}: &\quad   2\sigma + \pi^* (2 c_1 + \xi )   \,.
\ea
\ee


\subsection{Monodromies and $U(1)_{PQ}$}

We now discuss the monodromy structure induced by this factorization of the spectral cover and demonstrate that a suitable assignment of MSSM fields to matter curves leads to a realization of $U(1)_{PQ}$.  This will also allow us to demonstrate that the neutrino scenarios of \cite{Bouchard:2009bu} are in principle possible to realize in this setup.

When ${\cal{C}}$ factors into cubic and quadratic pieces, the monodromy group is a transitive subgroup of $S_3\times \mathbb{Z}_2$ where $S_n$ is the symmetric group acting on $n$ objects.  The allowed monodromy groups are therefore
\be
S_3 \times \mathbb{Z}_2 \ \supset   G^{(1)} \times G^{(2)} = S_3 \times \mathbb{Z}_2\,,\ \mathbb{Z}_3 \times \mathbb{Z}_2\,, \ D_3 \times \mathbb{Z}_2\,.
\ee
Recall that the  structure of matter curves and Yukawa couplings depends only on the action of the monodromy group on components of the fundamental and antisymmetric representations of $SU(5)_{\perp}$.  These are equivalent for all of the above choices so it is not necessary to distinguish these possibilities.

The $\mathbf{10}$'s that can arise carry $SU(5)_{\perp}$ weights $\lambda_I$ for $I=1,\ldots,5$ while the $\mathbf{\overline{5}}$'s carry $SU(5)_{\perp}$ weights $\lambda_I+\lambda_J$ for $I\ne J$.  The orbits of these fields under the monodromy group correspond to the distinct $\mathbf{10}$'s and $\mathbf{\overline{5}}$'s that are engineered.  We label these by
\begin{equation}
\ba
{\bf 10}^{(1)} &= \{ \lambda_1, \lambda_2, \lambda_3\} \cr
{\bf 10}^{(2)} &= \{\lambda_4, \lambda_5 \} \cr
{\bf \bar{5}}^{(1)}  &= \{ \lambda_i + \lambda_j \,, \  i, j = 1,2,3 \} \cr
{\bf \bar{5}}^{(2)}  &= \{ \lambda_4+ \lambda_5  \} \cr
{\bf \bar{5}}^{(1)(2)}  &= \{ \lambda_i + \lambda_a \,, \  i, j = 1,2,3\,,\ a=4,5 \}  \,.
 \ea
\end{equation}
These weight assignments allow us to directly determine the allowed Yukawa couplings
\be
\ba
&{\bf 10}^{(1)} \times {\bf \bar{5}}^{(1)} \times {\bf \bar{5}}^{(2)} \,,\qquad
{\bf 10}^{(2)} \times {\bf \bar{5}}^{(1)} \times {\bf \bar{5}}^{(1)(2)} \,,\qquad
{\bf 10}^{(1)} \times {\bf \bar{5}}^{(1)(2)} \times {\bf \bar{5}}^{(1)(2)} \,,\cr
&{\bf 10}^{(1)} \times {\bf 10}^{(1)} \times {\bf {5}}^{(1)} \,,\quad \
{\bf 10}^{(1)} \times {\bf 10}^{(2)} \times {\bf {5}}^{(1)(2)} \,,\quad \ \,
{\bf 10}^{(2)} \times {\bf 10}^{(2)} \times {\bf {5}}^{(2)} \,.
 \ea
\ee
We can also determine the charges of these fields under the $U(1)$ factor of the $U(1)^4$ Cartan subalgebra of $SU(5)_{\perp}$ that is not removed by the monodromy.  To do this, note that this $U(1)$ is generated by the element $q$ defined as
\begin{equation}q\sim \begin{pmatrix}-2 & 0 & 0 & 0 & 0 \\
0 & -2 & 0 & 0 & 0 \\
0 & 0 & -2 & 0 & 0 \\
0 & 0 & 0 & 3 & 0 \\
0 & 0 & 0 & 0 & 3\end{pmatrix}\,.\end{equation}
This makes it easy to read off the $U(1)$ charges of all fields.  We list these in the table below, where $i,j=1,2,3$ and $a,b=4,5$
\begin{equation}\begin{array}{c|c|c|c}
\text{Matter} & \text{Spectral Cover Origin} & \text{Weights} & U(1)\text{ Charge} \\ \hline
\mathbf{10}^{(1)} & {\cal{C}}^{(1)} & \lambda_i & -2 \\
\mathbf{10}^{(2)} & {\cal{C}}^{(2)} & \lambda_a & +3 \\
\mathbf{\overline{5}}^{(1)} & {\cal{C}}^{(1)}-{\cal{C}}^{(1)} & \lambda_i+\lambda_j & -4 \\
\mathbf{\overline{5}}^{(2)} & {\cal{C}}^{(2)}-{\cal{C}}^{(2)} & \lambda_a+\lambda_b & +6 \\
\mathbf{\overline{5}}^{(1)(2)} & {\cal{C}}^{(1)}-{\cal{C}}^{(2)} & \lambda_i+\lambda_a & +1\end{array}\end{equation}
To engineer the MSSM with a $U(1)_{PQ}$ symmetry, we identify these multiplets with the usual matter and Higgs fields according to
\begin{equation}\begin{split}\mathbf{10}_M &\leftrightarrow \mathbf{10}^{(2)} \\
\mathbf{5}_H & \leftrightarrow \mathbf{5}^{(2)} \\
\mathbf{\overline{5}}_H &\leftrightarrow \mathbf{\overline{5}}^{(1)(2)} \\
\mathbf{\overline{5}}_M &\leftrightarrow \mathbf{\overline{5}}^{(1)} \,.\end{split}\end{equation}
We will often refer to matter fields corresponding to $\mathbf{10}^{(1)}$ as $\mathbf{10}_{\text{other}}$ because they do not correspond to anything in the MSSM
\begin{equation}\mathbf{10}_{\text{other}}\leftrightarrow \mathbf{10}^{(1)}\,.
\end{equation}
Getting the proper number of zero modes of each type requires the introduction of $G$-fluxes, which we do not study until the next section.  For now, however, let us note that if we manage to get the MSSM and nothing else, the only allowed superpotential couplings are the standard ones
\begin{equation}
\mathbf{10}_M\times\mathbf{10}_M\times\mathbf{5}_H\,,\qquad \mathbf{10}_M\times\mathbf{\overline{5}}_M\times\mathbf{\overline{5}}_H,.
\end{equation}
The rest are forbidden by the $U(1)$ symmetry, with respect to which MSSM fields carry charges
\be
\begin{array}{c|c}
\text{Matter} & U(1)_{PQ}   \\ \hline
\mathbf{10}_M &  +3\\
\mathbf{5}_H &  -6  \\
\mathbf{\overline{5}}_H & + 1\\
\mathbf{\overline{5}}_M & -4\\
\mathbf{10}_{\text{other}} & -2
\end{array}
\ee
As indicated, this $U(1)$ is a $PQ$ symmetry, as desired.

\subsubsection{Neutrino Physics}

Let us now make a brief digression to review how this symmetry allows the Dirac neutrino scenario of \cite{Bouchard:2009bu} to be realized, at least in principle.  Details of neutrino physics depend on the presence or absence of various types of singlet fields that can play the role of right-handed neutrinos.  In semi-local $F$-theory models, singlets are associated to weights of the $SU(5)_{\perp}$ adjoint, namely $\lambda_I-\lambda_J$ for $I\ne J$.  In general, we get three kinds of singlets in the \textbf{3+2} factored model
\begin{equation}\begin{array}{c|c|c|c}
\text{Matter} & \text{Spectral Cover Origin} & \text{Weights} & U(1)_{PQ}\text{ Charge} \\ \hline
\mathbf{1}^{(1)} & {\cal{C}}^{(1)} & \lambda_i-\lambda_j & 0 \\
\mathbf{1}^{(2)} & {\cal{C}}^{(2)} & \lambda_a-\lambda_b & 0 \\
\mathbf{1}^{(1)(2)} & {\cal{C}}^{(1)}-{\cal{C}}^{(2)} & \lambda_a-\lambda_i & +5
\end{array}\end{equation}
where again $i,j=1,\ldots 3$ and $a,b=4,5$.  In these models, there is no singlet field $N_R$ with the right charges to allow the renormalizable operator
\begin{equation}\int\,d^2\theta\,H_u L N_R\,.
\label{Diractree}\end{equation}
This means that standard Majorana scenarios cannot be realized in these models.  Nevertheless, the absence of the tree-level Dirac mass \eqref{Diractree} makes possible a realization of the Dirac scenario of \cite{Bouchard:2009bu}.  This is because the singlet field $\mathbf{1}^{(1)(2)}$ carries the right charges to allow the dimension 5 operator
\begin{equation}\label{DiracNeutrino}
\frac{1}{\Lambda}\int\,d^4\theta\,H_d^{\dag} L\,\left( {\mathbf{1}^{(1)(2)}}\right)\,,
\end{equation}
which generates a small Dirac mass term because $H_d^{\dag}$ picks up a $\bar{\theta}^2$-component expectation value in the presence of a nonzero $\mu$ term \cite{Bouchard:2009bu}.  When the electroweak scale Dirac mass term \eqref{Diractree} is forbidden, this suppressed contribution can be the leading one and is capable of producing neutrinos that are naturally light.


\subsection{Matter curves and Yukawas in $S_{\rm GUT}$}

We now turn to a more explicit description of matter curves and Yukawas in models that exhibit the factorization \eqref{32fact}.  To gain some insight into the structure of matter curves, we will first study them directly in $S_{\rm GUT}$. For a detailed analysis including fluxes, we will require a more refined description of these inside the spectral cover, which we will provide in the next subsection.

Recall from section  \ref{subsec:Defs}, that the  ${\bf 10}$ matter curves are defined by the locus
\be
\Sigma_{\bf 10}:\qquad  b_5 = a_3 e_2 =0 \,.
\ee
Thus, there will be two types of ${\bf 10}$ matter curves, one from each factor of the spectral cover:
\be
{\bf 10}_M:\qquad e_2 =0  \qquad \hbox{or} \qquad a_3=0\,.
\ee
Likewise, the ${\bf 5}$ matter arises from the locus $P=0$, where $P$ was defined in the table in section  \ref{subsec:Defs}.
We expect $P$ to automatically factor into three components in the ${\bf 3+2}$ model: there are ${\bf \bar{5}}$ matter curves $\lambda_i + \lambda_j$, $\lambda_a + \lambda_b$ and $\lambda_i + \lambda_a$, where $i,j$ label sheets of the $\mathcal{C}^{(1)}$ component and $a,b,$ of $\mathcal{C}^{(2)}$.
Indeed, using the particular form of the coefficients $b_m$ in this model, as well as the solution (\ref{b1Sol}) for $b_1=0$ we find
\be
P =  {1728 }\prod_{i=1}^3 P_i \,,
\ee
where
\be\label{PFactors}
\ba
P_1& =  e_1\cr
P_2 &= \left(a_3 e_0+a_2 e_1\right) \cr
P_3 &= \left(a_3 e_1 \left(a_2-e_2 \alpha \right)+e_2 \left(a_2-e_2 \alpha \right){}^2+a_3^2 e_0\right) \,.
\ea
\ee
We will identify the origin of these matter curves in the spectral cover shortly.

Finally, the Yukawa couplings are characterized by the following loci, where we also specify in the brackets, which matter curves participate in the couplings:
\be\ba
SO(12): &\qquad \quad
\ba
e_2=a_3 e_0+a_2 e_1 =0: & \quad   \{ P_2, P_3, b_5 \} \cr
a_3=e_1= 0: & \quad    \{P_1 , P_2, b_5 \} \cr
\ea
\cr & \cr
E_6: &\qquad \quad
\ba
a_3 = a_2= 0 :&\qquad  \{P_2, b_5 \} \cr
e_2= e_1 =0 : & \qquad \{P_1, b_5 \} \cr
e_2 = a_3= 0 :&\qquad  \{P_3, b_5 \}\,.
\ea
\ea
\ee



\subsection{Matter Curves in the Spectral Cover}
\label{subsec:MatterinSC}

We can realize the matter curves (in particular the ${\bf 5}$ matter curves) in the spectral cover by the method of \cite{Marsano:2009gv}, where  we showed that they are components of  intersection of $\mathcal{C}$ with its image under the
action of $\tau:\ V\rightarrow -V$. This analysis was inspired by studies of matter curves in models with heterotic duals \cite{Donagi:2004ia, Blumenhagen:2006wj, Hayashi:2008ba}. 
In appendix \ref{app:MatterinSC} we study the various components of $\mathcal{C} \cap \tau \mathcal{C}$ and determine the  lift of the matter curves into the spectral cover.

In summary we obtained the following distribution of ${\bf 5}$
matter curves \be \ba \mathcal{C}^{(1)} \cap \tau\mathcal{C}^{(1)}
\,:&\qquad P_2 \qquad\hbox{in} \qquad 3c_1 -  t  \cr
\mathcal{C}^{(1)} \cap \tau\mathcal{C}^{(2)} \,:&\qquad P_3
\qquad\hbox{in} \qquad  4c_1 - 2 t -\xi \cr \mathcal{C}^{(2)} \cap
\tau\mathcal{C}^{(2)} \,:&\qquad P_1 \qquad\hbox{in} \qquad c_1 +
\xi, \ea \ee where in the last column we give the class of the
projection of the corresponding curve to the base $S_{\rm GUT}$. The
${\bf 10}$ matter curve is read off from $U=0$ \be \ba
\mathcal{C}^{(2)} \cap \tau\mathcal{C}^{(2)} \,:&\qquad e_2
\qquad\hbox{in} \qquad \xi \,. \ea \ee


\subsection{Embedding in a Global Model}

Having determined the general properties of the ${\bf 3+2}$ factored model, we will give an explicit realization of it in the geometry of \cite{Marsano:2009ym}.
For $B_3$ of \cite{Marsano:2009ym}, the GUT-surface is $S_{\rm GUT}=
dP_2$ and 
\be 
c_1 = 3 h - e_1 - e_2 \,,\qquad t= -c_1(N_{S_{\rm GUT}/B_3}) = h \,. 
\ee 
Recall further that 
\be
\eta = 6c_1 - t = 17 h- 6 e_1 - 6 e_2\,. \ee 
Finally, we have the freedom to choose the class
$\xi$. A convenient such choice is \be \xi = h - e_1 \,. \ee We make this
choice because it is not symmetric in the two exceptional classes
$e_1$ and $e_2$.  In order to realize a $U(1)_{PQ}$ symmetry, it is
necessary for MSSM matter fields associated to different components of
the spectral cover to carry a nontrivial restriction of the $U(1)_Y$ flux, which
is in the class
\be [F_Y] = e_2 - e_1 \,. \ee
If $\xi$ were symmetric in $e_1$ and $e_2$, we would not have any
matter curves to which $F_Y$ restrictricts nontrivially.

Key properties of the ${\bf 3+2}$ model realized in this geometry are listed in the table below:

\be
\begin{tabular}{c|c|c|c|c}
\hbox{Field} & \hbox{Section} & \hbox{Class in }$S_{\rm GUT}$ & \hbox{Special Choice of Classes} & \hbox{Restriction of }$\mathcal{L}_Y$\cr
\hline
${\bf 10}_M $       &$ e _2$&$  \xi$                    &  $ h-e_1$             & $-1$  \cr
${\bf 5}_H $        & $P_1$ &$ 3h - e_1 - e_2  + \xi$       & $4 h - 2 e_1 -  e_2$  &  $-1$  \cr
${\bf \bar{5}}_M $  & $P_2 $& $8 h - 3 e_1 - 3 e_2  $   & $8h - 3 e_1 - 3 e_2$      &  $0$\cr
${\bf \bar{5}}_H $  & $P_3 $& $10h - 4 e_1 -4 e_2 -\xi $        &$9 h -  3 e_1 - 4 e_2$ & $ +1$ \cr
${\bf 10}_{\rm other}$ & $a_3$ & $2h - e_1 - e_2 - \xi$ & $h-e_2$ & $+1$
\end{tabular}
\ee

Because the $U(1)_Y$ flux restricts nontrivially to the $\mathbf{10}_M$ and $\mathbf{10}_{\text{other}}$ matter curves, we will necessarily get some exotics there.  With these choices, we have managed to avoid $U(1)_Y$ flux on the $\mathbf{\overline{5}}_M$ matter curve so no exotics will live there.  It is important to note that with this choice, all holomorphic sections in \eqref{bmae} are well-defined.


\section{Chiral spectrum from G-fluxes}
In this section we describe the construction of consistent $G$-fluxes in the models of section \ref{sec:model} and find combinations which, together with $U(1)_Y$
flux $[F_Y]=e_2-e_1$ on $S_{GUT}$, give rise to models with three
generations plus the minimal number of vector-like pairs of exotics necessary to obtain a $U(1)_{PQ}$ symmetry{\footnote{Such exotics are also required for one proposed solution to the unification problem in $F$-theory GUTs  \cite{Blumenhagen:2008aw}, which we will discuss in more detail below.}}.
\subsection{G-fluxes from spectral cover and flux quantization}
Let us first clarify what we mean by G-fluxes. As we reviewed in
Section 2, $CY_4$ is locally described by a deformed $E_8$ singularity fibered
over $S_{GUT}$ with 2-cycles $\lambda_i$, $i=1,\ldots,5$, non-degenerate except over isolated curves in $S_{\rm GUT}$\footnote{In a standard
manner we abuse notations by using $\lambda_i$ to refer to both the 2-cycle in
the fiber and the curve in $S_{GUT}$ over which this 2-cycle
vanishes.}.

G-fluxes are $U(1)$ gauge fluxes that arise from the dual  M-theory
flux:
\be \label{gflux} G=\sum_{i=1}^5 \omega_i \wedge F_{(i)} \,,
\ee
where $F_{(i)}$ is a flux on $S_{\rm GUT}$ in  $U(1)\in
SU(5)_{\perp}$ and $\omega_i$ is  a harmonic $(1,1)$ form
 in the $E_8$ fiber dual to $\lambda_i$:
\be \label{harmon}
\int_{\lambda_i}\omega_j=\delta_{ij}\,.
\ee
Because the $\lambda_i$ are Cartan elements of $SU(5)_{\perp}$, they satisfy
\be \label{trace} \sum_{i=1}^5\lambda_i=0\,.
\ee
 In \cite{Donagi:2009ra} it was shown that an
efficient treatment of G-fluxes, taking into account monodromies that act on the $\lambda_i$,
is in terms of `spectral cover fluxes'.
Namely, one considers line bundles $\mathcal{L}_a$ over components
of the spectral surface $\mathcal{C}^{(a)}$ for $a=1,2$.  We will specify these line bundles by specifying holomorphic curves in ${\cal{C}}^{(a)}$ to which they are dual.

To be consistent, spectral cover fluxes must satisfy a number of important constraints.  These are
\begin{itemize}
\item $0 = c_1(p_{1*}{\cal{L}}_1) + c_1(p_{2*}{\cal{L}}_2)$
\item $c_1({\cal{L}}_a)\in H^{1,1}\left({\cal{C}}^{(a)},\mathbb{Z}\right)$
\item $c_1(p_{1*}{\cal{L}}_1)-c_1(p_{2*}{\cal{L}}_2)$ is a PD of a supersymmetric cycle in $S_{\rm GUT}$
\end{itemize}
where $p_a$ denotes the projection map from ${\cal{C}}^{(a)}$ to $S_{\rm GUT}$
\begin{equation}p_a:\quad {\cal{C}}^{(a)}\rightarrow S_{\rm GUT} \,.
\end{equation}
To understand these constraints, recall that we locally identify the five sheets of the cover with five elements $\lambda_i$ of the Cartan subalgebra of $SU(5)_{\perp}$ that satisfy the traceless constraint $\sum_i\lambda_i=0$.  The first condition above is then the statement that the flux is traceless so that it is really an $SU(5)_{\perp}$ flux rather than a $U(5)$ flux{\footnote{More specifically, for our factored example it is an $S[U(3)\times U(2)]$ flux.}}.  The second constraint is simply the statement that spectral cover fluxes must be properly quantized.  The third constraint is one that we noticed when checking consistency conditions for the induced D3-brane tadpole{\footnote{This constraint is not one that has appeared before in the literature.  We have not explicitly derived it so it should be viewed as a proposal for the time being.}}.  It arises because the difference $c_1(p_{1*}{\cal{L}}_1)-c_1(p_{2*}{\cal{L}}_2)$ has the interpretation of a flux associated to the $U(1)$ symmetry that is not projected out by the monodromy group.  In our models, this is simply the net $U(1)_{PQ}$ flux.  One might expect that because $U(1)_{PQ}$ is not affected by the monodromy group, it is possible to turn on any arbitrary $U(1)_{PQ}$ flux and indeed we will find such apparent freedom in the explicit construction of fluxes to follow.  Any such flux $F$, however, must satisfy the constraint of being supersymmetric by which we mean that
\begin{equation}\omega_{S_{\rm GUT}}\cdot_{S_{\rm GUT}} F=0\,,
\end{equation}
where $\omega_{S_{\rm GUT}}$ is the restriction of the K\"ahler form to $S_{\rm GUT}$.  We will not discuss K\"ahler moduli in detail in this paper but instead will simply require that the $U(1)_{PQ}$ flux be supersymmetric with respect to some element in the K\"ahler cone on $S_{\rm GUT}$ which, for us, is $dP_2$.  The K\"ahler cone on $dP_2$ is easily specified as consisting of $\omega$ of the form
\begin{equation}\omega = \alpha_1 e_1 + \alpha_2 e_2 + \beta (h-e_1-e_2)\,,\qquad\text{where}\qquad \alpha_1,\alpha_2 < \beta < \alpha_1+\alpha_2\,.
\label{Kform}\end{equation}
We will in fact require that $\omega$ be the restriction of a well-defined K\"ahler form from $B_3$ which, in turn, requires that $\alpha_1=\alpha_2$.

Let us now rewrite the above constraints in a way that is more familiar from previous work.  Following \cite{Donagi:2009ra}, it is conventional to use the Grothendieck-Riemann-Roch theorem to rewrite the trace condition as
\begin{equation}0 = p_{1*}(c_1({\cal{L}}_1)) - \frac{1}{2}p_{1*}r_1 + p_{2*}(c_1({\cal{L}}_2))-\frac{1}{2}p_{2*}r_2\,.
\label{FluxQ}\end{equation}
The objects $r_a$ here are the ramification divisors,
\begin{equation}r_a = p_a^*c_1 - c_1(T_{ {\cal{C}}^{(a)} })\,,\end{equation}
where again $c_1=c_1(S_{\rm GUT})$ and
\be c_1 (T_{\mathcal{C}^a}) = \left(c_1 (T_X) -
[\mathcal{C}^{(a)}] \right)\cdot [{\mathcal{C}^{(a)}}] \,.
\ee
To address the trace condition \eqref{FluxQ}, it is conventional to decompose
\be c_1 (\mathcal{L}_a) = \gamma_a +
{1\over 2} r_a \,, \ee where the $\gamma_a$ satisfy
\be p_{1*} \gamma_1 +   p_{2*} \gamma_2 =0 \,. \ee
To construct traceless spectral cover fluxes, then, we need only define a traceless combination of $\gamma_i$'s.

As for the quantization constraint, we need
\begin{equation}\gamma_a+\frac{1}{2}r_a \in H_2\left({\cal{C}}^{(a)},\mathbb{Z}\right)
\label{FluxQuant}\end{equation}
To study what this means for the $\gamma_a$ in our specific models, it is necessary to compute the ramification divisors $r_a$ explicitly.  Since $X=\mathbb{P}\left({\cal{O}}_{S_{\rm GUT}}\oplus K_{S_{\rm GUT}}\right)$ we have that
\begin{equation}c_1(T_X) = 2\sigma_{\infty}\equiv 2 (\sigma+\pi^*c_1)\,.\end{equation}
This allows us to
compute \be \ba
 c_1 (T_{\mathcal{C}^{(1)}}) &= \left.\left[ - (3\sigma + \pi^* [\eta - 2c_1- \xi] ) + 2 \sigma_{\infty} \right]\right|_{\mathcal{C}^{(1)}}\cr
  c_1 (T_{\mathcal{C}^{(2)}}) & = \left. \left[- (2\sigma + \pi^* [2c_1 + \xi] ) + 2 \sigma_\infty \right]\right|_{\mathcal{C}^{(2)}} \,.
\ea \ee
From this follows that the $r_a$ are given by
\be \label{RamDiv}\ba
r_1 =&
\left.\left[  \sigma + \pi^* (\eta - 3 c_1-\xi)
\right]\right|_{\mathcal{C}^{(1)}}\cr r_2 =&  \left.\left[  \pi^*
(c_1 + \xi)  \right]\right|_{\mathcal{C}^{(2)}} \,. \ea \ee

Finally, the supersymmetry condition that we impose is the statement that
\begin{equation}p_{1*}\gamma_1-p_{2*}\gamma_2 + \frac{1}{2}\left(p_{1*}r_1-p_{2*}r_2\right)\end{equation}
is a supersymmetric cycle in $S_{\rm GUT}$.  The object $p_{1*}r_1-p_{2*}r_2$ is simply the class $e_1-e_2$ in our models which is always supersymmetric because K\"ahler forms that descend from globally well-defined objects in $B_3$ are of the form \eqref{Kform} with $\alpha_1=\alpha_2$.  For this reason, we will require in the following that
\begin{equation}p_{1*}\gamma_1-p_{2*}\gamma_2\end{equation}
be supersymmetric{\footnote{It may be that this condition, rather than $c_1(p_{1*}{\cal{L}}_1)-c_1(p_{2*}{\cal{L}}_2)$ is the correct one to start with.  In our models, we cannot distinguish them but we believe the former seems more sensible.}} in $S_{\rm GUT}$.

\subsection{Constructing Fluxes}

We now turn to the construction of holomorphic curves that can be used to
specify the fluxes $\gamma_a$.  These curves, which we often refer
to directly as fluxes in an abuse of language, fall into two
classes.  The first class are the so-called universal fluxes, which
are present for any generic (factored) spectral cover.  These can be
written as the intersections of divisors in $X$ and their
restrictions to matter curves are easily computed.  To get nice
3-generation models with our base manifold, though, it will be
necessary to introduce non-universal fluxes, which are only present
when the spectral cover is sufficiently tuned.  In this subsection,
we will describe both types of fluxes in turn and present
intersection tables.  Intersections involving non-universal fluxes
are somewhat involved to compute and depend on our detailed
realization of the semi-local model in the global completion based
on $B_3$.  We therefore defer such computations to Appendix
\ref{app:fluxes}.


\subsubsection{Universal Fluxes}

Universal fluxes arise as intersections of divisors in $X$ with various components of the spectral surface.  In general, these take the form
\begin{equation}\gamma_i=\sigma\cdot {\cal{C}}^{(i)}\qquad\text{or}\qquad p_i^{\ast}\Sigma = \pi^{\ast}(\Sigma)\cdot {\cal{C}}^{(i)}\,,
\end{equation}
where $\Sigma$ refers to any curve inside $S_{\rm GUT}$.  From these building blocks, we can build a flux on ${\cal{C}}^{(1)}$ that is traceless and a flux on ${\cal{C}}^{(2)}$ that is traceless
\begin{equation}\tilde{\gamma}_1 = 3\gamma_1 - p_1^*p_{1*}\gamma_1\,,
\qquad\qquad \tilde{\gamma}_2 = 2\gamma_2-p_2^*p_{2*}\gamma_2\,.
\end{equation}
We can also consider combinations of fluxes on ${\cal{C}}^{(1)}$ and ${\cal{C}}^{(2)}$ that are not individually traceless in the sense that they do not satisfy $p_{1*}\gamma=0$ or $p_{2*}\gamma=0$ but whose combination satisfies the full traceless condition \eqref{FluxQ}.  Two fluxes of this type can be constructed as
\begin{equation}\delta_1 = 2\sigma\cdot {\cal{C}}^{(1)} - p_2^*p_{1*}\left(\sigma\cdot {\cal{C}}^{(1)}\right)\,,\qquad\qquad \delta_2 = 3\sigma\cdot {\cal{C}}^{(2)} - p_1^*p_{2*}\left(\sigma\cdot {\cal{C}}^{(2)}\right)\,.
\end{equation}
A third flux of this type takes the form
\begin{equation}\tilde{\rho} = 2p_1^*\rho - 3p_2^*\rho \,,
\end{equation}
for any $\rho\in H_2(S_{\rm GUT},\mathbb{R})$.  Note that $\rho$ does not have to be an effective class or even a combination of effective classes in general.  This is because we can build a flux of the form $\tilde{\rho}$ with any real linear combination $\tilde{\rho} = \sum_i a_i \tilde{\rho}_i$ of similar fluxes built from effective $\rho_i$.  Ultimately, any flux involving $\tilde{\rho}$ will have to satisfy the quantization condition (\ref{FluxQuant}), but this will depend on other types of fluxes present as well as the ramification divisors $r_1$ and $r_2$.

In total, then, any model in which the spectral surface factors into
a cubic and quadratic factor admits a universal flux of the form
\begin{equation}\gamma_u = \tilde{k}_1\tilde{\gamma}_1 + \tilde{k}_2\tilde{\gamma}_2 + \tilde{\rho} + \tilde{d}_1\delta_1 + \tilde{d}_2\delta_2\,.
\end{equation}
Using our expressions for the ramification divisors (\ref{RamDiv}), we can write the quantization conditions as
\begin{equation}\begin{split}\left(3\tilde{k}_1+2\tilde{d}_1+\frac{1}{2}\right)\sigma - \pi^*\left[\tilde{k}_1(\eta-5c_1-\xi)+\tilde{d}_2\xi -2\rho - \frac{1}{2}(\eta-3c_1-\xi)\right]&\in H_4(X,\mathbb{Z}) \\
\left(2\tilde{k}_2+3\tilde{d}_2\right)\sigma - \pi^*\left[\tilde{k}_2\xi + \tilde{d}_1(\eta-5c_1-\xi)+3\rho - \frac{1}{2}(c_1+\xi)\right]&\in H_4(X,\mathbb{Z})\,.
\end{split}\end{equation}
The supersymmetry condition, on the other hand, amounts to the requirement that
\be
\omega\cdot_{S_{\rm GUT}}\left[2\tilde{d}_1(\eta-5c_1-\xi)-3\tilde{d}_2\xi + 6\rho\right]=0\,,
\ee
for $\omega$ a suitable K\"ahler form on $S_{\rm GUT}$ \eqref{Kform}.

Now that we have summarized the consistency conditions, let us turn to the chiral spectrum induced by $\gamma_u$ on each of our matter curves.  In Appendix \ref{app:fluxes} we present general formulae for this in terms of the classes $c_1$, $\eta$, and $\xi$.  Below, we write the results for the specific model of section \ref{sec:model} with the notation that
\begin{equation}\rho = \tilde{X}h - \tilde{Y}e_1 - \tilde{Z}e_2 \,.
\end{equation}
The induced chiralities are then as follows:

\begin{equation}\begin{array}{c|c|c|c}
\text{Matter} & \text{Origin} & \text{Class in }S_{\rm GUT} & \text{Chirality induced by }\gamma_u \\ \hline
\mathbf{10}_M & 2-2 & h-e_1 & -4\tilde{k}_2-\tilde{d}_1-6\tilde{d}_2-3(\tilde{X}-\tilde{Y}) \\
\mathbf{5}_H & 2-2 & 4h-2e_1-e_2 & -6\tilde{d}_1+6\tilde{d}_2-6(4\tilde{X}-2\tilde{Y}-\tilde{Z}) \\
\mathbf{\overline{5}_H} & 1-2 & 9h-3e_1-4e_2 & -2\tilde{k}_1-4\tilde{k}_2-3\tilde{d}_1-3\tilde{d}_2-(9\tilde{X}-3\tilde{Y}-4\tilde{Z}) \\
\mathbf{\overline{5}}_M & 1-1 & 8h-3(e_1+e_2) & -4\tilde{k}_1+4\tilde{d}_1-10\tilde{d}_2+4(8\tilde{X}-3\tilde{Y}-3\tilde{Z}) \\
\mathbf{10}_{\text{other}} & 1-1 & h-e_2 & -6\tilde{k}_1-4\tilde{d}_1-\tilde{d}_2+2(\tilde{X}-\tilde{Z})
\end{array}\end{equation}
Unfortunately, it is impossible to simultaneously satisfy the quantization constraints while ensuring no net chirality on the $\mathbf{5}_H$ matter curve{\footnote{When $\gamma_u$ does not  induce any net chirality on the $\mathbf{5}_H$ curve, the $U(1)_Y$ flux ensures that a single $H_u$ doublet and no triplets are engineered.  On the other hand, if $\gamma_u$ does induce a net chirality then we will obtain exotics that cannot be lifted through the expectation value of a singlet field that carries $U(1)_{PQ}$ charge.}}.  For this reason, we turn now to non-universal fluxes.

\subsubsection{Non-universal Fluxes}

Under certain restrictions on complex structure of Calabi-Yau
4-fold, it is possible to construct additional fluxes that do not
arise as intersections of ${\cal{C}}^{(i)}$ with divisors in $X$.
Again, we explain all the details of the analysis in appendix
\ref{app:fluxes}, and summarize the results here.

We construct non-universal fluxes by choosing a curve $\psi$ in $S_{\rm GUT}$ and then lifting it to a well-defined curve that roughly sits on only a single sheet of either ${\cal{C}}^{(1)}$ or ${\cal{C}}^{(2)}$.  More specifically, we define the following curves in $X$
\begin{equation}\begin{split}\Psi_1 \quad\sim &\quad V+gU=0\ \text{ and }\ \psi=0 \\
\Psi_2\quad\sim &\quad V-gU=0\ \text{ and }\ \psi=0\,,\end{split}\end{equation}
and tune the coefficients of ${\cal{C}}^{(a)}$ so that $\Psi_a\subset {\cal{C}}^{(a)}$.  From $\Psi_1$ and $\Psi_2$ we can construct two fluxes that are independently traceless inside ${\cal{C}}^{(1)}$ and ${\cal{C}}^{(2)}$
\begin{equation}\tilde{\Psi}_1=\left(3 - p_1^{\ast}p_{1\,\ast}\right)\Psi_1\,,\qquad
 \tilde{\Psi}_2=\left(2-p_2^{\ast}p_{2\,\ast}\right)\Psi_2\,.\end{equation}
 The difference is a third flux that also satisfies the net trace constraint \eqref{FluxQ}
 \begin{equation}\Delta = \Psi_1-\Psi_2\,.
 \end{equation}
 The triple $\tilde{\Psi}_1$, $\tilde{\Psi}_2$, and $\Delta$ comprise a convenient basis for the non-universal fluxes that can be introduced when the ${\cal{C}}^{(a)}$ are tuned to contain the $\Psi_a$.

 We now present the chirality induced by $\tilde{\Psi}_1$, $\tilde{\Psi}_2$, and $\Delta$ on our various matter curves.  For this, we focus on the models of section \ref{sec:model} and denote the class $\psi$ by
 \begin{equation}\psi = \tilde{A}h - \tilde{B}e_1 - \tilde{C}e_2\,.\end{equation}
 Following appendix \ref{app:fluxes} we find
\begin{equation}\begin{array}{c|c|c|c|c|c}
\text{MC} & \text{Origin} & \text{Class in }S_{\rm GUT} & \tilde{\Psi}_1 & \tilde{\Psi}_2 & \Delta \\ \hline
\mathbf{10}_M & 22 & h-e_1& 0 & \tilde{B}-\tilde{A} & 0 \\
\mathbf{5}_H & 22 & 4h-2e_1-e_2 & 0 & 0 & -(4\tilde{A}-2\tilde{B}-\tilde{C}) \\
\mathbf{\overline{5}}_H & 12 & 9h-3e_1-4e_2 & -6\tilde{A}+4\tilde{C} & -\tilde{A}+\tilde{B} & -3\tilde{A}+2\tilde{C} \\
\mathbf{\overline{5}}_M & 11 & 8h-3(e_1+e_2) & 5\tilde{A}-3\tilde{C} & 0 & 7\tilde{A}-2\tilde{B}-3\tilde{C} \\
\mathbf{10}_{\text{other}} & 11 & h-e_2 & -\tilde{A}+\tilde{C} & 0 & 0
\end{array}\end{equation}

\subsubsection{Total Flux}

Now, we construct a total flux as

\begin{equation}\label{TotalGamma}
\Gamma = \tilde{k}_1\tilde{\gamma}_1 + \tilde{k}_2\tilde{\gamma}_2 + \tilde{\rho} + \tilde{d}_1\delta_1 + \tilde{d}_2\delta_2 + \tilde{m}_1\tilde{\Psi}_1 + \tilde{m}_2\tilde{\Psi}_2 + \tilde{q}\Delta\,.
\end{equation}
With this choice, we can identify the total fluxes, $\Gamma_1$ and $\Gamma_2$, on the components ${\cal{C}}^{(1)}$ and ${\cal{C}}^{(2)}$ as
\begin{equation}\begin{split}\label{Gamma1Gamma2}
\Gamma_1 &= \left(3\tilde{m}_1 + \tilde{q}\right)\Psi_1 + {\cal{C}}^{(1)}\cdot \left\{\left(3\tilde{k}_1 + 2\tilde{d}_1\right)\sigma - \pi^{\ast}\left[\tilde{k}_1(\eta-5c_1-\xi)+\tilde{d}_2\xi - 2\rho + \tilde{m}_1\psi \right]\right\} \\
\Gamma_2 &= \left(2\tilde{m}_2 - \tilde{q}\right)\Psi_2 + {\cal{C}}^{(2)}\cdot \left\{\left(2\tilde{k}_2 + 3\tilde{d}_2\right)\sigma - \pi^{\ast}\left[\tilde{k}_2\xi + \tilde{d}_1 (\eta-5c_1-\xi)+3\rho + \tilde{m}_2\psi \right]\right\} \,.
\end{split}\end{equation}
For the spectrum we find
\begin{equation}\begin{array}{c|c|c|c}
\text{MC} & \text{Origin} & \text{Class in }S_{\rm GUT} & \Gamma \\ \hline
\mathbf{10}_M & 22 & h-e_1 & -4\tilde{k}_2 - \tilde{d}_1 -6\tilde{d}_2 -3(\tilde{X}-\tilde{Y}) + \tilde{m}_2(\tilde{B}-\tilde{A}) \\
& & & \\
\mathbf{5}_H & 22 & 4h-2e_1-e_2 & -6\tilde{d}_1 + 6\tilde{d}_2 -6 (4\tilde{X}-2\tilde{Y}-\tilde{Z})-\tilde{q}(4\tilde{A}-2\tilde{B}-\tilde{C}) \\
& & & \\
\mathbf{\overline{5}}_H & 12 & 9h-3e_1-4e_2 & -2\tilde{k}_1 -4\tilde{k}_2 -3\tilde{d}_1 - 3\tilde{d}_2 - (9\tilde{X}-3\tilde{Y}-4\tilde{Z}) \\
& & & +\tilde{m}_1(4\tilde{C}-6\tilde{A}) + \tilde{m}_2 (\tilde{B}-\tilde{A}) + \tilde{q}(2\tilde{C}-3\tilde{A}) \\
& & & \\
\mathbf{\overline{5}}_M & 11 & 8h-3(e_1+e_2) & -4\tilde{k}_1+4\tilde{d}_1 - 10\tilde{d}_2 +4(8\tilde{X}-3\tilde{Y}-3\tilde{Z}) \\
& & & +\tilde{m}_1(5\tilde{A}-3\tilde{C}) + \tilde{q}(7\tilde{A}-2\tilde{B}-3\tilde{C}) \\
& & & \\
\mathbf{10}_{\text{other}} & 11 & h-e_2 & -6\tilde{k}_1 -4\tilde{d}_1 - \tilde{d}_2 + 2(\tilde{X}-\tilde{Z})+\tilde{m}_1(\tilde{C}-\tilde{A})
\end{array}\end{equation}
where we take
\begin{equation}\rho = \tilde{X}h-\tilde{Y}e_1-\tilde{Z}e_2\,,\qquad\qquad \psi = \tilde{A}h-\tilde{B}e_1-\tilde{C}e_2\,.
\end{equation}

The quantization conditions now read
\begin{equation}{\boxed{\begin{split} 3\tilde{m}_1 + \tilde{q}&\in\mathbb{Z}\\
2\tilde{m}_2 - \tilde{q}&\in\mathbb{Z} & \\
\left(3\tilde{k}_1 + 2\tilde{d}_1+\frac{1}{2}\right)\sigma - \pi^{\ast}\left[\tilde{k}_1(\eta-5c_1-\xi)+\tilde{d}_2\xi - 2\rho + \tilde{m}_1\psi -\frac{1}{2}(\eta-3c_1-\xi)\right] &\in H_4(X,\mathbb{Z})\\
\left(2\tilde{k}_2+3\tilde{d}_2\right)\sigma - \pi^{\ast}\left[\tilde{k}_2\xi + \tilde{d}_1(\eta-5c_1-\xi)+3\rho+\tilde{m}_2\psi -\frac{1}{2}(c_1+\xi)\right]&\in H_4(X,\mathbb{Z})
\end{split} }} \end{equation}

To study supersymmetry conditions, we write
\begin{equation}\begin{split}p_{1\,\ast}\Gamma_1 &= 2\tilde{d}_1(\eta-5c_1-\xi)-3\tilde{d}_2\xi + 6\rho + \tilde{q}\psi \\
p_{2\,\ast}\Gamma_2 &= -2\tilde{d}_1(\eta-5c_1-\xi)+3\tilde{d}_2\xi - 6\rho - \tilde{q}\psi\,.
\end{split}\end{equation}
This implies that the supersymmetry condition is that
\begin{equation}{\boxed{\omega\cdot_{S_{\rm GUT}} \left(2\tilde{d}_1(\eta-5c_1-\xi)-3\tilde{d}_2\xi + 6\rho + \tilde{q}\psi\right)=0}}\end{equation}
for some suitable K\"ahler form $\omega$ on $S_{\rm GUT}$ \eqref{Kform}.


\subsection{D3-brane tadpole}

As we are turning on non-trivial $G$-flux we need to carefully analyze, whether the D3-tadpole condition can be satisfied without requiring the introduction of anti-D3-branes, which would correspond to an uncontrolled source of supersymmetry breaking.
Recall that the tadpole condition for D3-branes is
\be
N_{D3} = {\chi (X_4) \over 24} - {1\over 2} \int_{X_{4}} G \wedge G \,,
\label{D3tad}\ee
where $X_4$ is the elliptically fibered CY fourfold with base $B_3$.
In particular, in order to have a positive number of D3-branes, it is crucial that the $G$-flux contribution does not overshoot the Euler characteristic.

\subsubsection{Geometric contribution}

The Euler characteristic for smooth elliptic fibrations over our base 3-fold $B_3$ \cite{Marsano:2009gv} is
\be
N_{D3, {\rm base}} = - {\chi^* (X_4) \over 24} = - 582 \,.
\ee
In the presence of non-abelian singularities, the Euler characteristic obtains additional contributions \cite{Blumenhagen:2009yv}\footnote{We are grateful to R. Blumenhagen, T. Grimm and T. Weigand for bringing this to our attention.}. If the local enhancement over the GUT surface $S_{\rm GUT}$ is to the gauge group  $H$, and we use  $G$ to denote the complement of $H$ in $E_8$, the geometric Euler characteristic is modified to
\be\label{RefinedChi}
\chi (X_4) = \chi^* (X_4) + \chi_G - \chi_{E_8} \,,
\ee
and the contribution to the D3-tadpole is
\be
N_{D3, {\rm geo}} = - {\chi(X_4)\over 24 }\,.
\ee
For us, the relevant values are
\be\label{ChiSUE8}
\ba
\chi_{SU(n)} &= \int_{S_{\rm GUT}} \left( c_1^2 \left(n^3-n\right)+3 n \eta  \left(\eta -n c_1 \right)  \right) \cr  
\chi_{E_8}    &=   \int_{S_{\rm GUT}}   120\left(-27 c_1 \eta +62 c_1^2+3 \eta ^2\right) \,.
\ea
\ee
For an unfactored spectral cover for $SU(5)$ we obtain 
\be
\underline{\hbox{Unfactored cover:}} \qquad \chi (X_4) = \chi^*(X_4) + \chi_{SU(5)} - \chi_{E_8} \,.
\ee
For the ${\bf 3+2}$ factored spectral cover, we specify a $S[U(3)\times U(2)]$ bundle so the refined Euler character is
\be
\underline{\bf 3+2}:\qquad \chi (X_4) =  \chi^*(X_4) + \chi^{(1)}_{SU(3)} + \chi^{(2)}_{SU(2)} - \chi_{E_8} \,,
\ee
where in  $\chi^{(i)}$ we have to replace $\eta$ by the respective class $\eta^{(i)}$ in the $i$th factor of the spectral cover $\mathcal{C}$, i.e.
\be
\eta^{(1)} =  \eta - (2 c_1 + \xi)\,,\qquad 
\eta^{(2)} = 2 c_1 + \xi \,.
\ee
In summary we obtain for the factored case 
\be\label{32Chi}
\chi (X_4) = \chi^*(X_4) +\int_{S_{\rm GUT}} \left(3 \left(c_1 \left(22 c_1-15 h-11 \xi \right)+3 h^2+6 h \xi +5 \xi
   ^2\right) \right)  - \chi_{E_8} \,.
\ee
Evaluating this for the base threefold \cite{Marsano:2009gv} with $\chi^* =  13968$, we obtain
\be\label{ND3geo}
\chi(X_4) = 10416 \,,\qquad
N_{D3, {\rm geo}} =  - 434 \,.
\ee

It is this refined Euler characteristic that should be used in $N_{D3, {\rm base}}$. We will see that this refinement will remove some of the models that would otherwise have also given rise to consistent supersymmetric models. 


\subsubsection{Flux contribution}

The induced D3-brane charge due to spectral cover fluxes, on the other hand, can be computed using the relation \cite{Blumenhagen:2009yv}
\begin{equation}N_{D3,\Gamma}=\frac{1}{2}\int_{X_4} G\wedge G = -\frac{1}{2}\Gamma^2 \,.
\end{equation}
where $\Gamma^2$ refers to the self-intersection number of $\Gamma$ inside ${\cal{C}}$.  In appendix \ref{app:Gamma}, we computed $\Gamma^2$ with the result
\begin{equation}
\begin{split}\Gamma^2 &= -2\left(3\tilde{k}_1+2\tilde{d}_1\right)^2 -2\left(2\tilde{k}_2+3\tilde{d}_2\right)^2 + \frac{5}{6}\left[6\rho + 2\tilde{d}_1(\eta-5c_1-\xi)-3\tilde{d}_2\xi+\tilde{q}\psi\right]^2\\
& -\left(3\tilde{m}_1+\tilde{q}\right)^2\psi(\eta-3c_1-\xi) - \left(2\tilde{m}_2-\tilde{q}\right)^2\psi(c_1+\xi) \\
& - \left(2\tilde{k}_2+3\tilde{d}_2\right)\left(2\tilde{m}_2-\tilde{q}\right)\xi\psi - \frac{2}{3}\left(3\tilde{k}_1+2\tilde{d}_1\right)\left(3\tilde{m}_1+\tilde{q}\right)\psi(\eta-5c_1-\xi) \\
 & + 2\left(3\tilde{m}_1^2 + \tilde{m}_2^2 + [2\tilde{m}_1 - \tilde{m}_2]\tilde{q}+\frac{7}{12}\tilde{q}^2\right)\psi^2 \,.
\end{split}\end{equation}
where the indicated curve intersections are computed inside $S_{\rm GUT}$.
Whenever possible, we will restrict to models for which the number of D3-branes that must be added to cancel the tadpole \eqref{D3tad} is positive
\begin{equation}N_{D3}>0\,.
\end{equation}
In the geometry that we discuss we will find plenty of solutions with this property. This is of course desirable as we do not wish to introduce uncontrolled sources of supersymmetry breaking.


\subsection{Three Generation Models}

We now turn to the construction of consistent spectral cover fluxes for 3-generation models.  Before getting to the fluxes, however, let us recall that asking for models with exactly 3 generations of MSSM matter and no exotics is too much when we have a $U(1)_{PQ}$ symmetry.  Rather, the nontrivial restriction of $U(1)_Y$ flux to $\Sigma_{{\bf 10}_M}$ and $\Sigma_{{\bf 10}_{other}}$ ,
which is unavoidable if we want to obtain $U(1)_{PQ}$, will lead to extra exotics.  To get models with 3 full generations and a minimal number of exotics, we will look for restrictions of the spectral cover fluxes that are different from what might naively expect.

To see this, recall that
\be
F_Y\vert_{\Sigma_{\bf 10_M}}=-1,\quad  F_Y\vert_{\Sigma_{\bf 10_{other}}}=1 \,,
\ee
This means that if the total spectral flux $\Gamma$ of (\ref{TotalGamma}) restricts as
\be
\Gamma\vert_{\Sigma_{\bf 10_M}}=M,\quad  \Gamma\vert_{\Sigma_{\bf 10_{other}}}=-M\,,
\ee
then we obtain the following chiral spectrum on the $\Sigma_{{\bf 10_M}}$ curve
\be
\ba
n_{({\bf 3},{\bf 2})_{+1/6}}-n_{({\bf {\bar 3}},{\bf 2})_{-1/6}} &=M \cr
n_{({\bf {\bar 3}},{\bf 1})_{-2/3}}-n_{({\bf 3},{\bf 1})_{+2/3}} &=M+1\cr
n_{({\bf 1},{\bf 1})_{+1}}-n_{({\bf 1},{\bf 1})_{-1}} & =M-1\,,
\ea\ee
and on the $\Sigma_{{\bf 10_{other}}}$ curve
\be
\ba
n_{({\bf 3},{\bf 2})_{+1/6}}-n_{({\bf {\bar 3}},{\bf 2})_{-1/6}}&=-M\cr
n_{({\bf {\bar 3}},{\bf 1})_{-2/3}}-n_{({\bf 3},{\bf 1})_{+2/3}}&=-M-1 \cr
n_{({\bf 1},{\bf 1})_{+1}}-n_{({\bf 1},{\bf 1})_{-1}} &=-M+1
\,.\ea\ee
Because we want to obtain at least 3 full generations on $\Sigma_{\mathbf{10}_M}$, let us write $M=3+Q$ with $Q\ge 1$.  With this notation, we get:

\begin{center}
$Q-1$ copies of $({\bf 1},{\bf 1})_{+1}$,  $Q$ copies of $({\bf 3},{\bf 2})_{+1/6}$,
and $Q+1$ copies of $({\bf {\bar 3}},{\bf 1})_{-2/3}$.
\end{center}
Meanwhile, on
$\Sigma_{{\bf 10_{other}}}$ we find conjugate exotics:

\begin{center}
$Q-1$ copies
of $({\bf 1},{\bf 1})_{-1},$ $Q$ copies of $({\bf {\bar 3}},{\bf
2})_{-1/6}$ and $Q+1$ copies of $({\bf 3},{\bf 1})_{+2/3}.$
\end{center}

Of course the number of exotics is limited by requiring perturbativeness of the gauge couplings up to the GUT-scale. Thus, we will not consider any values of $Q$ larger than 4.  Later, we will discuss how the exotics can in principle be removed through coupling to a charged singlet, ${\bf 10}_M {\overline {\bf 10}_{\rm other}}{\bf 1}$, provided that singlet picks up a nonzero bosonic expectation value.  For now, however,
we discuss two distinct choices of flux that give models of the sort that we are looking for.


\subsubsection{Models with $Q=1$}
\label{subsubsec:Q1}

To obtain a minimal number of exotics, we should set $Q=1$ and aim for a spectral cover flux $\Gamma$ with the following restrictions
\begin{equation}\begin{array}{c|c|c|c|c|c} & \mathbf{10}_M & \mathbf{5}_H & \mathbf{\overline{5}}_H & \mathbf{\overline{5}}_M & \mathbf{10}_{\text{other}} \\ \hline
\Gamma & 4 & 0 & 0 & 3 & -1
\end{array}\end{equation}
A two-parameter family that realizes this can be achieved by making the choice of classes
\be
\rho = {1\over 2} (h- e_1 - e_2) \,,\qquad
\psi = e_1+ e_2 \,,
\ee
as well as the parameters in $\Gamma$ given in (\ref{TotalGamma}) to be
\begin{equation}\begin{array}{c|c|c|c|c|c|c}
\tilde{k}_1 & \tilde{k}_2       & \tilde{d}_1   & \tilde{d}_2   & \tilde{m}_1   & \tilde{m}_2 & \tilde{q} \\ \hline
\frac{1}{2} & k_2 + {1\over 2 } & d_1       & -4-6 d_1  & 2(d_1+1)  & 18 -4k_2+35d_1 & -9-14d_1
\end{array}\end{equation}
where $k_2$ and $d_1$ are required to be integers
\begin{equation}k_2,d_1\in\mathbb{Z}\,.
\end{equation}

The condition on supersymmetry for this family of solutions is that
\begin{equation}\omega\cdot_{S_{\rm GUT}}(5h-8e_1-4e_2)=0\,.\end{equation}
This is true for $\omega$ of the form
\begin{equation}\omega = \alpha(e_1+e_2) + \beta (h-e_1-e_2)\,,\qquad \beta = \frac{12\alpha}{7}\,.\end{equation}
which parametrizes a 1-dimensional subset of the K\"ahler cone on $S_{\rm GUT}=dP_2$.

The D3-brane tadpole induced by $G$-fluxes in this family of solutions is
\begin{equation}N_{D3\,,\Gamma} = -\frac{1}{2}\Gamma^2 = -2604k_2d_1+124k_2^2-1395k_2+14248d_1^2+15351d_1+4159 \,.
\end{equation}
To have a supersymmetric background, we want the net induced D3-brane charge due to fluxes and the geometry (\ref{ND3geo}) to be negative so that it can be canceled by a positive number $N_{D3}$ of D3-branes.  We found various choices of flux that lead to a positive D3 and summarize them in the following table where  $N_{D3} = - (N_{D3, \Gamma} + N_{D3, {\rm geo}})$: 
\begin{equation}\label{Q1Flux}
\begin{array}{c|c|c|c}
k_2 & d_1 & N_{D3, \Gamma} & N_{D3}  \\ \hline
-6 & -1 & 266 & 168 \\
 -5 & -1 & 111 & 323 \\
 -4 & -1 & 204 & 230 \\ \hline
 5 & 0 & 284 & 150 \\
 6 & 0 & 253 & 181
\end{array}\end{equation}
It is promising that we can find vacua for which the contributions to the D3-brane tapdole from our choices of $G$-fluxes and base $B_3$ is negative.  Further, as discussed in \cite{Donagi:2009ra}, the $G$-fluxes that we have introduced manage to stabilize some complex structure moduli, namely those which would disrupt the 3+2 factorization and those which would deform the compactification in such a way that the non-universal fluxes $\Psi_1$ and $\Psi_2$ would cease to be well-defined.  Nevertheless, many moduli remain unstabilized in these models and any additional $G$-fluxes that must be switched on to remedy this will yield additional positive contributions to the D3-brane tadpole.  For this reason, declaring success in finding a completely consistent supersymmetric background would be somewhat premature.



\subsubsection{Models with general $Q$}
\label{subsubsec:Q}

More generally, we can find models with other values of $Q$, which arise when the induced chiralities from $\Gamma$ are given by

\begin{equation}\begin{array}{c|c|c|c|c|c} & \mathbf{10}_M & \mathbf{5}_H & \mathbf{\overline{5}}_H & \mathbf{\overline{5}}_M & \mathbf{10}_{\text{other}} \\ \hline
\Gamma & 3+Q & 0 & 0 & 3 & -Q
\end{array}\end{equation}
A one-parameter family of choices that puts $3+Q$ units of flux on $\mathbf{10}_M$ and $-Q$ on $\mathbf{10}_{\text{other}}$ arises by choosing $\Gamma$ as in (\ref{TotalGamma}) with
\be
\rho = {1\over 2} (h- e_1 - e_2) \,,\qquad
\psi = e_1+ e_2 \,,
\ee
as well as
\begin{equation}\begin{array}{c|c|c|c|c|c|c}
       \tilde{k}_1 & \tilde{k}_2          &\tilde{d}_1& \tilde{d}_2   & \tilde{m}_1 & \tilde{m}_2             & \tilde{q} \\ \hline
{1\over 2} & k_2 + {1\over 2 } & -1 & Q+1  & 0      & -4k_2-7Q-10 & 3+2Q
\end{array}\end{equation}
Here $k_2$ and $Q$ are both integers and we further require that $Q>0$, but of course not too large in order to keep the gauge couplings perturbative up to the GUT scale
\begin{equation}k_2\in\mathbb{Z}\,,\qquad Q\in\mathbb{N} \,.
\end{equation}
The supersymmetry condition here requires that
\begin{equation}\omega\cdot_{S_{\rm GUT}} \left[(2+3Q)h - (3+5Q)e_1 - 2(1+Q)e_2\right]=0 \,.
\end{equation}
This is true for
\begin{equation}\omega = \alpha(e_1+e_2) + \beta (h-e_1-e_2)\,,\qquad \beta = \left[\frac{5+7Q}{3+4Q}\right]\alpha \,,\end{equation}
which is an element of the K\"ahler cone on $S_{\rm GUT}=dP_2$ for arbitrary $Q\in\mathbb{N}$.

The D3-brane tadpole induced by $G$-fluxes in this family of solutions is
\begin{equation}
N_{D3\,,\Gamma} = -\frac{1}{2}\Gamma^2 = 496k_2Q + 124k_2^2+713 k_2 + 518 Q^2 + 1478 Q + 1060
\,.\end{equation}
In this case, we found various values of $Q$ and $k_2$ for which the number $N_{D3}$ of D3-branes that must be added to satisfy the D3-brane tadpole condition \eqref{D3tad} is positive. Again we use the contribution from the refined Euler characteristic (\ref{ND3geo}). We summarize these in the following table
\begin{equation}\label{QFlux}
\begin{array}{c|c|c|c}k_2 & Q &N_{D3, \Gamma}& N_{D3} \\ \hline
 -6 & 1 & 266 & 168 \\
 -5 & 1 & 111 & 323 \\
 -4 & 1 & 204 & 230 \\ \hline
-8 & 2 & 384 & 50 \\
 -7 & 2 & 229 & 205 \\
 -6 & 2 & 322 & 112 \\ \hline
-9 & 3 & 391 & 43 
\end{array}\end{equation}
The $Q=1$ solutions are equivalent to the $d_1=-1$ solutions of the previous subsection.  The only new ones, then, are the solutions with $Q=2$ and $Q=3$.  As before, we emphasize that the D3-brane tadpole will also get positive contributions from any $G$-fluxes that are introduced to stabilize complex structure moduli.




\section{F-enomenology: F-unification and gauge-mediation}

Finally, in this section we get to the physics of the F-theory models that we propose. Any geometry, realizing the ${\bf 3+2}$ semi-local model with the fluxes specified in the last section, such as the one of \cite{Marsano:2009gv}, will exhibit these phenomenological properties. The reader interested solely in the phenomenological implications can read this section relatively independently of the remainder of the paper.
We first summarize the low-energy field content of the ${\bf 3+2}$ model. Then we recall how the additional non-GUT multiplet exotics that are present in the ${\bf 3+2}$ can explain the generalized unification ("F-unification") in models that break the GUT group using hypercharge-flux. Finally, we use the exotics as messenger fields in a gauge-mediated model and compute the soft-masses.


\subsection{The Model}
\label{subsec:Model}

The model we obtained in the last section has the following chiral field content, where we already include the effect of the hypercharge flux
\be
\begin{array}{c|c|c|c}
\text{Field} & \text{Origin} & \text{Class in }S_{\rm GUT} & \text{Multiplicity}  \\ \hline
H_u & 22 & 4h-2e_1-e_2 & 1\\
H_d & 12 & 9h-3e_1-4e_2 &  1 \\
\mathbf{\overline{5}}_M & 11 & 8h-3(e_1+e_2) &  3 \\
\mathbf{10}_M & 22 & h-e_1 &  3 \\
\mathbf{10}_{\text{other}} & 11 & h-e_2 & 0 \\
\end{array}\ee
which is the standard MSSM content, and in addition we have following exoticts arising from the ${\bf 10}_M$ and ${\bf 10}_{\rm other}$:
\be\label{Exotics}
\begin{array}{l|c|c|c}
\text{Field} & \text{Origin} & \text{Class in }S_{\rm GUT} & \text{Multiplicity}  \\ \hline
({\bf 3}, {\bf 2})_{+1/6} & 22 & h-e_1 &  Q  \\
({\bf \bar{3}, {\bf 1}})_{-2/3} & 22 & h-e_1 &  Q +1\\
 ({\bf 1}, {\bf 1})_{+1} & 22 & h-e_1 &  Q -1\\
({\bf \bar{3}}, {\bf 2})_{-1/6} & 11 & h-e_2 &  Q \\
({\bf {3}, {\bf 1}})_{+2/3}  & 11 & h-e_2 & Q +1\\
 ({\bf 1}, {\bf 1})_{-1}& 11 & h-e_2 & Q-1
\end{array}
\ee
The non-GUT multiplet exotic fields arose from the non-trivial restriction of the hypercharge flux onto the matter curve $\Sigma_{\bf 10}$, thus giving different multiplicities to the constituents of ${\bf 10}$
\be
\ba
SU(5) \quad & \rightarrow \quad  SU(3) \times SU(2)\times U(1)\cr
{\bf 10} \quad &\rightarrow \quad ({\bf 3}, {\bf 2})_{1/6} + ({\bf \bar{3}, {\bf 1}})_{-2/3} + ({\bf 1}, {\bf 1})_{1}\,.
\ea
\ee
Note that in principle we have a choice
\be
Q\geq 1\,,
\ee
however, in order to keep the model as minimal as possible and to stay within the perturbative regime we will choose $Q=1$.

We will now discuss two features of this model in detail: the contributions to the beta-function and the resulting non-unification (F-unification) that we already encountered in (\ref{Fun}), and then to compute the soft-masses for a gauge mediated model, which makes use of the above {\bf 10}-exotics.


\subsection{F-unification}

This section is a summary of what we already pointed out in v2 of \cite{Marsano:2009gv}, namely, that additional non-GUT exotics arising from the non-trivial restriction of $U(1)_Y$ upon ${\bf 10}$ and ${\bf 5}$ matter curves, can precisely account for the generalized F-unification of gauge couplings that was pointed out by \cite{Blumenhagen:2008aw}.

In \cite{Blumenhagen:2008aw} it was observed that internal $U(1)_Y$ flux used to break the GUT group actually spoils gauge coupling unification in $F$-theory GUTs.  The reason for this is quite simple and we review it now.  The gauge group is realized on a stack of D7-branes whose worldvolume action contains a Chern-Simons term of the form
\begin{equation}
S_{CS} = \mu_7 \int_{\mathbb{R}^{3,1}\times S_{\rm GUT}}\, C_0\wedge\text{tr}(F^4)\,.
\end{equation}
A nontrivial background flux along the internal directions can lead to new quadratic couplings for 4d gauge fields that effectively split the couplings.  In \cite{Blumenhagen:2008aw}, an internal flux of the following form was considered
\begin{equation}F\sim f_a\begin{pmatrix} 1_{3\times 3} & 0 \\ 0 & 1_{2\times 2} \end{pmatrix}+ f_Y\begin{pmatrix}0_{3\times 3} & 0 \\ 0 & 1_{2\times 2}\end{pmatrix} \,.
\end{equation}
Here, $f_a$ should be identified with the bundle often referred to in the local-model literature as $L_Y^{1/6}\times V_{10}$, while $f_Y$ should be identified with the bundle $L_Y^{5/6}$.  Fluxes of the form $L_Y^{1/6}\otimes V_{10}$ are constructed using spectral covers as in \cite{Donagi:2009ra, Marsano:2009gv}, while $L_Y^{5/6}$ is conventionally taken to be ${\cal{O}}(e_1-e_2)$ where $e_1$ and $e_2$ are two exceptional classes of the underlying $dP_n$ surface.  By expanding the Chern-Simons action, Blumenhagen concluded in \cite{Blumenhagen:2008aw}  that, rather than unifying, the gauge couplings at $M_{GUT}$ only satisfy the weaker F-unification relation
\begin{equation}
\alpha_1^{-1}(M_{GUT})-\frac{3}{5}\alpha_2^{-1}(M_{GUT}) - \frac{2}{5}\alpha_3^{-1}(M_{GUT})= 0\,.
\label{blumenhagenconstraint}\end{equation}
We know that the MSSM matter spectrum is fairly consistent, to within a few percent, with complete gauge coupling unification.  To account for the fact that the couplings are not unified in $F$-theory GUTs, then, it is necessary to introduce some additional matter fields at some high scale which do not form complete GUT multiplets but whose net contribution $\delta b_i$ to the various $\beta$ function coefficients, $b_i$, nevertheless satisfies the relation
\begin{equation}\delta b_1 - \frac{3}{5}\delta b_2 - \frac{2}{5}\delta b_3=0 \,.
\end{equation}
It was argued in \cite{Blumenhagen:2008aw} that massive Higgs triplets, for instance, can do the job if they are included below the GUT scale.

From a different perspective, we have seen in \cite{Marsano:2009gv}, that it is impossible to realize the $U(1)_{PQ}$ symmetry necessary to implement the neutrino scenarios of \cite{Bouchard:2009bu} unless the hypercharge flux restricts nontrivially to more matter curves than just those associated to the Higgs fields.  This would lead to incorporation of exotic matter that does not comprise full GUT multiplets.  One way to remove this matter from the low energy spectrum is to ensure that pairs couple to one another via a charged singlet field as in
\be
W \supset \lambda \left(\overline{{\bf 1}^{(1)(2)}} \right) f \bar{f} \,,
\label{Gmed?}\ee
This is suggestive of gauge mediation with the exotics playing the role of messenger fields.  If we take this idea seriously, then we can be rid of the exotics without having to introduce any new scales beyond those that we already need for supersymmetry breaking.  Otherwise, we can view the bosonic expectation value of $X$ as an entirely new scale that must be generated.

What we showed in \cite{Marsano:2009gv}, and are going to summarize here, is that any exotics that are forced on us when we introduce $U(1)_{PQ}$ disrupt the gauge coupling unification at the GUT scale in such a way that \eqref{blumenhagenconstraint} is precisely satisfied!  This means that we can avoid the restrictions of \cite{Marsano:2009gv}, which seemed to cause problems for neutrino physics, while simultaneously addressing the problem of \cite{Blumenhagen:2008aw}, namely that the gauge couplings need to be slightly split before the GUT scale.


\subsubsection{$\beta$ Functions}

To summarize the $\beta$ function shifts, let us first set some conventions.  We implicitly define the standard $\beta$ coefficients $b_i$ through the RG equations
\begin{equation}\frac{d\alpha_i}{dt} = -\frac{b_i}{2\pi}\alpha_i^2\,.\label{betabdef}\end{equation}
The $\beta$ functions of $SU(N)$ gauge couplings in theories with ${\cal{N}}=1$ supersymmetry and $N_f$ fundamentals is given by the standard formula
\begin{equation}b_N = 3N-\frac{N_f}{2}\,.
\end{equation}
For $U(1)_Y$, on the other hand, the relevant coefficient is given by
\begin{equation}b_1 = -\frac{3}{5}\sum_{\text{flavors}} Y^2\,.
\end{equation}
Note that we have normalized the $U(1)_Y$ generator with the extra factor of $\sqrt{\frac{3}{5}}$ so that the condition of gauge coupling unification would take the form $\alpha_1=\alpha_2=\alpha_3$ at the GUT scale.

With these formulae, it is easy to reproduce the well-known result for the MSSM $\beta$ functions
\begin{equation}b_3=3\,,\qquad b_2=-1\,,\qquad b_1=-\frac{33}{5}\,.
\end{equation}

\subsubsection{$\beta$-function contributions from Exotics}

Let us now consider how the spectrum on a matter curve with nontrivial $U(1)_Y$ flux affects the $\beta$ function coefficients.  Rather than specifying to the specific setup of our present models, which attempt to minimize the number of exotics, we instead consider contributions from both $\mathbf{10}$ and $\mathbf{\overline{5}}$ matter curves in full generality.

We start with the $\mathbf{10}$ matter curve, $\Sigma_{\mathbf{10}}$, which houses three types of $SU(3)\times SU(2)\times U(1)_Y$ multiplets (and their conjugates), namely
\begin{equation}\left[(\mathbf{3},\mathbf{2})_{+1/6}\oplus (\mathbf{\overline{3}},\mathbf{1})_{-2/3}\oplus (\mathbf{1},\mathbf{1})_{1}\right]\oplus \left[cc\right]\,.
\end{equation}
The net chirality of each type of multiplet is given by the degree of certain line bundles restricted to $\Sigma_{\mathbf{5}}$.  More specifically, using a relatively naive notation from local models we have
\begin{equation}\begin{split}
n_{(\mathbf{3},\mathbf{2})_{+1/6}}-n_{(\mathbf{\overline{3}},\mathbf{2})_{-1/6}} &= \deg\left(L_Y^{1/6}\otimes V_{10}\right) \\
n_{(\mathbf{\overline{3}},\mathbf{1})_{-2/3}}-n_{(\mathbf{3},\mathbf{1})_{+2/3}} &= \deg\left(L_Y^{-2/3}\otimes V_{10}\right) \\
n_{(\mathbf{1},\mathbf{1})_{1}} - n_{(\mathbf{1},\mathbf{1})_{-1}} &= \deg\left(L_Y\otimes V_{10}\right)\,.
\end{split}\end{equation}
What appears here as $L_Y^{-1/3}\otimes V_5$ is a flux constructed from the spectral cover.  On the other hand, what appears here as $L_Y^{5/6}$ is precisely the object ${\cal{O}}(e_1-e_2)$ introduced to break the gauge group.
If we suppose that $L_Y^{5/6}$ has restriction of degree $N$ to $\Sigma_{\mathbf{10}}$ while $L_Y^{1/6}\otimes V_{10}$ has restriction of degree $M$ then these chiralities are
\begin{equation}\begin{split}
n_{(\mathbf{3},\mathbf{2})_{+1/6}}-n_{(\mathbf{\overline{3}},\mathbf{2})_{-1/6}} &= M \\
n_{(\mathbf{\overline{3}},\mathbf{1})_{-2/3}}-n_{(\mathbf{3},\mathbf{1})_{+2/3}} &= M-N \\
n_{(\mathbf{1},\mathbf{1})_1}-n_{(\mathbf{1},\mathbf{1})_{-1}} &= M+N\,.
\end{split}\end{equation}
In our models, we have $M=3+Q$, $N=Q$ on the $\mathbf{10}_M$ matter curve and $M=-Q$, $N=-Q$ on the $\mathbf{10}_{\text{other}}$ matter curve.

In the general case, the shift of the $\beta$ coefficients induced by the extra $(\mathbf{3},\mathbf{2})_{+1/6}$'s, $(\mathbf{\overline{3}},\mathbf{1})_{-2/3}$'s, and $(\mathbf{1},\mathbf{1})_{+1}$'s is given by
\begin{equation}\begin{split}
\delta b_3 &= -\frac{1}{2}\left(3M-N\right) \\
\delta b_2 &= -\frac{3}{2}M \\
\delta b_1 &= -\frac{1}{10}\left(15M-2N\right) \,.
\label{10shift}
\end{split}\end{equation}
Note that these shifts are all identical when $N=0$, as we expect.  When $N\ne 0$, the shifts differ and unification is spoiled.  Nevertheless, the shifts are consistent with the preservation of \eqref{blumenhagenconstraint} because they satisfy
\begin{equation}\delta b_1 - \frac{3}{5}\delta b_2 - \frac{2}{5}\delta b_3 = 0\,.
\end{equation}

We now turn to describe exotics from a $\mathbf{\overline{5}}$ matter curve, $\Sigma_{\mathbf{\overline{5}}}$, which houses two types of $SU(3)\times SU(2)\times U(1)_Y$ multiplet (and their conjugates), namely
\begin{equation}\left[(\mathbf{3},\mathbf{1})_{-1/3}\oplus (\mathbf{1},\mathbf{2})_{1/2}\right]\oplus\left[cc\right]\,.
\end{equation}
Note that we have managed to avoid introducing exotics of this type in the models of section \ref{sec:model}.  Nevertheless, it is nice to see that exotics of this type shift the $\beta$ functions in a way that satisfies \eqref{blumenhagenconstraint}.

Here, the net chiralities are computed by
\begin{equation}\begin{split}n_{(\mathbf{3},\mathbf{1})_{-1/3}}-n_{(\mathbf{\overline{3}},\mathbf{1})_{+1/3}}&=\deg\left(L_Y^{-1/3}\otimes V_5\right) \\
n_{(\mathbf{1},\mathbf{2})_{+1/2}}-n_{(\mathbf{1},\mathbf{2})_{-1/2}} &= \deg\left(L_Y^{1/2}\otimes V_5\right)\\
&= \deg\left(L_Y^{5/6}\otimes [L_Y^{-1/3}\otimes V_5]\right)\,.
\end{split}\end{equation}
Let us now suppose that the restriction of $L_Y^{-1/3}\otimes V_5$ to $\Sigma_{\mathbf{5}}$ has degree $M$ while the restriction of $L_Y^{5/6}$ to $\Sigma_{\mathbf{5}}$ has degree $N$.  In this case, we find
\begin{equation}\begin{split}
n_{(\mathbf{3},\mathbf{1})_{-1/3}}-n_{(\mathbf{\overline{3}},\mathbf{1})_{+1/3}} &= M \\
n_{(\mathbf{1},\mathbf{2})_{+1/2}}-n_{(\mathbf{1},\mathbf{2})_{-1/2}} &= M+N\,.
\end{split}\end{equation}
The shift of $\beta$ function coefficients induced by the extra $(\mathbf{3},\mathbf{1})_{-1/3}$'s and $(\mathbf{1},\mathbf{2})_{+1/2}$'s is given by
\begin{equation}\begin{split}
\delta b_3 &= -\frac{M}{2} \\
\delta b_2 &= -\frac{M+N}{2} \\
\delta b_1 &= -\frac{1}{10}\left(5M+3N\right)\,.
\label{5shift}
\end{split}\end{equation}
Again, these shifts are identical when $N=0$.  For $N\ne 0$, they are consistent with \eqref{blumenhagenconstraint} because they satisfy
\begin{equation}\delta b_1 - \frac{3}{5}\delta b_2 - \frac{2}{5}\delta b_3 = 0\,.
\end{equation}


\subsection{Soft masses for non-minimal gauge mediation}

We now become a bit more speculative and consider the possibility that the charged exotics that we obtain in $F$-theory GUTs that exhibit $U(1)_{PQ}$ symmetries play the role of gauge messenger fields.  As we have stated before, one motivation to consider this scenario is that the exotic mass becomes identified with the messenger mass, a scale that must already be introduced to realize gauge mediated supersymmetry breaking.

While messenger sectors that do not comprise complete GUT multiplets have received some prior attention in the literature, the models we propose here seem markedly different.  In this paper, we don't endeavour to describe the phenomenology of these models in general.  For now, we simply present well-known formulae for the leading gauge mediated soft masses that one obtains at the messenger scale \cite{Giudice:1997ni}.  To do this, we recall the results for the 1-loop gaugino masses
\begin{equation}M_{1/2}(M_{\text{Mess}}) = -\sum_{i=1}^3\delta b_i\left(\frac{\alpha_i(\mu)}{4\pi}\right)\left(\frac{F}{M_{\text{Mess}}}\right)\,,
\end{equation}
as well as the 2-loop squark and slepton masses
\begin{equation}m^2_Q(\text{Mess})=-\sum_{\text{relevant }i}\frac{c_i\,\delta b_i\,\alpha_i(\mu)^2}{8\pi^2}\left|\frac{F}{M_{\text{Mess}}}\right|^2\,.
\end{equation}
In these formulae, the $\delta b_i$ denote shifts of the $\beta$ function coefficients $b_i$ \eqref{betabdef} induced by the messenger sector, the $c_i$ denote quadratic Casimirs of the MSSM gauge groups, and the dimensionful quantities $F$ and $M_{\text{Mess}}$ arise as the bosonic and $F$-component expectation values of the singlet field $X$ in \eqref{Gmed?}
\begin{equation}\langle \phi\rangle = M_{\text{Mess}} + \theta^2 F\,.
\end{equation}
The results for different collections of exotics that we obtain when nontrivial $U(1)_Y$ flux threads various matter curves can be read off from the general formulae \eqref{10shift} and \eqref{5shift}.  For the models of this paper, the shifts are
\begin{equation}\begin{split}\delta b_3 &= -2Q \\
\delta b_2 &= -3Q \\
\delta b_1 &= -\frac{13}{5}Q \,.
\end{split}\end{equation}
One obvious feature of gauge mediated models such as this in which the messenger sector comprises non-GUT multiplets is that the 1-loop gaugino masses no longer unify at the GUT scale.

In the absence of a definite model for breaking supersymmetry with the singlet field in \eqref{Gmed?} with a sufficiently light gravitino mass for gauge mediation, this scenario may still be viewed as somewhat speculative.  Nevertheless, we consider it to be an intriguing possibility that arises from the study of semi-local $F$-theory GUTs and hope to study its phenomenology in more detail in future work.


\section{Singlets}

Because the dynamics of singlet fields play a crucial role in removing potential exotics within the class of models considered in this paper, we now make some brief comments regarding their study.  Ultimately, a detailed treatment of singlet fields will exhibit a strong dependence on global details of the compactification beyond those captured by the "semi-local" picture.  We do not endeavor to undertake such a study in the present paper but rather will try to clarify some basic properties of the curves on which various singlet fields localize as well as present a conjecture for how the number of such singlets might be counted in a "semi-local" framework.


\subsection{The Semi-Local Perspective}

\subsubsection{Singlets in the ${\bf 3+2}$ model}

Singlet fields arise in the same manner that charged $\mathbf{10}$'s and $\mathbf{\overline{5}}$'s do, namely as degrees of freedom that localize along curves where the singularity type of the elliptic fiber is enhanced.  Before addressing the singlet locus directly in the global context, however, let us first look at how singlets arise from a semi-local perspective.
The relevant physics here is that of an $E_8$ theory on $S_{\rm GUT}$ in which the $E_8$ group is explicitly broken to $SU(5)_{\rm GUT}$.  This is done by introducing a fixed vev of a scalar field that takes values in the adjoint of $SU(5)_{\perp}$, which is the commutant of $SU(5)_{\rm GUT}\subset E_8$
\begin{equation}\langle\phi_{adj}\rangle\sim \begin{pmatrix}\lambda_1 & 0 & 0 & 0 & 0 \\
0 & \lambda_2 & 0 & 0 & 0 \\
0 & 0 & \lambda_3 & 0 & 0 \\
0 & 0 & 0 & \lambda_4 & 0 \\
0 & 0 & 0 & 0 & \lambda_5
\end{pmatrix}\,,\qquad \sum_i \lambda_i=0 \,.
\end{equation}
As we know by now, the $\lambda_i$ are not individually well-defined on $S_{\rm GUT}$ but rather are mixed by nontrivial monodromies.  Nevertheless, we know that locally the equations of $\mathbf{10}$ and $\mathbf{\overline{5}}$ matter curves are given by $\lambda_i=0$ and $\lambda_j+\lambda_k=0$ for $j\ne k$, respectively.  The corresponding matter curves are given in terms of well-defined combinations as
\begin{equation}
\Sigma_{\mathbf{10}}:\quad 0=b_5\sim \prod_i\lambda_i\,,\qquad\qquad
\Sigma_{\mathbf{\overline{5}}}:\quad 0 = P\sim \prod_{j<k}(\lambda_j+\lambda_k)\,.\end{equation}
While singlet fields do not localize on $S_{\rm GUT}$ itself, we expect that some aspects of their physics should nevertheless be describable in the semi-local picture.  The number of singlet fields of a particular type, for instance is something that we might hope to address in this context because it naively should not depend on the magnitude of the $\lambda_i$.  Said differently, we can study the number of singlet fields in a limit where the deformation of our original $E_8$ singularity is almost completely turned off.  In that limit, one expects that a worldvolume description is appropriate, though we will point out later where this reasoning may break down.  At any rate, a semi-local approach leads us to expect that singlet fields localize on the locus $\lambda_i=\lambda_j$ for $i\ne j$.

To write this locus in terms of the globally well-defined objects $b_m$, we must express it as a symmetric polynomial in the $\lambda_i$.  The natural candidate $\prod_{i<j}(\lambda_i-\lambda_j)$ is asymmetric, but its square is symmetric so we turn to this.  We find
\begin{equation}\prod_{i<j}(\lambda_i-\lambda_j)^2\sim F\,,
\end{equation}
where $F$ is given by
\begin{equation}\begin{split}F = & 108b_2^5b_5^2 + 8b_2^4(2b_4^3-9b_3b_4b_5)+b_0b_2^2(-128b_4^4+560b_3b_4^2b_5+825b_3^2b_5^2) \\
&+4b_2^3(-b_3^2b_4^2+4b_3^3b_5-225b_0b_4b_5^2) \\
&+2b_0b_2(72b_3^2b_4^3-315b_3^3b_4b_5+1000b_0b_4^2b_5^2-1875b_0b_3b_5^3) \\
&+b_0(-27b_3^4b_4^2+108b_3^5b_5-1600b_0b_3b_4^3b_5+2250b_0b_3^2b_4b_5^2+b_0(256b_4^5+3125b_0b_5^4))\,,
\label{Fdef}\end{split}\end{equation}
and we have explicitly set $b_1=0$ to reflect the fact that $\sum_i\lambda_i=0$.  This suggests that singlet fields will localize on a curve in $B_3$ that projects down to the curve $F$ inside $S_{\rm GUT}$ under the projection $B_3\rightarrow S_{\rm GUT}$ that sets $z$ to zero.  As we shall see later, though, this is not quite the whole story as we should also consider the "component at $\infty$" where two $\lambda_i$'s simulaneously diverge as we approach $b_0=0$.

For the time being, our main objective is to determine an index for singlet fields as such a computation can be done without performing a detailed study of the zero mode wave functions.  Indices count net chiralities of zero modes, though, which in turn require some notion of charge in order to be defined.  For a generic monodromy group $G\cong S_5$, our singlet fields carry no charge whatsoever and do not in general come in well-defined vector-like pairs {\footnote{This is crucial for the Majorana neutrino scenario of \cite{Bouchard:2009bu}, which depends on KK masses being of Majorana type.}}.  We therefore expect that any attempt to count an index for such modes will yield a vanishing result.
{\footnote{This is analogous to the fact that the number of singlets on the heterotic side is determined by dimensions of specific homology groups that generically cannot be determined by an index computation.}}.

The case of interest for us, however, is one in which the monodromy group is a subgroup of $S_5$ that allows for the preservation of at least one $U(1)$ symmetry capable of forbidding dangerous dimension 4 proton decay operators.  As we reviewed in section \ref{sec:semilocal}, this structure is reflected by the fact that the associated spectral cover, ${\cal{C}}$, factors into two distinct components.  For definiteness, we will focus on the explicit 3+2 factorization described in section \ref{sec:model} though the generalization to other examples should be clear.  Adopting the notation of that model, the spectral cover factors according to
\begin{equation}\begin{split}{\cal{C}}&={\cal{C}}^{(1)}{\cal{C}}^{(2)}\\
&= \left(\alpha e_0 U^3 - \alpha e_1 U^2 V + a_2UV^2 + a_3V^3\right)\left(e_0 U^2 + e_1UV + e_2V^2\right)\\
&\sim  \left[\alpha e_0 \prod_{a=1}^3\left(V-U\lambda_a\right)\right]\left[e_0\prod_{m=1}^2\left(V-U\lambda_m\right)\right]\,.\end{split}\end{equation}

When ${\cal{C}}$ splits into two components, we anticipate that $F$ \eqref{Fdef} factors into three pieces corresponding to $\lambda_a=\lambda_b$ ($a\ne b)$, $\lambda_m=\lambda_n$ ($m\ne n$), and $\lambda_a=\lambda_m$.  We can see this explicitly in the 3+2 model by noting that
\begin{equation}F = F_1F_2F_3^2\,,
\end{equation}
where
\begin{equation}\begin{split}
F_1 &= 4 e_0 e_2-e_1^2 \\
F_2 &=   -4 a_3 e_1^3 \alpha _0^2+\alpha _0 \left(27 a_3^2 e_0^2+18 a_2 a_3 e_1 e_0-a_2^2 e_1^2\right)+4 a_2^3
   e_0 \\
F_3 &= a_2 \left(2 e_2 \left(e_1^2-e_0 e_2\right) \alpha _0-a_3 e_0 e_1\right)+a_3 e_1 \left(5 e_0 e_2-2
   e_1^2\right) \alpha _0+a_2^2 e_0 e_2+a_3^2 e_0^2+e_2^2 \left(2 e_1^2+e_0 e_2\right) \alpha _0^2 \,.
\end{split}\end{equation}
It is easy to identify each of these with the projection to $S_{\rm GUT}$ of various intersection loci of ${\cal{C}}^{(1)}$ and ${\cal{C}}^{(2)}$ as
\begin{equation}
{\cal{C}}^{(1)}\cap {\cal{C}}^{(1)}\rightarrow F_1\,,
\qquad {\cal{C}}^{(2)}\cap {\cal{C}}^{(2)}\rightarrow F_2\,,
\qquad {\cal{C}}^{(1)}\cap {\cal{C}}^{(2)}\rightarrow F_3\,.
\end{equation}
By $\rightarrow$ here we mean that $F_i$ is the projection of the specified intersection locus down to $S_{\rm GUT}$.


\subsubsection{Index computation for Singlets}

Singlets which carry a nontrivial $U(1)$ charge are associated with the locus $F_3=0$.  In order to compute the index associated to these singlets, we must understand all of the relevant gauge fluxes.  If we proceed by naive analogy to the counting of $\mathbf{10}$ and $\mathbf{\overline{5}}$ zero modes, the first step then is to identify the appropriate singlet locus inside the spectral cover, as it is in this context that we can construct $SU(5)_{\perp}$ fluxes that can be defined in the presence of the nontrivial monodromy group $G$.  The relevant curve for singlets associated to $F_3$ is the intersection ${\cal{C}}^{(1)}\cap {\cal{C}}^{(2)}$, which contains two components
\begin{itemize}
\item $\lambda_a=\lambda_m=$ finite
\item $\lambda_a,\lambda_m\rightarrow\infty$
\end{itemize}
The second component, which we refer to as the component at $\infty$, is defined by $e_0=0$, $V=0$.  Its presence reflects the fact that as we approach $e_0=0$ inside $S_{\rm GUT}$ one root of ${\cal{C}}^{(1)}$ and one root of ${\cal{C}}^{(2)}$ simultaneously approach $\infty$.

By analogy with the study of $\mathbf{\overline{5}}$ matter curves, one might wonder whether we have to be careful with the component at $\infty$.  Let us recall, however, the reason that the component at $\infty$ is explicitly thrown out in the $\mathbf{\overline{5}}$ case.   There, we are interested in the locus $\lambda_a+\lambda_m=0$ and identified it by studying the intersection ${\cal{C}}^{(1)}\cap \tau {\cal{C}}^{(2)}$, where $\tau$ is defined as a map that sends $V\rightarrow -V$.  While $\tau$ usually has the effect of sending $\lambda_i\rightarrow -\lambda_i$, this is not true for the "points at $\infty$" sitting at $V=0$, which are $\tau$-invariant.  As such, we get a component at $\infty$ in the intersection ${\cal{C}}^{(1)}\cap \tau {\cal{C}}^{(2)}$ that does not actually correspond to a locus where $\lambda_a=-\lambda_m$; rather it is a locus where $\lambda_a=\lambda_m\sim \infty$ that happens to contribute simply because the points at $V=0$ are invariant under $\tau$.  The component at $\infty$ in the $\mathbf{\overline{5}}$ case, then, represents an additional contribution to the intersection that has to be removed in order to obtain the specific curve that we want, which is the one locally given by $\lambda_a=-\lambda_m$.

In the case of singlets, there is no analogous reason to discard the component at $\infty$.  Its presence reflects the fact that $\lambda_a=\lambda_m$ admits solutions where both simultaneously diverge.  This does not  contradict anything that we know about singlets because we do not expect them to localize on curves contained inside $S_{\rm GUT}$ anyway.  We will see this in more detail in the next subsection from a global description of the singlet locus.

We conclude, then, that the correct "matter curve" for singlet fields in ${\cal{C}}$ is the entire locus ${\cal{C}}^{(1)}\cap {\cal{C}}^{(2)}${\footnote{As a nice check, it is easy to verify that the topological class of this curve takes the form $\sigma\cdot \pi^{\ast}([F_3]) + \pi^{\ast}\alpha\cdot \pi^{\ast}\beta$ where $[F_3]$ is the class of the curve $F_3=0$ inside $S_{\rm GUT}$.}}.  If we construct spectral fluxes on ${\cal{C}}^{(1)}$ and ${\cal{C}}^{(2)}$ of the form
\begin{equation}{\cal{N}}_1=\frac{r_1}{2}+\gamma_1\,,\qquad {\cal{N}}_2=\frac{r_2}{2}+\gamma_2\,.
\end{equation}
where $p_{1\,\ast}{\cal{N}}_1+p_{2\,\ast}{\cal{N}}_2=0$ and the $r_i$ are ramification divisors of $p_i:{\cal{C}}^{(i)}\rightarrow S_{\rm GUT}$, a natural guess for the net chirality of singlets is then{\footnote{Note that analogous formulae for uncharged singlets, $\int_{ {\cal{C}}^{(i)}\cap {\cal{C}}^{(i)} }(\gamma_i-\gamma_i)=0$ trivially vanishes as we expect.}}
\begin{equation}
n_{\mathbf{\overline{1^{(1)(2)}}}}-n_{\mathbf{1^{(1)(2)}}} =\int_{ {\cal{C}}^{(1)}\cap {\cal{C}}^{(2)}} \left(\gamma_1-\gamma_2\right)\,.
\label{singconj}\end{equation}
A very similar conjecture was made recently in \cite{Blumenhagen:2009yv}.  Ours differs only in the identification of the matter curve as the entire intersection ${\cal{C}}^{(1)}\cap {\cal{C}}^{(2)}$ rather than only the component at $\infty$.


\subsubsection{Cohomology computation for Singlets}
\label{subsec:Coh}

In fact in order to realize the Dirac neutrino scenarios as well as gauge-mediation with the non-GUT exotics we will require both types of singlets: ${\bf 1}^{(1)(2)}$ and $\overline{{\bf 1}^{(1)(2)}}$. 
Recall that 
\be
n_{\overline{{\bf 1}^{(1)(2)}}} - n_{{\bf 1}^{(1)(2)}} = 
h^0\left(\mathcal{C}^{(1)} \cap \mathcal{C}^{(2)}, {\cal{N}}_1\otimes {\cal{N}}_2^{-1}\otimes K_{S_{\rm GUT}} \right) - h^1\left(\mathcal{C}^{(1)} \cap \mathcal{C}^{(2)}, {\cal{N}}_1\otimes {\cal{N}}_2^{-1}\otimes K_{S_{\rm GUT}} \right) \,. 
\ee
We further conjecture that 
\be
\ba
n_{\overline{{\bf 1}^{(1)(2)}}}& = h^0\left(\mathcal{C}^{(1)} \cap \mathcal{C}^{(2)}, {\cal{N}}_1\otimes {\cal{N}}_2^{-1}\otimes K_{S_{\rm GUT}} \right) \cr
 n_{{\bf 1}^{(1)(2)}} & = h^1\left(\mathcal{C}^{(1)} \cap \mathcal{C}^{(2)}, {\cal{N}}_1\otimes {\cal{N}}_2^{-1}\otimes K_{S_{\rm GUT}} \right) \,,
\ea
\ee
although we have no a priori justification for it apart from the similarity to the heterotic formulae in the next subsection. 
To evaluate these bundle cohomologies over the curve $\Sigma=\mathcal{C}^{(1)} \cap \mathcal{C}^{(2)}$ we need to compute the genus of the curve and the degree of the line-bundle. Note that from adjunction from $\mathcal{C}^{(1)}$
\be
2 g -2 = \Sigma^2 + K_{\mathcal{C}^{(1)}} \cdot \Sigma \,.
\ee
The degree of the normal bundle of $\Sigma$ in $\mathcal{C}^{(1)}$ is 
\be
 \Sigma^2 = N_{\Sigma| \mathcal{C}^{(1)}} = \Sigma \cdot_{X} \mathcal{C}^{(2)}  =  \mathcal{C}^{(1)}  \cdot \mathcal{C}^{(2)}  \cdot\mathcal{C}^{(2)}\,.
\ee
Furthermore, we can obtain $K_{\mathcal{C}^{(1)}}$ is obtained from adjunction in $X$ 
\be
K_{\mathcal{C}^{(1)}} = K_X |_{\mathcal{C}^{(1)}} + \mathcal{C}^{(1)} \cdot \mathcal{C}^{(1)} \,.
\ee
Putting all this together and recalling that $K_X = -2 \sigma_\infty$ we find 
\be\label{2g2}
2 g -2 =  \mathcal{C}^{(1)}\cdot    \mathcal{C}^{(2)}\cdot    \mathcal{C}^{(2)}  - 2 \sigma_\infty \cdot   \mathcal{C}^{(1)} \cdot  \mathcal{C}^{(2)} +  \mathcal{C}^{(1)}\cdot  \mathcal{C}^{(1)} \cdot  \mathcal{C}^{(2)}  \,. 
\ee
Here, $g$ is the genus of the curve $\Sigma$.
The degree of the bundle is by Riemann-Roch
\be
\hbox{deg} \left({\cal{N}}_1\otimes {\cal{N}}_2^{-1}\otimes K_{S_{\rm GUT}} \right) =  g-1 + n_{\overline{{\bf 1}^{(1)(2)}}} - n_{{\bf 1}^{(1)(2)}} \,.
\ee
If we wish to have both ${{\bf 1}^{(1)(2)}}$ and $\overline{{\bf 1}^{(1)(2)}}$ we require that
\be\label{h1h0}
h^1 \not=0 \qquad \hbox{and} \qquad h^0 - h^1 \geq 0 \,.
\ee
A necessary condition for (\ref{h1h0}) is that 
\be\label{degBundle}
0  \leq \hbox{deg} \left({\cal{N}}_1\otimes {\cal{N}}_2^{-1}\otimes K_{S_{\rm GUT}} \right) \leq 2g -2 \,.
\ee
Evaluating this in a particular geometry may give us further constraints on the allowed fluxes. As we will check later on, in the concrete three-generation models that we studied, this does not impose any further restrictions.


\subsubsection{Comparison with Formulae from the Heterotic String}

This conjecture is partly motivated by its similarity to formulae for the $\mathbf{10}$ and $\mathbf{\overline{5}}$ fields but also from the perspective of analogous counting formulae in the Heterotic string.  To see this, let us imagine completing our semi-local model into a global K3 fibration that admits a standard Heterotic dual.  We will denote the Calabi-Yau 3-fold on the Heterotic side by $\hat{X}$.  The data of our spectral cover, ${\cal{C}}$, also determine a spectral surface $\hat{\cal{C}}$ embedded in $\hat{X}$ that does not factor in general, even when ${\cal{C}}$ does.  Nevertheless, factorization of ${\cal{C}}$ and the specification of two distinct spectral bundles ${\cal{N}}_1$ and ${\cal{N}}_2$, there, corresponds to the specification of an $S[U(3)\times U(2)]$ spectral bundle on $\hat{\cal{C}}$.
Let us consider for simplicity the case in which ${\cal{N}}_1$ and ${\cal{N}}_2$ are individually traceless so that $p_{1\,\ast}{\cal{N}}_1=p_{2\,\ast}{\cal{N}}_2=0$.  In this case, we have two spectral bundles on $\hat{\cal{C}}$, an $SU(3)$ bundle and an $SU(2)$ bundle.  These, in turn, give rise to two vector bundles, $V_3$ and $V_2$ on the Calabi-Yau 3-fold $\hat{X}$.

Let us recall the formulae for counting the numbers of $\mathbf{\overline{5}}$'s and $\mathbf{1}$'s on the heterotic side.  These are given by the dimensions of various cohomology groups as follows
\begin{equation}\begin{split}
n_{\mathbf{5}} &= h^1(X,\Lambda^2 V) \\
n_{\mathbf{\overline{5}}} &= h^1(X,\Lambda^2 V^{\ast})=h^2(X,\Lambda^2V) \\
n_{\mathbf{\overline{1}}} &= h^1(X,V_3\otimes V_2^{*}) \\
n_{\mathbf{1}} &= h^1(X,V_3^{*}\otimes V_2)=h^2(X,V_3\otimes V_2^*) \,,
\end{split}\end{equation}
where $V=V_2\oplus V_3$ and we made use of Serre duality and the fact that $\hat{X}$ is Calabi-Yau.  Because $h^0(X,\Lambda^2V)=h^3(X,\Lambda^2V)=0$ the difference $n_{\mathbf{\overline{5}}}-n_{\mathbf{5}}$ can be computed by a chiral index on $X$ \cite{Donagi:2004ia}
\begin{equation}n_{\mathbf{\overline{5}}}-n_{\mathbf{5}} = -\chi(X,\Lambda^2V)\,.
\end{equation}
This can be related to a computation on a matter curve through the relation \cite{Blumenhagen:2006wj}
\begin{equation}H^{i+1}(X,V_a\otimes V_b) = H^i(\tau {\cal{C}}_a\cap {\cal{C}}_b,\tau^{\ast}{\cal{N}}_a\otimes {\cal{N}}_b\otimes K_{S_{\rm GUT}})\,.
\label{hetF}\end{equation}
Note that if we apply this to $H^{i+1}(X,V^2)$ in order to study $\mathbf{\overline{5}}$'s, it is necessary to split the result into contributions to the symmetric and antisymmetric parts, $H^{i+1}(X,S^2V)$ and $H^{i+1}(X,\Lambda^2V)$, as described in detail in \cite{Blumenhagen:2006wj}.  In so doing, the curve $\tau {\cal{C}}_a\cap {\cal{C}}_b$ on the RHS of \eqref{hetF} is replaced by the usual $\mathbf{\overline{5}}$ matter curve with components along $V=0$ (the component at $\infty$) and $U=0$ removed.

As for singlet fields, we have that $h^0(X,V_3\otimes V_2^{\ast})=h^3(X,V_3\otimes V_2^{\ast})=0$ so that their net chirality can also be computed by an index.  Using \eqref{hetF} we can relate this to a computation on a matter curve
\begin{equation}n_{\mathbf{\overline{1}}}-n_{\mathbf{1}} = \chi({\cal{C}}^{(1)}\cap {\cal{C}}^{(2)},{\cal{N}}_1\otimes {\cal{N}}_2^{-1}\otimes K_{S_{\rm GUT}})\,,
\end{equation}
which evaluates to precisely the integral that we conjectured above \eqref{singconj}.  Note that, unlike the $\mathbf{\overline{5}}$ case, we don't have to split the contributions to the formula (\ref{hetF}) into multiple pieces analogous to the symmetric and antisymmetric parts of $V^2$.  As such, we see no reason that any part of ${\cal{C}}^{(1)}\cap {\cal{C}}^{(2)}$ should be removed in the computation.

Of course, to directly apply results from the Heterotic side we must either be working with a full $F$-theory compactification that admits a Heterotic dual or be able to argue that details of the UV completion of our semi-local model do not  affect the computation.  In the case of $\mathbf{10}$'s and $\mathbf{\overline{5}}$'s, it is obvious that the latter should be true because the charged fields localize entirely on $S_{\rm GUT}$.  Singlets are not localized on $S_{\rm GUT}$, however, so it seems that they can probe details of the compactification beyond those directly captured by the semi-local picture.  For that reason, it is necessary to obtain a more global understanding of singlet matter curves in order to proceed.

\subsection{Singlets in the Three-generation Models}

Finally, we can now compute the number of singlets for the fluxes that we discussed and  in particular for the models discussed in sections \ref{subsubsec:Q1} and \ref{subsubsec:Q}.  Recall that the general form of the  fluxes on the spectral cover factors $\mathcal{C}^{(1)}$ and $\mathcal{C}^{(2)}$ are given in (\ref{Gamma1Gamma2}) and have the general structure
\be
\ba
\Gamma_1 &= (3 \tilde{m}_1 + \tilde{q}) \Psi_1 +  \mathcal{C}^{(1)} \cdot \mathcal{D}_1 \cr
\Gamma_2 &= (2 \tilde{m}_2 - \tilde{q}) \Psi_2 +  \mathcal{C}^{(2)} \cdot \mathcal{D}_2 \,,
\ea
\ee
with $\mathcal{C}_i$ as specified in (\ref{Gamma1Gamma2}).
To obtain the chiral number of singlets we need to evaluate
\be
n_{\bf 1_{-5}} - n_{\bf 1_{+5}} = \int_{\mathcal{C}^{(1)} \cap \mathcal{C}^{(2)}}  \left( \Gamma_1 - \Gamma_2\right) \,.
\ee
Note that the intersection with the non-universal fluxes $\Psi_i$ is easily obtained by
\be
\ba
\int_{\mathcal{C}^{(1)} \cap \mathcal{C}^{(2)}}  \Psi_1 &  = (2 c_1 + \xi) \cdot_{S_{\rm GUT}} \psi \cr
\int_{\mathcal{C}^{(1)} \cap \mathcal{C}^{(2)} } \Psi_2 &  = (\eta -2 c_1 - \xi) \cdot_{S_{\rm GUT}} \psi \,.
\ea
\ee
Inserting the expressions for the universal fluxes in terms of $\mathcal{D}_i$ we obtain
\be
\ba
n_{\bf 1_{-5}} - n_{\bf 1_{+5}}=
&
c_1 \left(14 \eta  \left(\tilde{k}_1-\tilde{d}_1\right)+\xi  \left(-11 \tilde{d}_1+19 \tilde{d}_2-14
   \tilde{k}_1+6 \tilde{k}_2\right)+10 \tilde{m}_1 \psi -20 \rho \right)    
    \cr
&  +20 c_1^2
   \left(\tilde{d}_1-\tilde{k}_1\right)  +\eta  \left(\xi  \left(\tilde{d}_1-5 \tilde{d}_2+4
   \tilde{k}_1\right)+\psi  \left(\tilde{q}-2 \tilde{m}_1\right)+10 \rho \right)
   +2 \eta ^2 \left(\tilde{d}_1-\tilde{k}_1\right) \cr
&   +\xi ^2 \left(-3 \tilde{d}_1+2 \tilde{d}_2-2 \tilde{k}_1+3
   \tilde{k}_2\right)+\left(2 \tilde{m}_1+3 \tilde{m}_2\right) \xi  \psi +5 \xi  \rho  \,.
\ea
\ee
With the specific classes for $\eta, \xi, \psi$ this simplifies to
\be
n_{\bf 1_{-5}} - n_{\bf 1_{+5}}= 17 \tilde{d}_1-17 \tilde{d}_2-12 \tilde{k}_1+12 \tilde{k}_2-2 \tilde{m}_1+3 \tilde{m}_2+12 \tilde{q}+15 \,.
\ee
For the fluxes of sections \ref{subsubsec:Q1} and \ref{subsubsec:Q} which gave three-generations of MSSM matter and were consistent with the D3-tadpole cancellation with only D3-branes, we obtain
\be\label{SingletCount}
\ba
Q=1 :&\qquad  n_{\bf 1_{-5}} - n_{\bf 1_{+5}} =  25 + 52 d_1 \cr
Q\in \mathbb{N}: & \qquad  n_{\bf 1_{-5}} - n_{\bf 1_{+5}} = -13 - 14 Q \,.
\ea
\ee
In particular these are independent of the choices for $k_2$.

Recall, that in the ${\bf 3+2}$ models we were only able to realize Dirac neutrino scenarios, which were generated by the coupling (\ref{DiracNeutrino}). This requires singlets with $U(1)_{PQ}$ charge $+5$. On the other hand in order to give masses to the non-GUT exotics, e.g. in order to use them as messengers in a gauge-mediation scenario, we require singlets of $U(1)_{PQ}$ charge $-5$. As we  discussed in section  \ref{subsec:Coh},  a necessary requirement for this is (\ref{h1h0}).
Explicit computation yields
\be
2g-2 = 188 \,,
\ee
and so together with (\ref{SingletCount}) we find that in fact all flux choices that satisfied the D3-tadpole constraint automatically fall into the bound (\ref{degBundle}). To determine whether both types of singlets are present, one will have to compute $h^1$ and $h^0$ explicitly, which may further constrain the allowed models. 
In summary we find:
\be\label{SummaryModel}
\boxed{
$D3-tadpole, 3 generations  $ \\
\Rightarrow \ $Fluxes in (\ref{Q1Flux}) and (\ref{QFlux}), and  (\ref{degBundle}) is met$}\,.
\ee



\subsection{The Global Perspective}

We now turn to a study of the singlet locus in the global picture of a full, honest $F$-theory compactification.  To start, recall that for $SU(5)$ models, the discriminant locus of our 4-fold takes the form
\begin{equation}\Delta \sim z^5 P_m(b_n,z) \,.
\end{equation}
where $z$ is a holomorphic section on the base $B_3$ whose vanishing defines the GUT divisor, $S_{\rm GUT}$.  The section $P_m$ here is a polynomial in the holomorphic sections $b_n$ and $z$ along which the singularity type is $U(1)$.  This generically irreducible component of the discriminant locus, ${\cal{D}}$, is the $F$-theoretic analog of type IIB "matter branes" whose intersection with $S_{\rm GUT}$ gives rise to charged fields.  GUT singlet fields, on the other hand, localize on curves of codimension 1 in ${\cal{D}}$ where the singularity type enhances from $U(1)$ to $SU(2)$.

Details of ${\cal{D}}$ and the $SU(2)$ enhancement locus depend largely on the global structure of the fibration.  For now, let us focus on CY 4-folds of the type that we have constructed explicitly here and in \cite{Marsano:2009gv}, where the holomorphic sections $f$ and $g$ appearing in the Weierstrass form
\begin{equation}y^2 = x^3 + fx + g\end{equation}
are written as explicit polynomials in the holomorphic section $z$ of degree 3 and 5, respectively
\begin{equation}f = \sum_{i=0}^3 f_i z^i\,,
\qquad g = \sum_{j=0}^5 g_j z^j\,.
\end{equation}
The coefficients $f_i$ and $g_j$ are $z$-independent and are related to the $b_n$ by
\begin{equation}
\begin{split}
f_0 &= -b_5^4 \\
f_1 &= -8b_4b_5^2 \\
f_2 &= 8(3b_3b_5-2b_4^2)\\
f_3 &= 48b_2 \\
g_0 &= b_5^6 \\
g_1 &= 12 b_4b_5^4 \\
g_2 &= 12b_5^2(4b_4^2-3b_3b_5)\\
g_3 &= 8(8b_4^3-18b_3b_4b_5-9b_2b_5^2)\\
g_4 &= 72(3b_3^2-4b_2b_4)\\
g_5 &= 864b_0\,.
\end{split}\end{equation}

Since $z=0$ is a rigid divisor in the base $B_3$, it is expected that an expansion of $f$ and $g$ as polynomials in $z$ will terminate at some order and this choice corresponds to the minimal one for which all singularities are of ADE type.  In this case, the polynomial $P_m(b_n,z)$ is of degree 5 and begins
\begin{equation}P_5(b_n,z)\sim b_0^2 z^5 + \text{lower order in }z\,.
\end{equation}
We can roughly think of $P_5(b_n,z)$ as defining a divisor that covers $S_{\rm GUT}$ five times.

To study the singularity structure, let us begin by studying patches that contain (parts of) $S_{\rm GUT}$.  Within any such patch, we can use the holomorphic section $z$ to define an affine coordinate, $\hat{z}$.  As long as we are away from the locus $b_0=0$, we can locally factor $P_5(b_n,z)$ into five sheets explicitly
\begin{equation}P_5(b_n,z) \sim \prod_{i=1}^5 (\hat{z}-z_i)\,.
\end{equation}
On each sheet, we can use $\hat{z}$ as a local normal coordinate for determining the singularity type.  An enhancement from $U(1)$ to $SU(2)$ will arise whenever $z_i=z_j$ for some $i\ne j$.  This, in turn, occurs whenever the discriminant $\tilde{\Delta}$ of $P_5(b_n,z)$ vanishes
\begin{equation}\tilde{\Delta}\sim \prod_{i<j}(z_i-z_j)=0\,.
\end{equation}
It is easy to compute this discriminant explicitly, with the result
\begin{equation}\tilde{\Delta}\sim b_5^5\times F\times G^3\,,
\end{equation}
where
\begin{equation}
\begin{split}
G=&-32b_3^3b_4^2(9b_2b_3^2-4b_2^2b_4+12b_0b_4^2) \\
&+3b_3^2(9b_2b_3^4+204b_2^2b_3^2b_4-64b_2^3b_4^2+204b_0b_3^2b_4^2+192b_0b_2b_4^3)b_5 \\
&-6b_3b_5^2(48b_2^3b_3^2+9b_0b_3^4-16b_2^4b_4+300b_0b_2b_3^2b_4+48b_0b_2^2b_4^2)\\
&+4b_5^3(-4b_2^5+261b_0b_2^2b_3^2+12b_0b_2^3b_4+270b_0^2b_3^2b_4)\\
&-1080b_0^2b_2b_3b_5^4+324b_0^3b_5^5\,.
\end{split}\end{equation}
and $F$ is the section that entered into our description of the singlet locus in the semi-local picture \eqref{Fdef}.

Based on our semi-local reasoning, we expect that our singlet fields localize along the intersection of $F=0$ with ${\cal{D}}=0$.  To be completely sure, however, we should understand the nature of the $b_5$ and $G$ factors in $\tilde{\Delta}$.
When $b_5=0$, for instance, two roots of $P_5(b_n,\hat{z})$ collide with one another at $\hat{z}=0$.  This is responsible for enhancing the $SU(5)$ singularity at $z=0$ to an $SO(10)$ singularity so does not indicate the presence of any new singlets.  The factor of $G$, on the other hand, signals the presence of the "cusp curve" studied by Andreas and Curio \cite{Andreas:2009uf}.  This is a curve of intrinsic cusp singularities along $f=g=0$ above which the fiber exhibits a type II singularity.  To see why, note that we can determine an equation for this locus in terms of the $b_m$ from a ratio of discriminants
\begin{equation}\frac{\Delta_{fg}}{\Delta_f\Delta_g}=\prod_{i,m}(z_i-z_m)\,.
\label{discratio}\end{equation}
where the $z_i$ denote zeroes of $f$, the $z_m$ denote zeroes of $g$, and $\Delta_h$ denotes the  discriminant of $h$ viewed as a polynomial in $z$.  The ratio \eqref{discratio} is easily evaluated with the result
\begin{equation}\frac{\Delta_{fg}}{\Delta_f\Delta_g}\sim G^2\,.
\end{equation}
The $b_5$ and $G$ components of $\tilde{\Delta}$ are therefore well-understood and do not serve as obvious sources for singlet fields.  This leaves us essentially with the $F=0$ component, which is what we expected at the outset from our semi-local reasoning.

An important caveat to this is that we have not  accounted for possible contributions to the singlet locus from $b_0=0$.
As we approach $b_0=0$ within a coordinate patch containing $S_{\rm GUT}$, the degree of $P_5(b_n,\hat{z})$ in $\hat{z}$ drops by one, signifying that one of the roots $z_i$ seems to be approaching $\infty$.  We must be careful about this region because, as we follow the diverging sheet toward $b_0=0$, $\hat{z}$ no longer serves as a good local normal coordinate for studying the singularity type.  A nice choice to make for this part of the geometry is instead $b_0$ as the troublesome locus is sitting at the fixed value $b_0=0$.  Viewed as a polynomial in $b_0$, $P_5(b_n,\hat{z})$ is in fact quadratic so that $b_0(\hat{z})$ exhibits two different sheets.  Further, because the quadratic term is precisely given by
\begin{equation}P_5(b_n,z)\sim b_0^2 \hat{z}^5 + \text{ lower order in }b_0\end{equation}
it is easy to see that these two sheets both approach $b_0=0$ as $\hat{z}\rightarrow \infty${\footnote{These two sheets also intersect one another at finite $z$ along the cusp curve of \cite{Andreas:2009uf}.}}.  Of course, in the limit that $\hat{z}\rightarrow\infty$ we exit our coordinate patch so holomorphic sections must be transformed with the appropriate transition functions.  This means that holomorphic sections $b_m$ will now depend on the affine coordinate $\hat{z}$ but the most important observation is that the limit $\hat{z}\rightarrow\infty$ will land us on a "divisor at $\infty$" just beyond the reach of our original coordinate patch, which contained $S_{\rm GUT}${\footnote{In the explicit compact models of this paper, this "divisor at $\infty$" is defined by $Z_4=0$.}}.  By continuity of the $\hat{z}\rightarrow\infty$ limit we expect that the new contribution to the singlet locus is the intersection of this divisor with $b_0=0$.


This contribution is what we expect is being captured by the "component at $\infty$" of the singlet matter curve in the semi-local picture.  Setting $b_0=0$ in the spectral cover ${\cal{C}}$ reduces the degree from 5 to 3, meaning that two of the sheets, and hence two of the $\lambda_i$, have simultaneously moved off to $\infty${\footnote{The traceless condition $\sum_j\lambda_j=0$ makes it impossible for one $\lambda_i$ to move to $\infty$ by itself.}}.  In the case of our factored spectral cover, $b_0=e_0^2\alpha$ with $e_0=0$ corresponding to the case in which one sheet of ${\cal{C}}^{(1)}$ and one sheet of ${\cal{C}}^{(2)}$ simultaneously move off to $\infty$.



\subsubsection{Fluxes in the Global Picture}

To count singlet zero modes correctly, then, it is necessary to construct fluxes in the full 4-fold as opposed to in the semi-local description only.  A natural candidate for this is to proceed as follows.  First, we "disentangle" the monodromy structure by constructing an auxiliary 4-fold as a projective bundle over our 3-fold base $B_3$.  This is easily done by introducing projective coordinates $U$ and $V$.  We then define a 5-fold cover $\tilde{\cal{C}}$ of $B_3$ by the polynomial
\begin{equation}\sum_{m=1}^5 V^m U^{5-m}b_m=0\,,
\end{equation}
where, as in the rest of this subsection, the $b_m$ here are interpreted as holomorphic sections on $B_3$.  All we have done is extend the spectral cover construction over $S_{\rm GUT}$ to a similar cover over $B_3$.  We can specify the 2-forms $F_i$ for our fluxes by constructing divisors in this cover.  Indeed, we can use the same equations as before by simply reinterpreting the $b_m$'s again as holomorphic sections on $B_3$.  $G$-fluxes are built by analogy with the semi-local picture by wedging the $F_i$ with harmonic 2-forms dual to the appropriate holomorphic $\mathbb{P}^1$'s that grow when the $E_8$ singularity is unfolded.  While this picture seems very simplistic, we see no obvious obstacle to constructing $G$-fluxes in this way.  It is also easy to see that, as with the relationship between ${\cal{C}}$ and the heterotic spectral cover, $\tilde{\cal{C}}$ provides a local description of a divisor in the full $Y_4$ in which the five sheets covering $B_3$ all live in the elliptic fiber, rather than a $\mathbb{P}^1$.

For the purposes of constructing fluxes, what we have done is replace a 2-cycle on a sheet of ${\cal{C}}$ with a 4-cycle on the corresponding sheet of $\tilde{\cal{C}}$.  Because the 2-cycle on ${\cal{C}}$ is a lift of a curve $\rho$ in $S_{\rm GUT}$, the 4-cycle on $\tilde{\cal{C}}$ will be a similar lift of the divisor in $B_3$ obtained by pulling $\rho$ back from $S_{\rm GUT}$ to $B_3$.   In a coordinate patch containing $S_{\rm GUT}$, the 4-cycle in $\tilde{\cal{C}}$ is roughly just the product of $\rho$ with the $z$-direction, "$\rho\times \{z\}$".  This means that for a matter curve  at finite $z$ constructed as a lift of a curve $\Sigma$ in $S_{\rm GUT}$, intersections with a flux "divisor" (either a 2-cycle in ${\cal{C}}$ or a 4-cycle in $\tilde{\cal{C}}$) are essentially computed by $\rho_{\cdot S_{\rm GUT}}\Sigma$ in both the semi-local picture and the global picture described here.  This is the agreement that we anticipated for matter curves within the vicinity of $S_{\rm GUT}$.

The only subtlety arises from the "component at $\infty$" of the singlet matter curve.  This is the place where semi-local reasoning has a chance of breaking down.  Nevertheless, it naively seems that the intersection of the "component at $\infty$" of the singlet matter curve should really be no different because it sits at a fixed value of the affine coordinate $\hat{z}$.  Again, the computation in both the semi-local and global pictures will boil down to an intersection inside $S_{\rm GUT}$.  While this is far from a proof, it seems to indicate that the conjecture \eqref{singconj} is not completely unreasonable.

\section{Outlook}

In this paper we F-theoretically realized $SU(5)$ GUTs with a gauged $U(1)_{PQ}$ symmetry. The PQ symmetry is crucial for many model building questions, including the prevention of a tree-level $\mu$-term and the realization of neutrino scenarios of \cite{Bouchard:2009bu} in F-theory GUTs. 

From the analysis in \cite{Marsano:2009gv} it followed that if one requires embeddability into a compact CY four-fold, then the existence of a $U(1)_{PQ}$ symmetry always comes at the price of  non-GUT exotics in the model. On the other hand, these exotics are required in order to explain the generalized gauge coupling F-unification, and are phenomenologically distinct from minimal GUT models, as they give rise to non-standard gauge-mediated soft masses, which we discussed in this paper. 
We presented a detailed description of these models in the semi-local regime, which a priori is independent of a specific CY four-fold,  and then specializing to the geometry of \cite{Marsano:2009ym}, analyzed $G$-fluxes, which give rise to realistic three-generation models and at the same time satisfy the D3-tadpole constraint. Finally, the models that satisfied the D3-tadpole constraint automatically also satisfy the necessary condition for the existence of 
 singlets with  $+5$ PQ charge, which are candidate right-handed neutrinos $N_R^i$ that can participate in (\ref{DiracNeutrino}), as well as singlets with $-5$ PQ charge, to which the non-GUT vector-like pairs of exotics can couple. 
These models are characterized by the flux choices in (\ref{Q1Flux}) and (\ref{QFlux}). 
%
Thus, embedding into semi-local F-theory compactifications already highly reduces the allowed number of models, which is a more than welcome property given the vastness of local models. 

We would like to stress once more, that  semi-local models provide a very constraining, and thus phenomenologically desirable framework for string model building. The most appealing feature is the generality with which they allow one to address the constraints arising from embedding into F-theory compactifications.

One interesting question is whether we have provided a general analysis of all possible semi-local models of F-theory GUTs -- we commented on this question already in section \ref{subsec:Consts}.
  Requiring no exotics, i.e. a realization of a minimal $SU(5)$  GUT in a compact CY four-fold, implied that the spectral cover factors into ${\bf 4+1}$ \cite{Marsano:2009gv} and no gauged $U(1)_{PQ}$. Requiring a  gauged $U(1)_{PQ}$ in a semi-local model always implies additional exotics. Allowing for such exotics, there are both ${\bf 4+1}$ and 
   ${\bf 3+2}$ factorizations of the spectral cover. The ${\bf 3+2}$ model has the advantage, that no further tuning of the spectral cover coefficients is required in order to realize the GUT model, which is why we considered it here.  
The remaining cases, which in principle could realize $SU(5)$ GUTs with the correct Yukawas and a $U(1)_{PQ}$ are ${\bf 2+2+1}$, which would have a $U(1)_{PQ}$ symmetry and an additional $U(1)$, as well as ${\bf 2+1+1+1}$ with a  $U(1)^3$. We have briefly analyzed the case of ${\bf 2+2+1}$ and found that solving the $b_1=0$ constraint usually introduces singularities of a non-Kodaira type, and thus renders the spectral cover singular. The case of a fully factored spectral cover has  four additional independent $U(1)$ gauge groups, (as in the local models), but it is clear that one cannot realize the GUT Yukawa couplings in this scenario. 

In this paper we presented explicit ${\bf 3+2}$ factored models with fluxes that give rise to three-generation $SU(5)$ GUTs (and satisfy the D3-tadpole condition), with $Q$ additional non-GUT vector-like exotics, arising from GUT multiplets as specified in (\ref{Exotics}). Furthermore, there are singlets, with multiplicities depending on $Q$, given in (\ref{SingletCount}). It would be nice to explictly realize a gauge-mediated SUSY-breaking scenario where the exotics play the role of gauge messengers, and the singlets obtain an F-term from some SUSY-breaking mechanism in a hidden sector -- much in the same spirit as the local constructions in \cite{Marsano:2008jq}. Of course, this question cannot be answered without at the same time addressing the issue of moduli stabilization.


\section*{Acknowledgements}

We thank F.~Quevedo, T.~Watari, and T.~Weigand for helpful discussions. We are in particular grateful to R.~Blumenhagen, T.~Grimm and T.~Weigand for bringing to our attention the modified D3-tadpole analysis in their paper. 
The work of JM and SSN was partially supported by John A. McCone Postdoctoral
Fellowships. The research of JM was supported also in part by the DOE grant DE-FG02-90ER-40560.  The          
research of SSN was supported also in part by the National Science Foundation under     
Grant No. PHY05-51164. 
 The work of NS was supported in part by the DOE-grant DE-FG03-92-ER40701. SSN thanks the Aspen Center for Physics, Brandeis University, University of Michigan and  Perimeter Institute for hospitality.  NS thanks the Simons Workshop on Geometry and Physics
for hospitality during the course of this work.  JM is grateful to the theory groups at the University of Illinois and Ohio State University, the Perimeter Institute, the Yukawa Institute for Theoretical Physics at Kyoto University, and the organizers of the workshop "Branes, Strings, and Black Holes" for their hospitality.

\newpage

\appendix


\section{Matter curves in the Spectral Cover}
\label{app:MatterinSC}

In this appendix we provide a detailed analysis of the matter curves and their origin in the spectral cover of the ${\bf 3+2}$ factored cover, that we refered to in section \ref{subsec:MatterinSC}. Recall that ${\bf 5}$ matter curves arise from the intersection of $\mathcal{C} \cap \tau \mathcal{C}$, where $\tau$ is the involution $V\rightarrow -V$ \cite{Marsano:2009gv}.


\subsection{$\mathcal{C}^{(2)}\cap \tau \mathcal{C}^{(2)}$}

For the intersection $\mathcal{C}^{(2)}\cap \tau \mathcal{C}^{(2)}$, we consider the invariant locus under $\tau$ inside $\mathcal{C}^{(2)}$.
The net class of this intersection is
\begin{equation}
{\cal{C}}^{(2)}\cdot {\cal{C}}^{(2)} = 4\sigma\cdot \pi^{\ast}(c_1+\xi) + \left[\pi^{\ast}(2c_1+\xi)\right]^2 \,.
\end{equation}

The equations for this intersection are

\begin{equation}\begin{split}
0 &= e_0 U^2 + e_2V^2 \\ 0
&= e_1 UV\,.
\end{split}\end{equation}
We find solutions with
\begin{equation}U=e_2=0\,,\qquad V=e_0=0\,,
\end{equation}
and
\begin{equation}e_1=0\,,\qquad e_0U^2+e_2V^2=0\,.
\end{equation}
These correspond to a $\mathbf{10}$ matter curve in the class
\begin{equation}\sigma\cdot \pi^{\ast}\xi\,,
\end{equation}
a component at infinity in the class
\begin{equation}\sigma_{\infty}\cdot \pi^{\ast}(2c_1+\xi)\,,
\end{equation}
and a $\mathbf{5}$ matter curve in the class
\begin{equation}\left[2\sigma + \pi^{\ast}(2c_1+\xi)\right]\cdot \pi^{\ast}(c_1+\xi)\,.
\end{equation}
We can summarize this in the table
\begin{equation}\begin{array}{c|c|c}\text{Part of Surface} & \text{Equations} & \text{Class} \\ \hline
\mathbf{10}\text{ MC} & U=e_2=0 & \sigma\cdot \pi^{\ast}\xi \\
\mathbf{5}\text{ MC} &  e_1=e_0U^2+e_2V^2=0 &\left[2\sigma+\pi^{\ast}(2c_1+\xi)\right]\cdot \pi^{\ast}(c_1+\xi)\\
\cap\text{ at }\infty &V=e_0=0 &  \sigma_{\infty}\cdot \pi^{\ast}(2c_1+\xi) \\ \hline
\mathbf{\text{Total}} & & 4\sigma\cdot \pi^{\ast}(c_1+\xi) + [\pi^{\ast}(2c_1+\xi)]^2
\end{array}\end{equation}



\subsection{$\mathcal{C}^{(1)}\cap \tau \mathcal{C}^{(1)}$}

The net class of this intersection is
\begin{equation}{\cal{C}}^{(1)}\cdot {\cal{C}}^{(1)} = 3\sigma\cdot \pi^{\ast}\left(2\eta - 7c_1-2\xi\right) + \left[\pi^{\ast}(\eta-2c_1-\xi)\right]^2 \,.\end{equation}

For this intersection we obtain the two equations
\be
\ba
 U \left(a_0 U^2+a_2 V^2\right) &=0 \cr
 V \left(a_1 U^2+a_3 V^2\right)& =0 \,.
\ea
\ee
We get a piece from
\begin{equation}U=a_3=0\,,\end{equation}
in the class
\begin{equation}\sigma\cdot \pi^{\ast}(\eta-5c_1-\xi)\,.
\end{equation}
This is a $\mathbf{10}$ matter curve.  We also get a piece from
\begin{equation}V=a_0=0\,,
\end{equation}
in the class
\begin{equation}\sigma_{\infty}\cdot \pi^{\ast}(\eta-2c_1-\xi)\,.
\end{equation}
This is a piece at $\infty$.  What remains are simultaneous solutions to $a_0U^2+a_2V^2=0$ and $a_1U^2+a_3V^2=0$.  A necessary condition for any such solutions is
\begin{equation}a_0a_3=a_1a_2 \,,
\end{equation}
or
\begin{equation}0=\alpha(a_3e_0+a_2e_1) = \alpha P_2\,.
\end{equation}
We recognize $P_2$ as a $\mathbf{5}$ matter curve but not $\alpha$.  The reason for this is that $\alpha=0$ implies $a_0=a_1=0$ so that, above $\alpha=0$, ${\cal{C}}^{(1)}$ is given by $a_2V^2=a_3V^2=0$.  Generically we have no solutions to $\alpha=a_2=a_3=0$ so this part of ${\cal{C}}^{(1)}$ is a piece at infinity given by $\alpha=V=0$.  This is in the class
\begin{equation}
2\sigma_{\infty}\cdot \pi^{\ast}(\eta-4c_1-2\xi)\,,
\end{equation}
where the 2 comes about because this component appears with multiplicity 2.  What remains is
\begin{equation}
2\sigma\cdot \pi^{\ast}(\eta-3c_1)+ [\pi^{\ast}\eta]^2+14[\pi^{\ast}c_1]^2+[\pi^{\ast}\xi]^2-7\pi^{\ast}c_1\cdot\pi^{\ast}\eta+9\pi^{\ast}c_1\cdot\pi^{\ast}\xi-2\pi^{\ast}\eta\cdot\pi^{\ast}\xi\,.
\end{equation}
We can summarize everything in the following table
\begin{equation}\begin{array}{c|c|c}\text{Part of Surface} & \text{Equations} & \text{Class} \\ \hline
\mathbf{10}\text{ MC} & U=a_3=0 & \sigma\cdot\pi^{\ast}(\eta-5c_1-\xi) \\
\mathbf{5}\text{ MC} & a_0U^2+a_2V^2=0 & 2\sigma\cdot\pi^{\ast}(\eta-3c_1) \\
& a_1U^2+a_3V^2=0 & +[\pi^{\ast}\eta]^2+14[\pi^{\ast}c_1]^2+[\pi^{\ast}\xi]^2\\
& \text{(less the }V=\alpha=0\text{ piece)} & -7\pi^{\ast}c_1\cdot\pi^{\ast}\eta+9\pi^{\ast}c_1\cdot\pi^{\ast}\xi-2\pi^{\ast}\eta\cdot\pi^{\ast}\xi\\
\cap\text{ at }\infty & V=a_0=0 & \sigma_{\infty}\cdot \pi^{\ast}(\eta-2c_1-\xi) \\
& V=\alpha=0 & 2\sigma_{\infty}\cdot\pi^{\ast}(\eta-4c_1-2\xi) \\ \hline
\mathbf{\text{Total:}} & & 3\sigma\cdot\pi^{\ast}(2\eta-7c_1-2\xi)+[\pi^{\ast}(\eta-2c_1-\xi)]^2
\end{array}\end{equation}


\subsection{$\mathcal{C}^{(1)}\cap \tau \mathcal{C}^{(2)}$}

The net class of this intersection is
\begin{equation}
{\cal{C}}^{(1)}\cdot {\cal{C}}^{(2)} = \sigma\cdot \pi^{\ast}(2\eta-4c_1+\xi)+\pi^{\ast}(\eta-2c_1-\xi)\cdot \pi^{\ast}(2c_1+\xi) \,.
\end{equation}
The relevant intersection is
\be
\ba
a_0 U^3  + a_1 U^2 V + a_2 UV^2 + a_3 V^3 & =0 \cr
e_0 U^2 - e_1 UV + e_2 V^2 &=0 \,.
\ea
\ee
We have no full component at $U=0$ since this would also require $e_2=a_3=0$, which is a finite set of points.  On the other hand, we get a component at $V=0$ from $V=e_0=0$.  Moreover, when we set $e_0=0$ we have
\begin{equation}
e_1\alpha U^2 V = a_2UV^2+a_3V^3\quad\text{and}\quad e_1UV=e_2V^2\,.
\end{equation}
Plugging the second into the first we find that a $V^2$ comes out, meaning that we get the $V=e_0=0$ solution with multiplicity 2.  This is in the class
\begin{equation}2\sigma_{\infty}\cdot \pi^{\ast}(2c_1+\xi)\,.
\end{equation}
This satisfies two consistency checks.  First, it means that the $\mathbf{5}$ matter curve is the remainder
\begin{equation}\sigma\cdot \pi^{\ast}(2\eta-8c_1-\xi)+\pi^{\ast}(\eta-4c_1-\xi)\cdot \pi^{\ast}(2c_1+\xi)\,,
\end{equation}
which is of the form $\sigma\cdot \pi^{\ast}([P_3])+\ldots$ as expected.  Second, if we combine this with the other components at $\infty$ then the net is $3\sigma_{\infty}\cdot \pi^{\ast}\eta$ as we expect from a general analysis of the full spectral surface.

Anyway, we can summarize the contributions from ${\cal{C}}^{(1)}\cap \tau{\cal{C}}^{(2)}$ in the table
\begin{equation}\begin{array}{c|c|c}\text{Part of Surface} & \text{Equations} & \text{Class} \\ \hline
\mathbf{10}\text{ MC} & \cdot & \cdot \\
\mathbf{5}\text{ MC} & \alpha(e_0U^3-e_1U^2V)+a_2UV^2+a_3V^3=0 & \sigma\cdot \pi^{\ast}(2\eta-8c_1-\xi) \\
& e_0U^2-e_1UV+e_2V^2=0\text{ (less }V=0\text{ pieces)} & +\pi^{\ast}(\eta-4c_1-\xi)\cdot \pi^{\ast}(2c_1+\xi) \\
\cap\text{ at }\infty & V=e_0=0 & 2\sigma_{\infty}\cdot \pi^{\ast}(2c_1+\xi) \\ \hline
\mathbf{\text{Total}} & & \sigma\cdot \pi^{\ast}(2\eta-4c_1+\xi) \\
&& + \pi^{\ast}(\eta-2c_1-\xi)\cdot\pi^{\ast}(2c_1+\xi)\\
&
\end{array}\end{equation}

This makes clear that we have a component at infinity.  In particular, we have such a component along
\begin{equation}V=e_0=0\,.\end{equation}
This is in the class
\begin{equation}\sigma_{\infty}\cdot \pi^{\ast}(2c_1+\xi)\,.\end{equation}



\subsection{Summary}

In summary we obtain the following matter curves:
\be
\ba
\mathcal{C}^{(1)} \cap \tau\mathcal{C}^{(1)} \,:&\qquad P_2 \qquad\hbox{in} \qquad 3 c_1 -  t  \cr
\mathcal{C}^{(1)} \cap \tau\mathcal{C}^{(2)} \,:&\qquad P_3 \qquad\hbox{in} \qquad  4 c_1 - 2 t  - \xi \cr
\mathcal{C}^{(2)} \cap \tau\mathcal{C}^{(2)} \,:&\qquad P_1 \qquad\hbox{in} \qquad  c_1   + \xi\,.
\ea
\ee
The ${\bf 10}$ matter curve is read off from  $U=0$
\be
\ba
\mathcal{C}^{(2)} \cap \tau\mathcal{C}^{(2)} \,:&\qquad e_2  \qquad\hbox{in} \qquad   \xi \,.
\ea
\ee


\section{Review of Three-fold base in the compact model}
\label{app:review}

In \cite{Marsano:2009ym} we constructed compact
almost Fano three-folds $X$ and $\tilde X$ which can be used as a
base of elliptically fibered Calabi-Yau four-fold. In the present paper, we make use only of $\tilde{X}$ and to avoid confusion with the auxiliary space used to construct spectral covers we refer to $\tilde{X}$ as $B_3$ in the main text.

Here we briefly
review this construction and summarize the topology of $X$ and
$\tilde X=B_3$.

\subsection{Construction}
Let $Z=\mathbb{P}^3$ with homogenous coordinates
$[Z_0,Z_1,Z_2,Z_3]$.  The canonical class is given in terms of the
hyperplane class $H$ as \be K_Z = - 4 H \,. \ee Inside
$\mathbb{P}^3$, we consider the nodal curve ${\cal{C}}$ defined by
the equations \be \ba Z_4Z_1Z_2+(Z_1+Z_2)^3&=0 \cr
            Z_3 &=0\,.
\ea \ee Alternatively, this can be written in affine coordinates
$z_i$ as
\begin{equation}
{\cal{C}}=\left\{[z_1,z_2,0,1]\,|\,
z_1z_2+(z_1+z_2)^3=0\right\}\cup\left\{[1,-1,0,0]\right\}\,.
\label{Ceqns}\end{equation}
In what follows, we will typically consider the affine patch
$[z_1,z_2,z_3,1]$ of $\mathbb{P}^3$ since this contains all of
${\cal{C}}$ except for a single ``point at infinity".  As clear from
\eqref{Ceqns}, ${\cal{C}}$ exhibits a singular point at $[0,0,0,1]$
which is of the form $z_1z_2=z_3=0$.

The first step in constructing our three-fold is to blow up along
${\cal{C}}$ to obtain the three-fold $Y$ with the blow-down map \be
\psi: Y \rightarrow Z \,. \ee In coordinates this can be described
by considering $\mathbb{C}^3\times\mathbb{P}^1$ in the $Z_4=1$ patch
with homogeneous coordinates $[V_0,V_1]$ on the new $\mathbb{P}^1$,
which we shall hereafter denote by $\mathbb{P}^1_V$.  The blow-up is
then defined in this patch by the equation
\begin{equation}
Y\,:\qquad V_0\left(z_1z_2+(z_1+z_2)^3\right)=V_1z_3 \,.
\label{blowup1}
\end{equation}
From \eqref{blowup1}, we see that the resulting three-fold exhibits
a singular point at $\{(z_1,z_2,z_3),[V_0,V_1]\}=\{(0,0,0),[1,0]\}$.
Let us  pass to an affine patch covering the north pole $v_0\not=0$
of $\mathbb{P}^1_V$. Then defining again $u = v_1 /v_0$ the equation
\eqref{blowup1} in fact becomes
\begin{equation}
[z_1z_2+(z_1+z_2)^3]=uz_3\,,
\end{equation}
so that near the singular point it behaves like
\begin{equation}
z_1z_2=u  z_3\,.
\end{equation}
We recognize this as a conifold singularity.

The divisor classes in $Y$ are the exceptional divisor $Q$, which is
a $\mathbb{P}^1$-bundle over ${\cal C},$
 and $\psi^* (H)  = Q +(H-Q)$. The canonical class is
\be K_Y = \psi^* (K_Z) + Q = - 4 H + Q \,. \ee

The final step is to blow-up the conifold singularity in $Y$ by \be
\phi: X\rightarrow Y \,. \ee To do this, we move to a local patch
covering the north pole of $\mathbb{P}^1_V$ with coordinates
$(z_1,z_2,z_3,u=v_1/v_0).$  Let us blow up the origin of this
$\mathbb{C}^4$ by gluing in a $\mathbb{P}^3_W$ with homogeneous
coordinates $[W_1,W_2,W_3,W_4]$ and restrict to $z_1z_2=z_3u$ and
its smooth continuation, $W_1W_2=W_3W_4$, at the origin.  In the
end, the three-fold takes the following form in this local patch
\be\label{XDef} \ba X_1 = &
\left\{(z_1,z_2,z_3,v_1;W_1,W_2,W_3,W_4)\in \mathbb{C}^4\times
\mathbb{P}^3_W\,:\quad \right.\cr & \left.  \qquad \qquad
(z_1,z_2,z_3, u) \in [W_1,W_2,W_3,W_4] \,, \quad
 z_1 z_2 = z_3  u \,,\quad W_1 W_2 = W_3 W_4
  \right\}\,.
  \ea
\ee

We can identify the two $\mathbb{P}^1$'s with the submanifolds
\be\label{P1s} \mathbb{P}^1_{(1)}:\quad W_2=W_4=0\,,\qquad
\mathbb{P}^1_{(2)}:\quad W_2=W_3=0\,. \ee Note that in this local
patch it is not possible to see that these $\mathbb{P}^1$s are in
the same class in $X$. It is however clear from the global topology
of $X$ since their intersections with all divisors are equivalent.
The canonical class of $X$ is \be K_X = - 4 H + (D + E )+  E \,, \ee
where the exceptional divisor is \be \phi^* Q = D + E  \,. \ee

\begin{figure}
\begin{center}
\epsfig{file=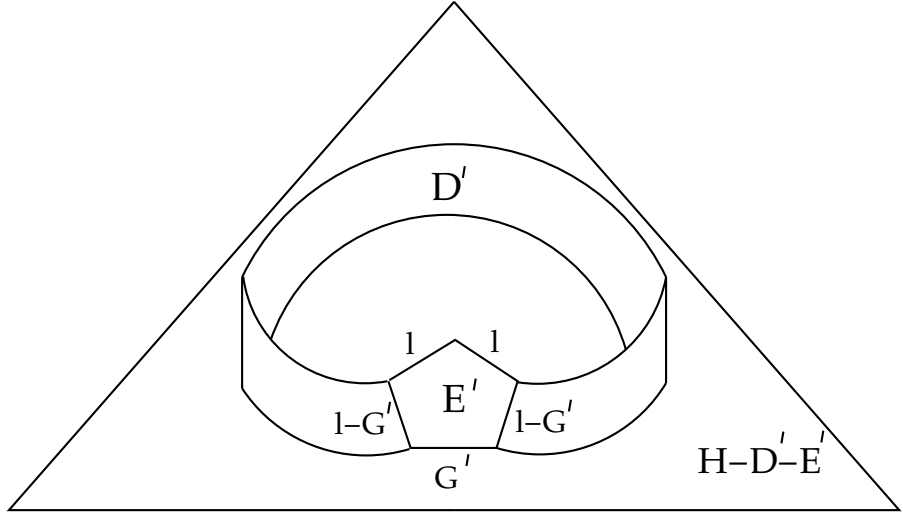,width=.7\textwidth}
\caption{Final  three-fold $\tilde{X}$} \label{fig:GlobalThree}
\end{center}
\end{figure}

The curve $G$ is a $(-1,-1)$ curve because it is an exceptional
$\mathbb{P}^1$ so that we can   flop it to obtain a new three-fold,
$\tilde{X}$, depicted in figure \ref{fig:GlobalThree}.  The divisors
$D$ and $E$ of $X$ carry over to new divisors $D'$ ad $E'$ in
$\tilde{X}$.  The canonical class also follows simply from $K_X$ as
\be\label{KXtilde} K_{\tilde{X}} = - 4 H + D' + 2 E' \,. \ee The
resulting three-fold $\tilde{X}$ has  the desired property that the
two curves $\ell-G'$ are distinct in $H_2(E',\mathbb{Z})$ but are
nonetheless equivalent in $H_2(\tilde{X},\mathbb{Z})$ so that they
satisfy the condition for existence of a suitable hypercharge flux.
\begin{figure}
\begin{center}
\epsfig{file=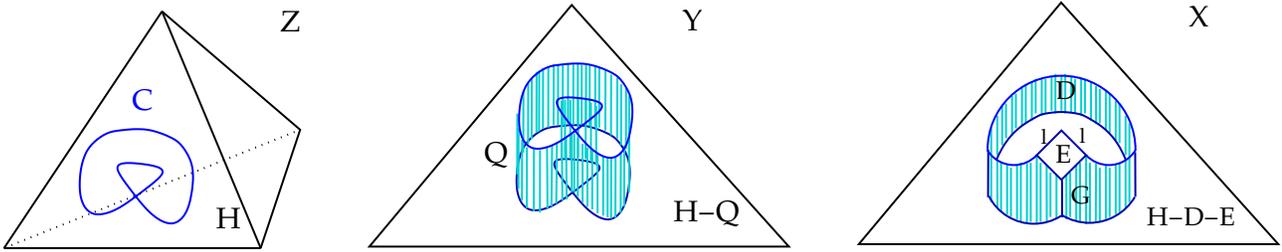,width=1\textwidth}
\caption{Global Construction of Threefold: blowups.}
\label{fig:Global12}
\end{center}
\end{figure}

\subsection{Topology of $X$}
Let us first summarize the topology of $X.$ As a basis of
$H_2(X,\mathbb{Z})$, we take the curve $\ell_0$, which descends from
the unique generator of $H_2(\mathbb{P}^3,\mathbb{Z})$, as well as
the curves $\ell$ and $G$ depicted in figure \ref{fig:Global12}. A
useful basis of divisors is $H$, $E$ and $H-D-E$. Their topology is
\be \ba H &\cong dP_3\cr E &\cong \mathbb{P}^1 \times \mathbb{P}^1
\cr H-D-E &\cong \mathbb{P}^2 \,. \ea \ee The intersection numbers
with various divisors are given by the following table
\begin{center}
\begin{tabular}{|c||r|r|r|r|}
\hline
        &$H$    & $D$   &   $E$  \cr\hline\hline
$\ell_0$    & $+1$& $0$& $0$ \cr\hline $\ell$      & $0$&$ +1$ &
$-1$ \cr\hline $G$     & $0$& $-2$ & $1$  \cr\hline
\end{tabular}
\end{center}

The intersections of divisors with one another is furthermore
\begin{equation}\label{DoubleIntBF}
\begin{array}{|c||c|c|c|}
\hline & H & E & D \\ \hline\hline H & \ell_0 & 0 & 3(\ell+G)
\cr\hline E & 0 & -2\ell & 2\ell \cr\hline D & 3(\ell+G) & 2\ell &
-3\ell_0 + 12(\ell+G) - 2\ell \cr\hline H-D-E & \ell_0-3(\ell+G) & 0
& 3\left(\ell_0-3(\ell+G)\right)\cr\hline
\end{array}
\end{equation}
from which the following non-vanishing triple-intersections follow
\be \ba H^3 &= 1 \cr
D^3 &= -14 \cr E^3 &= 2 \cr D^2H &= -3 \cr D^2E &= 2 \cr
E^2 D &= -2 \,.
\ea \ee
Let us further recall the basis of holomorphic sections for $X$:

\be\label{SectionsX}
\begin{array}{|l||l| }\hline
\text{Holomorphic Section} & \hbox{Divisor class} \cr\hline\hline
Z_4 & H\cr\hline Z_{1,2} & (H-E) + E =H \cr\hline Z_3 & (H-D-E) +
(D+E)= H \cr\hline W_{1,2,3} & H-E\cr\hline W_4 & 3H -D-2E \cr\hline
V_1 & ( 3H - D -2E) + E = 3 H - D - E \cr\hline V_0 & H-D-E\cr\hline
\end{array}
\ee

Note that for $S_{\rm GUT}=X$ we find $t=-c_1({\cal N}S_{\rm
GUT})=-E^2\vert_{E}=l_1+l_2$ where we use that ${\cal N}S_{\rm
GUT}=E$ and the fact that the class $2l$ in $X$ restricts to the
class $l_1+l_2$ in $E=\mathbb{P}^1\times\mathbb{P}^1.$

\subsection{Topology of $\tilde X$}
Let us review the topology of $\tilde{X}$, including the topology of
various divisors and the intersection tables for divisors and
curves.  We start with a discussion of several interesting divisor
classes.  The divisor $H$, which was a $dP_3$ before the flop,
remains a $dP_3$ because it is unaffected by the flop.  From the
viewpoint of $H-D-E=\mathbb{P}^2$, however, the flop corresponds to
blowing up a point so that $H-D'-E'$ becomes a $dP_1$.  Similarly,
from the viewpoint of $E=\mathbb{P}^1\times\mathbb{P}^1$, the flop
effectively blows up a point so that $E'$ is simply $dP_2$. Finally
the divisor $D'$ is the Hirzebruch surface $\mathbb{F}_4$.
\be\label{DivBasis} \ba H &\cong dP_3 \cr E' &\cong dP_2\cr D'
&\cong\mathbb{F}_4 \cr H-D'-E' &\cong dP_1 \,. \ea \ee

As a basis of $H_2(\tilde{X},\mathbb{Z})$, we take the curves
$\ell_0$ and $\ell$ along with the flopped curve $G'$ as depicted in
figure \ref{fig:GlobalThree}.  The intersection numbers of these
curves with various divisors are presented in the following table

\begin{center}
\begin{tabular}{|r||r|r|r|r|}
\hline ${}$        & $H$ & $E'$ & $H-D'-E'$ & $D'$ \cr\hline\hline
$\ell_0$    & $1$ & $0$ & $+1$  & $0$ \cr\hline $\ell$      & $0$ &
$-1$ & $0$  & $1$\cr\hline $G'$    & $0$ & $-1$ & $-1$ &
$2$\cr\hline $\ell-G'$   & $0$ & $0$ & $+1$ & $-1$\cr\hline
\end{tabular}
\end{center}

\noindent The intersections of the divisors with one another are as
follows

\begin{center}
\begin{tabular}{|r||c|c|c|}
\hline
        &$H$    & $E'$  &   $H-D'-E'$   \cr\hline\hline
$H$     & $\ell_0   $& $0 $& $\ell_0 - 3l+ 3 G' $    \cr\hline $E'$
& $0$   &$ - 2\ell+ G'$ & $G'$  \cr\hline $H-D'-E'$   & $\ell_0 -
3\ell + 3 G'$ & $G'$ & $- 2 \ell_0+ 6 \ell - 5 G' $  \cr\hline $D'$
& $3 \ell- 3 G'$ &  $2\ell - 2 G' $ & $3 \ell_0 - 9 \ell+ 7 G'$
\cr\hline
\end{tabular}
\end{center}

\noindent It is useful to distinguish the two $\mathbb{P}^1$'s of
$E'$ that are equivalent to $\ell$ inside $\tilde{X}$.  Denoting
these by $\ell_1$ and $\ell_2$, we find that \be\label{NewEll}
{E'}^2 = G' - \ell_1 -\ell_2 \,,\qquad D'.E' = (\ell_1 - G') +
(\ell_2 - G') \,. \ee

The non-vanishing triple intersection numbers are easily computed
from the above data with the following results \be \ba H^3 &= 1 \cr
E^{\prime\,3}&= 1 \cr D^{\prime\,3}&= -6 \cr D^{\prime\,2}H &= -3
\cr D^{\prime\,2}E' &= -2 \,. \ea \ee

In the previous section  we listed various divisors and their
corresponding holomorphic sections on $X$. Each of these carries
over to a divisor or section after the flop. We will abuse notation
in what follows and continue to use the labels $Z_i,W_j,V_k$ of
\eqref{SectionsX} for the corresponding holomorphic sections on
$\tilde{X}$.

We use the standard basis for $S_{GUT}=dP_2$ consisting of the
hyperplane class, $h$, and the two exceptional curves, $e_1$ and
$e_2$ \be H_2(E',\mathbb{Z}) = \langle h,e_1,e_2\rangle\,. \ee From
the intersection form \be h^2=1\,,\qquad e_i\cdot
e_j=-\delta_{ij}\,, \ee it is easy to obtain the relation of these
classes to $\ell_1$, $\ell_2$, and $G'$, \be\label{CurveIdent} \ba
\ell_1 &= h-e_1 \\
\ell_2 &= h-e_2 \\
G' &= h-e_1-e_2\,. \ea \ee

Finally note that $t=-c_1({\cal N}S_{\rm GUT})=-E'^2\vert_{E'}=h$
where we use that ${\cal N}S_{\rm GUT}=E'$ and the fact that the
class $2l-G'$ in $\tilde X$ restricts to the class $h$ in $E'=dP_2.$


\section{Explicit Construction of the  Spectral Surface}
\label{app:ss}

Let us start by trying to write our factored spectral
surface in terms of globally well-defined holomorphic sections.
This is tricky because we must account for the class $\xi=h-e_1$
which is not specified by the vanishing of a single globally
well-defined holomorphic section but rather the simultaneous
vanishing of two of them as in $W_2=W_4=0$.  Before getting started,
we need to recall in what classes various sections $W_i$ intersect $S_{\rm GUT}$
\begin{equation}\begin{array}{c|c} \text{Section} & \text{Class in }S_{\rm GUT} \\ \hline
W_1 & (h-e_2) + e_2 \\
W_2 & (h-e_1) + e_1 \\
W_3 & (h-e_1-e_2) + (e_1) + (e_2) \\
W_4 & (h-e_1) + (h-e_2)
\end{array}\end{equation}
We were a bit cavalier about $W_3$ in the past so let us specify this more carefully.  Recall that, before the flop, when we blew up the node in our construction we had the equation
\begin{equation}V_0 z_1z_2 = z_3 V_1\end{equation}
In the patch $V_0=1$ this became $z_1z_2=z_3v_1$ and to blow up the conifold point we promoted $z_i\rightarrow W_i$ and $v_1\rightarrow W_4$.  Because of this, $W_3=0$ extends beyond the conifold point to $z_3=0$, which can occur when $z_1=0$, $z_2=0$, or $V_0=0$.  Usually by $W_3$ we are only describing the regime close to the conifold point but if we extend away from that the holomorphic section "$W_3$" also vanishes along the $V_0=0$ locus.  After the flop, this means that $W_3=0$ picks up an additional piece in the class $h-e_1-e_2$.  This is not seen by $W_1W_2=W_3W_4$ because this equation describes $\mathbb{P}^1\times\mathbb{P}^1$ before the flop, which does not intersect this additional piece ($V_0=0$).  The flop forces these pieces to intersect.

To simplify presentation let us change notation and use $W_3$ to refer to the holomorphic section that we would have previously called `$W_3/V_0$'.  This object is also nicely holomorphic and is a section of $(H-E)-(H-D-E)=D$ before the flop and $D'$ after the flop.  The restriction of this to $S_{\rm GUT}$ is in the class $e_1+e_2$.  So, we will use the following
\begin{equation}\begin{array}{c|c} \text{Section} & \text{Class in }S_{\rm GUT} \\ \hline
W_1 & (h-e_2) + e_2 \\
W_2 & (h-e_1) + e_1 \\
W_3 & (e_1) + (e_2) \\
W_4 & (h-e_1) + (h-e_2)
\end{array}\end{equation}

It is also helpful to remember how various cycles in $dP_2$ are described by the vanishing of various sections
\begin{equation}\begin{array}{c|c}\text{Class in }H_2(dP_2,\mathbb{Z}) & \text{Equation} \\ \hline
e_1 & W_2=W_3=0 \\
e_2 & W_1=W_3=0 \\
h-e_1 & W_2=W_4=0 \\
h-e_2 & W_1=W_4=0
\end{array}\end{equation}

Now, to construct a suitably factored ${\cal{C}}$ from globally well-defined holomorphic sections, we will instead write the following
\begin{equation}\left(\tilde{a}_0 U^3 + \tilde{a}_1U^2V + \tilde{a}_2UV^2 + \tilde{a}_3V^3\right)\left(\tilde{e}_0 U^2 + \tilde{e}_1UV + \tilde{e}_2V^2\right)\end{equation}
where the $\tilde{a}_m$ and $\tilde{e}_n$ are sections of
\begin{equation}\begin{array}{c|c}\text{Section} & \text{Bundle} \\ \hline
\tilde{a}_m & \eta-(n+2)c_1 \\
\tilde{e}_n & (2-n)c_1
\end{array}\end{equation}
We will take these to be meromorphic sections, though.  The idea is then that the $\tilde{e}_n$ will have a pole in the class $\xi$ while the $\tilde{a}_m$ will have a zero in the class $\xi$.  The entire surface will be holomorphic because the pole will cancel off of the zero.  This means, however, that the net class of the cubic piece will decrease by $\pi^{\ast}\xi$ while the net class of the quadratic piece will increase by this amount.  This will allow us to recover the desired classes.  We can think of the $a_m$ as the $\tilde{a}_m$ less this particular part of the zero locus while we can think of the $e_n$ as the $\tilde{e}_n$ less this particular pole.

We must be careful, however, that when constructing $\tilde{a}_m$ and $\tilde{e}_n$ that only the desired zero or pole can come out common. We will use that $\frac{W_1}{W_4}$ has a pole along $W_4=W_2=0$ (the one at $W_4=W_1=0$ is obviously cancelled). So, to construct sections with poles along $(h-e_1)$ we write
\begin{equation}\tilde{e}_n\sim X_n(W_i) + \frac{W_1}{W_4}Y_n(W_i)\end{equation}
where $X_n$ and $Y_n$ are polynomials in $W_i.$
As for zeroes, this is easy to deal with.  There are two manifestly different objects that exhibit zeroes along $(h-e_1)$ -- these are $W_2$ and $W_4$.  So, the general form we should take for the $\tilde{a}_m$ is
\begin{equation}\tilde{a}_m\sim W_2 A_m(W_i) + W_4 B_m(W_i)\end{equation}

We should then think of $e_n$ as $\tilde{e}_n$ with the $W_2=W_4$ pole removed and $a_m$ as $\tilde{a}_m$ with the $W_2=W_4$ zero removed.

It remains to readdress the $b_1=0$ constraint.  Here it takes the form
\begin{equation}\tilde{a}_1\tilde{e}_0+\tilde{a}_0\tilde{e}_1=0\end{equation}
Before, we solved it by requiring
\begin{equation}\frac{a_0}{e_0} = -\frac{a_1}{e_1} = \alpha\end{equation}
This implies that
\begin{equation}\frac{\tilde{a}_0}{\tilde{e}_0} = -\frac{\tilde{a}_1}{\tilde{e}_1}=\tilde{\alpha}\end{equation}
In going from $a_0$ to $\tilde{a}_0$ we must add a zero and in going from $e_0$ to $\tilde{e}_0$ we must add a pole so in going from $\alpha$ to $\tilde{\alpha}$ we have added a zero of multiplicity 2 along $h-e_1$.  Said differently, we solve $b_1=0$ by setting
\begin{equation}\tilde{a}_0=\tilde{\alpha}\tilde{e}_0\qquad \tilde{a}_1=-\tilde{\alpha}\tilde{e}_1\end{equation}
and then interpreting $\alpha$ as $\tilde{\alpha}$ less a zero of multiplicity 2 along $W_2=W_4=0$.

The ratio $\tilde{a}_m/\tilde{e}_n$ generically has the form
\begin{equation}\frac{W_2A_m+W_4B_m}{X_n+\frac{W_1}{W_4}Y_n}\end{equation}
The easiest way to ensure that such a ratio is holomorphic is to take $W_4X_n+W_1Y_n$ to be a factor of both $A_m$ and $B_m$.

A nice proposal for a general form of our embedding is then to take
\begin{equation}\label{def_e}\tilde{e}_n = X_n + \frac{W_1}{W_4}Y_n\end{equation}
\begin{equation}\label{def_a}\tilde{a}_0 = \left(W_2P_1 + W_4 Q_1\right)\left(W_4\tilde{e}_0\right)\qquad \tilde{a}_1 = -\left(W_2P_1 + W_4Q_1\right)\left(W_4\tilde{e}_1\right)\end{equation}
\begin{equation}\label{def_aa}\tilde{a}_m = W_2A_m + W_4B_m\qquad m=2,3\end{equation}
In this case
\begin{equation}\tilde{\alpha} = W_4\left(W_2P_1 + W_4Q_1\right)\end{equation}
and $\alpha$ is understood to be $\tilde{\alpha}$ less the degree two zero along $W_2=W_4=0$.

\begin{equation}\begin{array}{c|c} \text{Object} & \text{Bundle} \\ \hline
P_1 & 2h-e_1-e_2 \\
Q_1 & h \\
A_2 & 4h-2(e_1+e_2) \\
A_3 & h-e_1-e_2 \\
B_2 & 3h-e_1-e_2 \\
B_3 & {\cal{O}} \\
X_0 & 6h-2(e_1+e_2) \\
X_1 & 3h-(e_1+e_2) \\
X_2 & {\cal{O}} \\
Y_0 & 7h-3(e_1+e_2) \\
Y_1 & 4h-2(e_1+e_2) \\
Y_2 & h-e_1-e_2
\end{array}\end{equation}
Projectivity then tells us that
\begin{equation}P_1,Q_1\text{ are linear polynomials in the }W_i\end{equation}
Further, we can set
\begin{equation}B_3=1\end{equation}
which means that we must have
\begin{equation}A_3\sim V_0\end{equation}
which we replace by a constant "$c$" for our $W_i$ considerations.  Finally,
\begin{equation}A_2,B_2\text{ are quadratic polynomials in the }W_i\end{equation}

Similarly,
\begin{equation}X_1,Y_1\text{ are quadratic polynomials in the }W_i\end{equation}
while
\begin{equation}X_0,Y_0\text{ are quartic polynomials in the }W_i\end{equation}
In each case, we have to make sure that the power of $W_4$ does not become too large.  This leads to
\begin{equation}(W_4\tilde{e}_0)\text{ is a degree 5 polynomial in the }W_i\qquad (W_4\tilde{e}_1)\text{ is a degree 3 polynomial in the }W_i\end{equation}

The form of ${\cal{C}}$ with this ansatz is therefore
\begin{equation}\begin{split}\left\{\left(W_2P_1+W_4Q_1\right)\right.&\left.\left[(W_4\tilde{e}_0) U^3 - (W_4\tilde{e}_1)U^2V\right] + W_2\left(A_2UV^2+cV^3\right)+W_4\left(B_2UV^2+V^3\right)\right\}\\
&\qquad \times\left\{ \tilde{e}_0 U^2 + \tilde{e}_1 UV + \tilde{e}_2 V^2\right\}\end{split}\end{equation}


\section{Explicit construction of $G$-Fluxes}
\label{app:fluxes}

 Here we explicitly construct $G$-fluxes, both
universal and non-universal, for 3+2 model. These fluxes are used in
Section 4.2 to obtain chiral spectrum.


\subsection{Universal Fluxes}

We start by constructing a generic universal class of the form
\begin{equation}\Gamma_u = \tilde{k}_1\tilde{\gamma}_1 + \tilde{k}_2\tilde{\gamma}_2 + \tilde{\rho} + \tilde{d}_1\delta_1 + \tilde{d}_2 \delta_2 \,,
\end{equation}
To do this, we describe each of the objects that enter $\Gamma_u$.

\subsubsection{$\tilde{\gamma}_i$ Fluxes}

We start with two types of universal flux.  They are constructed from
\begin{equation}\gamma_1 = {\cal{C}}^{(1)}\cdot \sigma \,, \qquad \gamma_2= {\cal{C}}^{(2)}\cdot \sigma \,,
\end{equation}
and take the form
\begin{equation}\tilde{\gamma}_1 = 3\gamma_1 - p_{1}^{\ast}p_{1\,\ast}\gamma_1\,,
\qquad\qquad \tilde{\gamma}_2 = 2\gamma_2 - p_2^{\ast}p_{2\,\ast}\gamma_2\,.
\end{equation}
Each of these is the restriction of a well-defined flux inside $X$ to the relevant ${\cal{C}}^{(i)}$.  In particular, we have that
\begin{equation}\tilde{\gamma}_1 = {\cal{C}}^{(1)}\cdot \left[3\sigma - \pi^{\ast}(\eta-5c_1-\xi)\right]\,,\qquad \tilde{\gamma}_2 = {\cal{C}}^{(2)}\cdot \left[2\sigma - \pi^{\ast}\xi\right]\,.\end{equation}
This allows us to compute intersections of $\tilde{\gamma}_i$ inside ${\cal{C}}^{(i)}$ using intersection data in $X$.  In particular, we have that
\begin{equation}\begin{split}\tilde{\gamma}_1\cdot_{{\cal{C}}^{(1)}} \Sigma_1 &=
\left[3\sigma - \pi^{\ast}(\eta-5c_1-\xi)\right]\cdot_X \Sigma_1 \\
\tilde{\gamma}_2\cdot_{{\cal{C}}^{(2)}} \Sigma &= \left[2\sigma - \pi^{\ast}\xi\right]\cdot_X \Sigma_2
\end{split}\end{equation}
where $\Sigma_i\subset {\cal{C}}^{(i)}$.  Using this, we can easily evaluate the intersection table
\begin{equation}\begin{array}{c|c|c|c|c}\text{MC} & \text{Origin} & \text{Class in $X$} & \tilde{\gamma}_1 & \tilde{\gamma}_2 \\ \hline
\mathbf{10}_M & 22 & \sigma\cdot\pi^{\ast}\xi & 0 & - \xi\cdot_{S_{\rm GUT}}(2c_1+\xi) \\
\mathbf{10}_{\text{Other}} & 11 & \sigma\cdot \pi^{\ast}(\eta-5c_1-\xi) & -(\eta-2c_1-\xi)\cdot_{S_{\rm GUT}} (\eta-5c_1-\xi) & 0 \\
\mathbf{5}_H &22 & [2\sigma + \pi^{\ast}(2c_1+\xi)]\cdot \pi^{\ast}(c_1+\xi)& 0 & 0 \\
\mathbf{\overline{5}}_H & 12 & 2\left[\sigma\cdot\pi^{\ast}(2\eta-8c_1-\xi)\right. &-2(\eta-4c_1-2\xi)\cdot_{S_{\rm GUT}} (\eta-5c_1-\xi)& -\xi\cdot_{S_{\rm GUT}} (2c_1+\xi)\\
&  & \left. +\pi^{\ast}(\eta-4c_1-\xi)\cdot \pi^{\ast}(2c_1+\xi)\right] & & \\
\mathbf{\overline{5}}_M & 11 & 2\sigma\cdot \pi^{\ast}(\eta-3c_1) & (\eta-5c_1-\xi)\cdot_{S_{\rm GUT}}(\eta-6_1-3\xi) & 0 \\
& & + \pi^{\ast}(\eta)^2 + 14\pi^{\ast}c_1^2 + \pi^{\ast}\xi^2 & & \\
& & +9\pi^{\ast}c_1\cdot\pi^{\ast}\xi - 2\pi^{\ast}\eta\cdot \pi^{\ast}\xi & & \\
& &  - 7\pi^{\ast}c_1\cdot \pi^{\ast}\eta & &
\end{array}\end{equation}
We have expressed triple intersections in $X$ in terms of
intersections in $S_{\rm GUT}$ by using:
\be
\sigma^2=-\sigma\cdot \pi^{\ast}c_1,\quad \pi^{\ast}(\ldots)\cdot \pi^{\ast}(\ldots)\cdot \pi^{\ast}(\ldots) =0,\quad
 \sigma \cdot \pi^{\ast}(a)\cdot \pi^{\ast}(b)=a\cdot_{S_{\rm GUT}}b \,.
\ee

Plugging in $\xi=h-e_1$ we find for these
\begin{equation}\begin{array}{c|c|c|c|c} \text{MC} & \text{Origin} & \text{Class in }S_{\rm GUT} & \tilde{\gamma}_1 & \tilde{\gamma}_2 \\ \hline
\mathbf{10}_M & 22 & h-e_1 & 0 & -4\\
\mathbf{10}_{\text{Other}} & 11 &h-e_2 & -6 & 0\\
\mathbf{5}_H & 22 & 4h-2e_1-e_2 & 0 & 0\\
\mathbf{\overline{5}}_H & 12 & 9h-3e_1-4e_2 & -2 & -4\\
\mathbf{\overline{5}}_M & 11 & 8h-3e_1-3e_2 & -4 & 0
\end{array}\end{equation}

\subsubsection{$\rho$-flux}

There is a third kind of universal flux that we can write.  This is
\begin{equation}\tilde{\rho} = 2p_1^{\ast}\rho - 3p_2^{\ast}\rho\end{equation}
where $\rho$ is a class in $S_{\rm GUT}$.

Intersections of $\tilde{\rho}$ with various matter curves are easy to compute
\begin{equation}\begin{array}{c|c|c|c}
\text{Curve} & \text{Origin} & \text{Class in }X & \tilde{\rho} \\ \hline
\mathbf{10}_M & 22 & \sigma\cdot\pi^{\ast}\xi & -3\rho\cdot_{S_{\rm GUT}} \xi \\
\mathbf{10}_{\text{other}} & 11 & \sigma\cdot\pi^{\ast}(\eta-5c_1-\xi) & 2\rho\cdot_{S_{\rm GUT}} (\eta-5c_1-\xi) \\
\mathbf{5}_H & 22 & \left[2\sigma + \pi^{\ast}(2c_1+\xi)\right]\cdot\pi^{\ast}(c_1+\xi) & -6\rho\cdot_{S_{\rm GUT}} (c_1+\xi) \\
\mathbf{\overline{5}}_H & 12 & 2\left[\sigma\cdot\pi^{\ast}(2\eta-8c_1-\xi)\right. & -\rho\cdot_{S_{\rm GUT}} (2\eta-8c_1-\xi) \\
& & \left. + \pi^{\ast}(\eta-4c_1-\xi)\cdot\pi^{\ast}(2c_1+\xi)\right] & \\
\mathbf{\overline{5}}_M & 11 & 2\sigma\cdot\pi^{\ast}(\eta-3c_1) & 4\rho\cdot_{S_{\rm GUT}} (\eta-3c_1) \\
& & + (\pi^{\ast}\eta)^2 + 14(\pi^{\ast}c_1)^2 + (\pi^{\ast}\xi)^2 & \\
& & + 9 \pi^{\ast}c_1\cdot\pi^{\ast}\xi - 2\pi^{\ast}\eta\cdot\pi^{\ast}\xi & \\
& & - 7\pi^{\ast}c_1 \cdot \pi^{\ast}\eta \\
\end{array}\end{equation}


\subsubsection{$\delta$-fluxes}

We finally consider two more kinds of flux
\begin{equation}\delta_1 = 2\sigma\cdot {\cal{C}}^{(1)} - p_2^{\ast}p_{1\,\ast}(\sigma\cdot {\cal{C}}^{(1)}) \,,\qquad \qquad \delta_2 = 3\sigma\cdot {\cal{C}}^{(2)}-p_1^{\ast}p_{2\,\ast}(\sigma\cdot {\cal{C}}^{(2)})\,,
\end{equation}
which we can alternatively write as
\begin{equation}\delta_1 = 2\sigma\cdot {\cal{C}}^{(1)} - \pi^{\ast}(\eta-5c_1-\xi)\cdot {\cal{C}}^{(2)}\,,\qquad\qquad \delta_2 = 3\sigma\cdot {\cal{C}}^{(2)} - \pi^{\ast}\xi\cdot {\cal{C}}^{(1)}\,.\end{equation}

It is a simple matter to determine how each of these restricts to our matter curves
\begin{equation}\begin{array}{c|c|c|c|c}
\text{Curve} & \text{Origin} & \text{Class in }X & \delta_1 & \delta_2 \\ \hline
\mathbf{10}_M & 22 & \sigma\cdot\pi^{\ast}\xi & -\xi (\eta-5c_1-\xi) & -3c_1\xi \\

\mathbf{10}_{\text{other}} & 11 & \sigma\cdot\pi^{\ast}(\eta-5c_1-\xi) & -2c_1(\eta-5c_1-\xi) & -\xi(\eta-5c_1-\xi)\\

\mathbf{5}_H & 22 & \left[2\sigma + \pi^{\ast}(2c_1+\xi)\right]\cdot\pi^{\ast}(c_1+\xi) & -2(c_1+\xi)(\eta-5c_1-\xi) & 3\xi(c_1+\xi)\\

\mathbf{\overline{5}}_H & 12 & 2\left[\sigma\cdot\pi^{\ast}(2\eta-8c_1-\xi)\right. & -(\eta-5c_1-\xi)(2\eta-8c_1-3\xi) & \xi(\eta-7c_1-2\xi)\\
& & \left. + \pi^{\ast}(\eta-4c_1-\xi)\cdot\pi^{\ast}(2c_1+\xi)\right] & & \\

\mathbf{\overline{5}}_M & 11 & 2\sigma\cdot\pi^{\ast}(\eta-3c_1) & 2(\eta-\xi)^2-18c_1(\eta-\xi)+40c_1^2 & -2\xi(\eta-3c_1) \\
& & + (\pi^{\ast}\eta)^2 + 14(\pi^{\ast}c_1)^2 + (\pi^{\ast}\xi)^2 & & \\
& & + 9 \pi^{\ast}c_1\cdot\pi^{\ast}\xi - 2\pi^{\ast}\eta\cdot\pi^{\ast}\xi & & \\
& & - 7\pi^{\ast}c_1 \cdot \pi^{\ast}\eta & & \\
\end{array}\end{equation}
where the intersections in the right two columns are computed in $S_{\rm GUT}$.

Plugging in $\xi=h-e_1$ we find for these
\begin{equation}\begin{array}{c|c|c|c|c} \text{MC} & \text{Origin} & \text{Class in }S_{\rm GUT} & \delta_1 & \delta_2 \\ \hline
\mathbf{10}_M & 22 & h-e_1 & -1 & -6\\
\mathbf{10}_{\text{Other}} & 11 &h-e_2 & -4 & -1\\
\mathbf{5}_H & 22 & 4h-2e_1-e_2 & -6 & 6\\
\mathbf{\overline{5}}_H & 12 & 9h-3e_1-4e_2 & -3 & -3\\
\mathbf{\overline{5}}_M & 11 & 8h-3e_1-3e_2 & 4 & -10
\end{array}\end{equation}

\subsubsection{Total universal flux}

We now construct a total flux $\Gamma_u$ as a linear combination of $\tilde{\gamma}_i$, $\tilde{\rho}$, and $\delta_j$ of the form
\begin{equation}\Gamma_u = \tilde{k}_1\tilde{\gamma}_1 + \tilde{k}_2\tilde{\gamma}_2 + \tilde{\rho} + \tilde{d}_1\delta_1 + \tilde{d}_2 \delta_2\,.\end{equation}

We must satisfy several conditions, though.  First, the flux must be suitably quantized.  This means that if we split up $\Gamma_u$ into its pieces $\Gamma_{u,1}$ and $\Gamma_{u,2}$ on ${\cal{C}}^{(1)}$ and ${\cal{C}}^{(2)}$, respectively, we need that
\begin{equation}
\Gamma_{u,1} + \frac{1}{2}\left[\sigma+\pi^{\ast}(\eta-3c_1-\xi)\right]\cdot {\cal{C}}^{(1)} \qquad \hbox{and }
\qquad \Gamma_{u,2}+\frac{1}{2} \pi^{\ast}(c_1+\xi)\cdot {\cal{C}}^{(2)}\end{equation}
are integer classes in ${\cal{C}}^{(1)}$ and ${\cal{C}}^{(2)}$, respectively.  Note that
\begin{equation}\begin{split}\Gamma_{u,1} &= {\cal{C}}^{(1)}\cdot\left\{ \left(3\tilde{k}_1 + 2\tilde{d}_1\right)\sigma - \pi^{\ast}\left[\tilde{k}_1(\eta-5c_1-\xi) + \tilde{d}_2 \xi - 2\rho\right]\right\} \\
\Gamma_{u,2} &= {\cal{C}}^{(2)}\cdot \left\{ \left(2\tilde{k}_2 + 3\tilde{d}_2\right)\sigma - \pi^{\ast}\left[\tilde{k}_2\xi + \tilde{d}_1(\eta-5c_1-\xi) + 3\rho\right]\right\}\\
\end{split}\end{equation}
and
\begin{equation}\begin{split}p_{1\,\ast}\Gamma_{u,1} &= 2\tilde{d}_1(\eta-5c_1-\xi)-3\tilde{d}_2\xi+6\rho\\
p_{2\,\ast}\Gamma_{u,2} &= - 2\tilde{d}_1(\eta-5c_1-\xi)+3\tilde{d}_2\xi-6\rho\,,
\end{split}\end{equation}
which sum to zero as they should.

The quantization conditions boil down to requiring that the following be integer divisor classes in $X$
\begin{equation}{\boxed{\begin{split}\left(3\tilde{k}_1 + 2\tilde{d}_1+\frac{1}{2}\right)\sigma - \pi^{\ast}\left[\tilde{k}_1(\eta-5c_1-\xi)+\tilde{d}_2\xi - 2\rho -\frac{1}{2}(\eta-3c_1-\xi)\right] &\in H_4(X,\mathbb{Z})\\
\left(2\tilde{k}_2+3\tilde{d}_2\right)\sigma - \pi^{\ast}\left[\tilde{k}_2\xi + \tilde{d}_1(\eta-5c_1-\xi)+3\rho-\frac{1}{2}(c_1+\xi)\right]&\in H_4(X,\mathbb{Z})
\end{split} }} \end{equation}

We must also impose "supersymmetry", which in this case amounts to the requirement that the following class is supersymmetric inside $S_{\rm GUT}$, i.e. that
\begin{equation}{\boxed{\omega\cdot_{S_{\rm GUT}} \left(2\tilde{d}_1(\eta-5c_1-\xi)-3\tilde{d}_2\xi + 6\rho\right)=0}}\end{equation}
for
\begin{equation}\omega = Ae_1 + Be_2 + C(h-e_1-e_2)\end{equation}
with
\begin{equation}A,B,C>0 \qquad A,B<C<A+B \,.
\end{equation}

Now, if we write $\rho$ as
\begin{equation}\rho = \tilde{X}h - \tilde{Y}e_1 - \tilde{Z}e_2\end{equation}
then we can write the spectrum that our $\Gamma_u$ yields as
\begin{equation}\begin{array}{c|c|c|c}
\text{MC} & \text{Origin} & \text{Class in }S_{\rm GUT} & \Gamma_u \\ \hline
\mathbf{10}_M & 22 & h-e_1 & -4\tilde{k}_2 - \tilde{d}_1 -6\tilde{d}_2 -3(\tilde{X}-\tilde{Y}) \\
\mathbf{5}_H & 22 & 4h-2e_1-e_2 & -6\tilde{d}_1 + 6\tilde{d}_2 -6 (4\tilde{X}-2\tilde{Y}-\tilde{Z}) \\
\mathbf{\overline{5}}_H & 12 & 9h-3e_1-4e_2 & -2\tilde{k}_1 -4\tilde{k}_2 -3\tilde{d}_1 - 3\tilde{d}_2 - (9\tilde{X}-3\tilde{Y}-4\tilde{Z}) \\
\mathbf{\overline{5}}_M & 11 & 8h-3(e_1+e_2) & -4\tilde{k}_1+4\tilde{d}_1 - 10\tilde{d}_2 +4(8\tilde{X}-3\tilde{Y}-3\tilde{Z}) \\
\mathbf{10}_{\text{other}} & 11 & h-e_2 & -6\tilde{k}_1 -4\tilde{d}_1 - \tilde{d}_2 + 2(\tilde{X}-\tilde{Z})
\end{array}\end{equation}
To study the constraints, let us write the explicit class that must be supersymmetric
\begin{equation}\left(6\tilde{X} + 2\tilde{d}_1-3\tilde{d}_2\right)h - \left(\tilde{Y}-3\tilde{d}_2\right)e_1 - \left(\tilde{Z}+2\tilde{d}_1\right)e_2\,.
\end{equation}
We also note that $\Gamma_{u,1}$ is properly quantized when
\begin{equation}\tilde{k}_1-\frac{1}{2}\in\mathbb{Z}\,,\qquad 2\tilde{d}_1\in\mathbb{Z}\,,
\end{equation}
and
\begin{equation}h\left[\tilde{d}_2 - 2\tilde{X}\right] - e_1\left[\tilde{d}_2 - 2\tilde{Y}\right] +2e_2\tilde{Z}\in H_2(dP_2,\mathbb{Z})\,.\end{equation}
Similarly, $\Gamma_{u,2}$ is properly quantized when
\begin{equation}2\tilde{k}_2+3\tilde{d}_2\in\mathbb{Z}\end{equation}
and
\begin{equation}h\left[\tilde{k}_2+\tilde{d}_1+3\tilde{X}-2\right]-e_1\left[\tilde{k}_2+3\tilde{Y}-1\right]-e_2\left[\tilde{d}_1+3\tilde{Z}-\frac{1}{2}\right]\in H_2(dP_2,\mathbb{Z})\,.\end{equation}


\subsection{Non-universal Flux}

We now discuss the construction of non-universal fluxes.


\subsubsection{Non-universal Flux in $\mathcal{C}^{(1)}$}

To start, let us try to build a non-universal fluxin $\mathcal{C}^{(1)}$.  More specifically, we consider
\begin{equation}V+gU=0\,,\qquad \psi=0\,,\end{equation}
where $\psi=0$ defines a curve in $S_{\rm GUT}$.  We assume that $\psi$ is not proportional to either $W_2$ or $W_4$ for now.  To ensure that this belongs to our cubic curve we can set
\begin{equation}\begin{split}W_4\tilde{e}_0 &= G\psi + {\cal{F}}\left[B_2+HQ_1+c^{-1}J\psi\right] \\
W_4\tilde{e}_1 &= -{\cal{F}} + K\psi + H\left[B_2+HQ_1+c^{-1}J\psi\right] \\
A_2 &= B_2 c - H(P_1-cQ_1)+J\psi\end{split}\end{equation}
at which point our flux belongs to the cubic for
\begin{equation}g=B_2+HQ_1\,.\end{equation}

One of the features of this solution is that
\begin{equation}\begin{split}W_4\tilde{e}_0+B_2W_4\tilde{e}_1 &= H\left[B_2^2+{\cal{F}}Q_1+B_2HQ_1\right] + c^{-1}\psi\left[({\cal{F}}+B_2H)J+c(G+B_2K)\right]\\
cW_4\tilde{e_0} + A_2W_4\tilde{e}_1&= H\left[{\cal{F}}P_1+(B_2+HQ_1)(B_2c-H(P_1-cQ_1))\right]+{\cal{O}}(\psi)\,.\end{split}\end{equation}
This is important if we recall the expression for $P_{\mathbf{\overline{5}}_M}$
\begin{equation}\begin{split}P_{\mathbf{\overline{5}}_M}&=\tilde{a}_3\tilde{e}_0+\tilde{a}_2\tilde{e}_1 \\
&= W_2\left(c\tilde{e}_0 + A_2\tilde{e}_1\right)+W_4\left(\tilde{e}_0 + B_2\tilde{e}_1\right)\\
&=\frac{1}{W_4}\left[W_2\left(c(W_4\tilde{e}_0)+A_2(W_4\tilde{e}_1)\right)+W_4\left((W_4\tilde{e}_0)+B_2(W_4\tilde{e}_1)\right)\right]\,.
\end{split}\end{equation}
In the last line we wrote everything in terms of the holomorphic objects $(W_4\tilde{e}_m)$ (recall that $\tilde{e}_m$ themselves are meromorphic).  There is no net pole because the $W_2=W_4=0$ pole is manifestly canceled while the $W_1=W_4=0$ pole is canceled because $(W_4\tilde{e}_m)$ vanish at $W_1=W_4=0$ by construction.  What is interesting here is that for our choices above which were made to ensure that our non-universal flux could be constructed inside ${\cal{C}}^{(1)}$, $P_{\mathbf{\overline{5}}_M}$ takes the form
\begin{equation}P_{\mathbf{\overline{5}}_M}\sim \frac{1}{W_4}\left[ H\times P_{\mathbf{\overline{5}}_M}^{(H)}+ \psi \times P_{\mathbf{\overline{5}}_M}^{(\psi)}\right]\,.\end{equation}
When studying intersections of our flux with $\mathbf{\overline{5}}_M$, only the first term here will be relevant.  To simplify things, we will assume that the $\frac{1}{W_4}$ pole is canceled independently in the first and second terms.  This is not necessary but it is easy to arrange.  In fact, it is automatic provided we require the ${\cal{O}}(\psi^0)$ and ${\cal{O}}(\psi^1)$ pieces of $W_4\tilde{e}_m$ to separately vanish at $W_1=W_4=0$.  This, in turn, can be accomplished provided one of the following pairs of objects are both proportional to either $W_1$ or $W_4$
\begin{itemize}
\item ${\cal{F}}$ and $H$
\item ${\cal{F}}$ and $B_2+HQ_1$
\end{itemize}
In the second situation, both the $W_4=W_1=0$ and the $W_4=W_2=0$ pole are canceled by corresponding zeroes in $P_{\mathbf{\overline{5}}_M}^{(H)}$ while, in the first, the $W_4=W_1=0$ pole is canceled by a zero in $H$ and only the $W_4=W_2=0$ pole is canceled by a zero in $P_{\mathbf{\overline{5}}_M}$.

The intersections of our flux with $\mathbf{\overline{5}}_M$ can now be put into two categories
\begin{enumerate}
\item The lift of an intersection of $\psi$ with what remains of $P_{\mathbf{\overline{5}}_M}^{(H)}$ after any zeroes needed to cancel the $\frac{1}{W_4}$ are removed.
\item The lift of an intersection of $\psi$ with what remains of $H$ after any zeroes needed to cancel the $\frac{1}{W_4}$ are removed.
\end{enumerate}
What we remove from $P_{\mathbf{\overline{5}}_M}^{(H)}$ and $H$ depends on whether $H$ has a zero along $W_1=W_4=0$ or not.

What we would like to establish now is that intersections of $\psi$ with $P_{\mathbf{\overline{5}}_M}^{(H)}$ lift to honest intersections in ${\cal{C}}^{(1)}$ while intersections of $\psi$ with $H$ do not.  To do this, recall that the $\mathbf{\overline{5}}_M$ matter curve is given by the intersection of
\begin{equation}\tilde{a}_0U^2+\tilde{a}_2V^2=0\qquad \text{with}\qquad \tilde{a}_1U^2+\tilde{a}_3V^2=0\end{equation}
with the $(W_2=W_4=0)\times \mathbb{P}^1$ and $V=\alpha=0$ components removed.  To study the latter recall that $\alpha$ is simply $\tilde{\alpha}$,
\begin{equation}\tilde{\alpha}=W_4(P_1W_2+Q_1W_4)\end{equation}
with the double zero at $W_2=W_4=0$ removed.  In other words, $\alpha=0$ is the locus $(P_1W_2+Q_1W_4)=0$ for $W_2,W_4$ generically nonzero along with $W_1=W_4=0$.

To compute the intersection of $\mathbf{\overline{5}}_M$ with $\Psi$, we should combine the above two quadratics with $V+gU=0$ and $\psi=0$.  Plugging both of these into the above quadratics we find
\be
\ba
\tilde{a}_0U^2+\tilde{a}_2V^2 &\rightarrow U^2(B_2+HQ_1)P_{\mathbf{\overline{5}}_M}^{(H)}\cr
 \tilde{a}_1U^2+\tilde{a}_3V^2 &\rightarrow U^2P_{\mathbf{\overline{5}}_M}^{(H)}\,.
\ea\ee
This means that the intersections of $\psi=0$ with $P_{\mathbf{\overline{5}}_M}^{(H)}$ are lifted to true intersections in ${\cal{C}}^{(1)}$.  Note that just as we always have to remove the $W_2=W_4=0$ component from $P_{\mathbf{\overline{5}}_M}^{(H)}$ downstairs, we must do so here as well because this component is not part of the $\mathbf{\overline{5}}_M$ matter curve in ${\cal{C}}^{(1)}$ as we reviewed above.

We also have to remove a $W_1=W_4$ component if it exists.  In general, if $H$ vanishes at $W_1=W_4=0$ then $B_2$ is not required to do so in order for the $W_4\tilde{e}_m$ to vanish there.  This means that a generic choice of $B_2$ will not vanish at $W_1=W_4=0$ and hence $g=B_2+HQ_1$ will not vanish there either.  Because of this, the $\mathbf{\overline{5}}_M$ matter curve does not have a "component at infinity" along $W_1=W_4=0$ so we do not need to subtract it.  This is consistent with what we saw downstairs; if $H$ had a zero at $W_1=W_4=0$ then we did not  have to remove any zero of $P_{\mathbf{\overline{5}}_M}^{(H)}$ at $W_1=W_4=0$ in order to cancel the $\frac{1}{W_4}$ out front.

On the other hand, if $H$ does not vanish at $W_1=W_4=0$ then $g=B_2+HQ_1$ is required to vanish there.  In this case, we do have a "component at infinity" where $g=B_2+HQ_1$, $W_1$, and $W_4$ all vanish.  This must be subtracted from the $\mathbf{\overline{5}}_M$ matter curve, resulting effectively in the removal of the $W_1=W_4=0$ locus from $P_{\mathbf{\overline{5}}_M}$ (which it will have in this case).  This is consistent with what we saw downstairs in $S_{\rm GUT}$; if $H$ could not cancel the $W_1=W_4=0$ pole from the $\frac{1}{W_4}$ out front, then a zero had to be removed from $P_{\mathbf{\overline{5}}_M}$ in order to do so.

What happened to the intersections of $\psi$ with $H$ downstairs in $S_{\rm GUT}$ when we lifted them to ${\cal{C}}^{(1)}$?  To see what happened, let us plug $\psi=H=0$ into the cubic equation from which we obtain ${\cal{C}}^{(1)}$.  In this case, we get
\begin{equation}(V+B_2U)\left[V^2(cW_2+W_4)+{\cal{F}}U^2(P_1W_2+Q_1W_4)\right]\,.\end{equation}
The first factor is simply
\begin{equation}(V+B_2U)=\left[V+gU\right]_{\psi,H\rightarrow 0}\,,\end{equation}
while the second factor is simply
\begin{equation}V^2(cW_2+W_4)+{\cal{F}}U^2(P_1W_2+Q_1W_4) = \left[\tilde{a}_1U^2+\tilde{a}_3V^2\right]_{\psi,H\rightarrow 0}\,.\end{equation}
Apparently, $\psi=0$ does not lift to either of the sheets that $H=0$ does in the neighborhood of any of the $\psi=H=0$ intersection points in $S_{\rm GUT}$.  This is why these intersections do not appear when computing $\Psi\cdot \mathbf{\overline{5}}_M$.

In the end, the intersection $\Psi\cdot \mathbf{\overline{5}}_M$ depends on whether $H$ vanishes at $W_1=W_4=0$ or not.  We can summarize this as
\begin{equation}\Psi\cdot \mathbf{\overline{5}}_M = \left\{\begin{array}{cc} \psi\cdot_{S_{\rm GUT}} ([P_{\mathbf{\overline{5}}_M}] - [W_2=W_4=0]) &\qquad H\rightarrow 0\text{ as }W_1=W_4=0\\
\psi\cdot_{S_{\rm GUT}} ([P_{\mathbf{\overline{5}}_M}] - [W_2=W_4=0] - [W_1=W_4=0]) &\qquad H(W_1=W_4=0)\ne 0\end{array}\right.\end{equation}
where we used the fact that
\begin{equation}[P_{\mathbf{\overline{5}}_M}] = [P_{\mathbf{\overline{5}}_M}^{(H)}]\,.\end{equation}




An important condition on $\psi$ arises if we note that it is one component of the projected curve $p_{1\,\ast}\left(\sigma_{\infty}\cdot {\cal{C}}^{(1)}\right) = \eta-2c_1-\xi$.  In particular, this means that
\begin{equation}\boxed{\psi\text{ and }\eta-2c_1-\xi-\psi\text{ must both be effective curves in }S_{\rm GUT}}\end{equation}




\subsubsection{Non-Universal Flux in ${\cal{C}}^{(2)}$}

It will also be helpful to have a nonuniversal flux in ${\cal{C}}^{(2)}$ that is the lift of $\psi$.  To arrange this, we simply set
\begin{equation}\tilde{e}_2 = -\frac{H}{W_4}\,.\end{equation}

In this case, when $\psi=0$ ${\cal{C}}^{(2)}$ factors as
\begin{equation}{\cal{C}}^{(2)}\rightarrow (V-gU)\left[\frac{1}{W_4}\left(HV + {\cal{F}}U\right)\right]\,,\end{equation}
namely a factor in the class $\sigma_{\infty}$ and a factor in the class $\sigma_{\infty}+\pi^{\ast}\xi$ once we recall that we have to remove a $W_2=W_4=0$ part of the pole of the second factor by hand.  I think we also need $H$ and ${\cal{F}}$ to vanish at $W_1=W_4=0$ to cancel the remaining pole part of the $1/W_4$ in general.  This means we must specialize to
\begin{equation}\epsilon=0\,.\end{equation}

We can explicitly compute intersections of this $\psi$ because, unlike the matter curves in ${\cal{C}}^{(1)}$, the matter curves in ${\cal{C}}^{(2)}$ can all be written as the intersection of a divisor in $X$ with ${\cal{C}}^{(2)}$
\begin{equation}\ba
\Sigma_{\mathbf{5}_H} &= {\cal{C}}^{(2)}\cdot \pi^{\ast}(c_1+\xi)\cr
 \Sigma_{\mathbf{\overline{5}}_H} &= {\cal{C}}^{(2)}\cdot \left( {\cal{C}}^{(1)}-2\sigma_{\infty}\right) = {\cal{C}}^{(2)}\cdot \left(\sigma+\pi^{\ast}(\eta-4c_1-\xi)\right) \cr
\Sigma_{\mathbf{10}_{\text{other}}} &= {\cal{C}}^{(2)}\cdot
\ea
\end{equation}
Denoting this flux by $\Psi_2$ then we have
\begin{equation}\begin{array}{c|c|c}
\text{Matter Curve} & \text{Class in }$X$ & \Psi_2 \\ \hline
\mathbf{10}_{\text{other}} & \sigma\cdot \pi^{\ast}\xi & 0 \\
\mathbf{5}_H & 2\sigma\cdot \pi^{\ast}(c_1+\xi) + \pi^{\ast}(c_1+\xi)\cdot\pi^{\ast}(2c_1+\xi) & \psi\cdot_{S_{\rm GUT}}(c_1+\xi) \\
\mathbf{\overline{5}}_H & \sigma\cdot\pi^{\ast}(2\eta-4c_1+\xi)+\pi^{\ast}(\eta-2c_1-\xi)\cdot\pi^{\ast}(2c_1+\xi) & \psi\cdot_{S_{\rm GUT}} (\eta-4c_1-\xi)
\end{array}\end{equation}

\subsubsection{Summary of Non-Universal Fluxes}

Let us call the object $\Psi$ inside ${\cal{C}}^{(1)}$ by $\Psi_1$, to distinguish it from $\Psi_2$ in $\mathcal{C}^{(2)}$.  In that case, we have the following intersection data
\begin{equation}\begin{array}{c|c|c}
\text{Matter Curve} & \Psi_1 & \Psi_2 \\ \hline
\mathbf{10}_M & 0 & 0 \\
\mathbf{10}_{\text{other}} & 0 & 0 \\
\mathbf{5}_H & 0 & \psi\cdot_{S_{\rm GUT}} (4h-2e_1-e_2) \\
\mathbf{\overline{5}}_H & \psi\cdot_{S_{\rm GUT}} (h-e_1) & \psi\cdot_{S_{\rm GUT}}\left(4h-e_1-2e_2\right)  \\
\mathbf{\overline{5}}_M & \psi\cdot_{S_{\rm GUT}} \left[7h - 2e_1 - 3e_2\right] & 0
\end{array}\end{equation}
where we deduced the restriction of $\Psi_1$ to $\mathbf{\overline{5}}_H$ by using the fact that
\begin{equation}\left(3\Psi_1-p_1^{\ast}p_{1\,\ast}\Psi_1\right)\cdot_{ {\cal{C}}^{(1)} } \left(\mathbf{\overline{5}}_M+\mathbf{\overline{5}}_H - \mathbf{10}_{\text{other}}\right) = 0\,.
\end{equation}


\section{Details of D3-tadpole analysis}
\label{app:Gamma}

In this appendix we provide the details for the computation of the D3-tadpole arising from the $G$-flux. In terms of the spectral cover description of the fluxes, this is obtain by evaluating
\begin{equation}N_{D3\text{ induced}} = -\frac{1}{2}\left[\Gamma_1\cdot_{ {\cal{C}}^{(1)}}\Gamma_1 + \Gamma_2\cdot_{ {\cal{C}}^{(2)}} \Gamma_2 \right]\end{equation}


\subsection{Universal Fluxes}

We start with universal fluxes $\Gamma_{u,1}$ and $\Gamma_{u,2}$.  We have that
\begin{equation}\begin{split}\Gamma_{u,1}\cdot_{ {\cal{C}}^{(1)}} \Gamma_{u,1} &= {\cal{C}}^{(1)}\cdot \left\{\left(3\tilde{k}_1 + 2\tilde{d}_1\right)\sigma-\pi^{\ast}\left[\tilde{k}_1(\eta-5c_1-\xi)+\tilde{d}_2\xi - 2\rho\right]\right\}^2 \\
&= -2\left(3\tilde{k}_1 + 2\tilde{d}_1\right)^2 - 4\tilde{d}_1\tilde{d}_2 + 4\left[ 2\tilde{d}_1 (\eta-5c_1-\xi) - 3\tilde{d}_2\xi\right]\cdot_{S_{\rm GUT}}\rho + 12 \rho\cdot_{S_{\rm GUT}}\rho
\end{split}\end{equation}
and
\begin{equation}\begin{split}
\Gamma_{u,2}\cdot_{{\cal{C}}^{(2)}}\Gamma_{u,2} &= {\cal{C}}^{(2)}\cdot \left\{\left(2\tilde{k}_2 + 3\tilde{d}_2\right)\sigma - \pi^{\ast}\left[\tilde{k}_2\xi + \tilde{d}_1 (\eta-5c_1-\xi)+3\rho\right]\right\}^2 \\
&= -2\left(2\tilde{k}_2+3\tilde{d}_2\right)^2-6\tilde{d}_1\tilde{d}_2+6\left[2\tilde{d}_(\eta-5c_1-\xi)-3\tilde{d}_2\xi\right]\cdot_{S_{\rm GUT}}\rho + 18\rho\cdot_{S_{\rm GUT}}\rho
\end{split}\end{equation}
so that the total contribution from univeral fluxes is
\begin{equation}\begin{split}\Gamma_u^2 &= -2\left[(3\tilde{k}_1+2\tilde{d}_1)^2 + (2\tilde{k}_2 + 3\tilde{d}_2)^2\right] - 10\tilde{d}_1\tilde{d}_2 + 10\left[2\tilde{d}_1(\eta-5c_1-\xi)-3\tilde{d}_2\xi\right]\cdot_{S_{\rm GUT}}\rho + 30\rho\cdot_{S_{\rm GUT}}\rho\\
&= -2\left(3\tilde{k}_1+2\tilde{d}_1\right)^2-2\left(2\tilde{k}_2+3\tilde{d}_2\right)^2 + \frac{5}{6}\left[6\rho + 2\tilde{d}_1(\eta-5c_1-\xi) - 3\tilde{d}_2\xi\right]^2 \end{split}\end{equation}
We recognize the object in $[\,]$'s in the second line as the class that must be supersymmetric for $\Gamma_u$ to be consistent in the absence of non-universal fluxes.  Because any supersymmetric class has negative self-intersection in $dP_2$ we see that $\Gamma_u^2$ is negative definite, as we expect from consistency.

\subsection{Non-universal Fluxes}

We now incorporate non-universal fluxes.
Among other things, we will need the self-intersections $\Psi_1^2$ and $\Psi_2^2$.  For $\Psi_1\cdot_{ {\cal{C}}^{(1)} }\Psi_1$ we proceed by using repeated applications of adjunction.  For starters, we have
\begin{equation}c_1({\cal{C}}^{(1)})|_{\Psi_1} = c_1(\Psi_1) + \Psi_1\cdot_{ {\cal{C}}^{(1)} }\Psi_1\,.
\end{equation}
A nice thing about $c_1(\Psi)$ is that it is a line bundle of the same degree as $c_1(\psi)$, which we can get from adjunction in $S_{\rm GUT}$
\begin{equation}c_1\cdot_{S_{\rm GUT}} \psi = c_1(\psi) + \psi\cdot_{S_{\rm GUT}}\psi\,.
\end{equation}
Note here that we continue to use $c_1$ as shorthand for $c_1(S_{\rm GUT})$.  As for $c_1({\cal{C}}^{(1)})$, we can compute it using adjunction on $X$
\begin{equation}c_1(X)|_{ {\cal{C}}^{(1)} }=c_1({\cal{C}}^{(1)} ) + {\cal{C}}^{(1)}\cdot_X {\cal{C}}^{(1)}\,.
\end{equation}
Because we know that $c_1(X) = 2\sigma_{\infty}$ we find
\begin{equation}c_1({\cal{C}}^{(1)})|_{\Psi_1} = (2\sigma_{\infty}-{\cal{C}}^{(1)})\cdot\Psi_1\,.
\end{equation}
Putting everything together, we find that
\begin{equation}\begin{split}\Psi_1\cdot_{ {\cal{C}}^{(1)} }\Psi_1 &= \left(2\sigma_{\infty}- {\cal{C}}^{(1)}\right)\cdot \Psi_1- \psi\cdot_{S_{\rm GUT}} (c_1 - \psi)\\
&= -\psi\cdot_{S_{\rm GUT}} (\eta-3c_1-\xi-\psi) \,.
\end{split}\end{equation}
For $\Psi_2\cdot_{ {\cal{C}}^{(2)}} \Psi_2$ we proceed in the same way to find
\begin{equation}\begin{split}\Psi_2\cdot_{ {\cal{C}}^{(2)} }\Psi_2 &= \left(2\sigma_{\infty}- {\cal{C}}^{(2)}\right)\cdot \Psi_2 - \psi\cdot_{S_{\rm GUT}} (c_1-\psi) \\
&=-\psi\cdot_{S_{\rm GUT}} (c_1+\xi-\psi)\,.
\end{split}\end{equation}

We can now compute
\begin{equation}\begin{split}
\Gamma_{u,1}\cdot_{ {\cal{C}}^{(1)} }\Psi_1 &= - \psi\cdot_{S_{\rm GUT}} \left[\tilde{k}_1(\eta-5c_1-\xi) + \tilde{d}_2\xi - 2\rho\right] \\
\Gamma_{u,1}\cdot_{ {\cal{C}}^{(1)}}p_1^{\ast}\psi &= \psi\cdot_{S_{\rm GUT}}\left[2\tilde{d}_1(\eta-5c_1-\xi)-3\tilde{d}_2\xi + 6\rho\right] \\
\Psi_1\cdot_{ {\cal{C}}^{(1)}} p_1^{\ast}\psi &= \psi\cdot_{S_{\rm GUT}}\psi\\
\Gamma_{u,2}\cdot_{ {\cal{C}}^{(2)} }\Psi_2 &= -\psi\cdot_{S_{\rm GUT}} \left[\tilde{k}_2\xi + \tilde{d}_1(\eta-5c_1-\xi)+3\rho\right] \\
\Gamma_{u,2}\cdot_{ {\cal{C}}^{(2)} }p_2^{\ast}\psi &= -\psi\cdot_{S_{\rm GUT}} \left[2\tilde{d}_1(\eta-5c_1-\xi)-3\tilde{d}_2\xi + 6\rho\right] \\
\Psi_2\cdot_{ {\cal{C}}^{(2)} }p_2^{\ast}\psi &= \psi \cdot_{S_{\rm GUT}} \psi\,.
\end{split}\end{equation}
which is useful when we write our net fluxes as
\begin{equation}\Gamma_1 = \Gamma_{u,1} + (3\tilde{m}_1+\tilde{q})\Psi_1 - \tilde{m}_1p_1^{\ast}\psi\qquad \Gamma_2 = \Gamma_{u,2} + (2\tilde{m}_2 - \tilde{q})\Psi_2 - \tilde{m}_2p_2^{\ast}\psi\,.
\end{equation}

We now find
\begin{equation}\begin{split}\Gamma_1^2 &= \Gamma_{u,1}^2 - (3\tilde{m}_1+\tilde{q})^2\psi(\eta-3c_1-\xi-\psi) + 3\tilde{m}_1^2\psi^2 \\
&\qquad - 2(3\tilde{m}_1+\tilde{q})\psi\left[\tilde{k}_1(\eta-5c_1-\xi)+\tilde{d}_2\xi-2\rho\right]-2\tilde{m}_1\psi\left[2\tilde{d}_1(\eta-5c_1-\xi)-3\tilde{d}_2\xi+6\rho\right]\\
&\qquad\qquad-2\tilde{m}_1(3\tilde{m}_1+\tilde{q})\psi^2\\
\Gamma_2^2 &= \Gamma_{u,2}^2 -(2\tilde{m}_2-\tilde{q})^2\psi(c_1+\xi-\psi) + 2\tilde{m}_2^2\psi^2 \\
&\qquad - 2(2\tilde{m}_2-\tilde{q})\psi\left[\tilde{k}_2\xi + \tilde{d}_1(\eta-5c_1-\xi)+3\rho\right] + 2\tilde{m}_2\psi\left[2\tilde{d}_1(\eta-5c_1-\xi)-3\tilde{d}_2\xi+6\rho\right]\\
&\qquad\qquad-2\tilde{m}_2(2\tilde{m}_2-\tilde{q})\psi^2
\end{split}\end{equation}
so that
\begin{equation}\begin{split}\Gamma^2 &=
\Gamma_u^2 - (3\tilde{m}_1+\tilde{q})^2\psi(\eta-3c_1-\xi) - (2\tilde{m}_2-\tilde{q})^2\psi(c_1+\xi) +2\left(3\tilde{m}_1^2+\tilde{m}_2^2+[2\tilde{m}_1-\tilde{m}_2]\tilde{q}+\tilde{q}^2\right)\psi^2\\
&\quad -2(3\tilde{m}_1+\tilde{q})\psi\left[\tilde{k}_1(\eta-5c_1-\xi)+\tilde{d}_2\xi - 2\rho\right] - 2(2\tilde{m}_2-\tilde{q})\psi\left[\tilde{k}_2\xi + \tilde{d}_1(\eta-5c_1-\xi)+3\rho\right] \\
&\quad -2(\tilde{m}_1-\tilde{m}_2)\psi\left[2\tilde{d}_1(\eta-5c_1-\xi)-3\tilde{d}_2\xi+6\rho\right]\,.
\end{split}\end{equation}
This can be rewritten as
\begin{equation}\begin{split}\Gamma^2 &= -2\left(3\tilde{k}_1+2\tilde{d}_1\right)^2 -2\left(2\tilde{k}_2+3\tilde{d}_2\right)^2 + \frac{5}{6}\left[6\rho + 2\tilde{d}_1(\eta-5c_1-\xi)-3\tilde{d}_2\xi+\tilde{q}\psi\right]^2\\
& -\left(3\tilde{m}_1+\tilde{q}\right)^2\psi(\eta-3c_1-\xi) - \left(2\tilde{m}_2-\tilde{q}\right)^2\psi(c_1+\xi) \\
& - \left(2\tilde{k}_2+3\tilde{d}_2\right)\left(2\tilde{m}_2-\tilde{q}\right)\xi\psi - \frac{2}{3}\left(3\tilde{k}_1+2\tilde{d}_1\right)\left(3\tilde{m}_1+\tilde{q}\right)\psi(\eta-5c_1-\xi) \\
 & + 2\left(3\tilde{m}_1^2 + \tilde{m}_2^2 + [2\tilde{m}_1 - \tilde{m}_2]\tilde{q}+\frac{7}{12}\tilde{q}^2\right)\psi^2\,.
\end{split}\end{equation}
This is the main result that we need to discuss D3-tadpole cancellation in the main text.

\newpage

\bibliographystyle{JHEP}
\renewcommand{\refname}{Bibliography}

\begin{thebibliography}{10}

\bibitem{Marsano:2009gv}
J.~Marsano, N.~Saulina, and S.~Schafer-Nameki, {\it {Monodromies, Fluxes, and
  Compact Three-Generation F-theory GUTs}},  {\em JHEP} {\bf 08} (2009) 046,
  [\href{http://xxx.lanl.gov/abs/0906.4672}{{\tt 0906.4672}}].

\bibitem{Marsano:2009ym}
J.~Marsano, N.~Saulina, and S.~Schafer-Nameki, {\it {F-theory Compactifications
  for Supersymmetric GUTs}},  \href{http://xxx.lanl.gov/abs/0904.3932}{{\tt
  0904.3932}}.

\bibitem{Donagi:2008ca}
R.~Donagi and M.~Wijnholt, {\it {Model Building with F-Theory}},
  \href{http://xxx.lanl.gov/abs/0802.2969}{{\tt 0802.2969}}.

\bibitem{Beasley:2008dc}
C.~Beasley, J.~J. Heckman, and C.~Vafa, {\it {GUTs and Exceptional Branes in
  F-theory - I}},  {\em JHEP} {\bf 01} (2009) 058,
  [\href{http://xxx.lanl.gov/abs/0802.3391}{{\tt 0802.3391}}].

\bibitem{Beasley:2008kw}
C.~Beasley, J.~J. Heckman, and C.~Vafa, {\it {GUTs and Exceptional Branes in
  F-theory - II: Experimental Predictions}},  {\em JHEP} {\bf 01} (2009) 059,
  [\href{http://xxx.lanl.gov/abs/0806.0102}{{\tt 0806.0102}}].

\bibitem{Donagi:2008kj}
R.~Donagi and M.~Wijnholt, {\it {Breaking GUT Groups in F-Theory}},
  \href{http://xxx.lanl.gov/abs/0808.2223}{{\tt 0808.2223}}.

\bibitem{Marsano:2008jq}
J.~Marsano, N.~Saulina, and S.~Schafer-Nameki, {\it {Gauge Mediation in
  F-Theory GUT Models}},  \href{http://xxx.lanl.gov/abs/0808.1571}{{\tt
  0808.1571}}.

\bibitem{Heckman:2008qt}
J.~J. Heckman and C.~Vafa, {\it {F-theory, GUTs, and the Weak Scale}},
  \href{http://xxx.lanl.gov/abs/0809.1098}{{\tt 0809.1098}}.

\bibitem{Font:2008id}
A.~Font and L.~E. Ibanez, {\it {Yukawa Structure from U(1) Fluxes in F-theory
  Grand Unification}},  {\em JHEP} {\bf 02} (2009) 016,
  [\href{http://xxx.lanl.gov/abs/0811.2157}{{\tt 0811.2157}}].

\bibitem{Heckman:2008qa}
J.~J. Heckman and C.~Vafa, {\it {Flavor Hierarchy From F-theory}},
  \href{http://xxx.lanl.gov/abs/0811.2417}{{\tt 0811.2417}}.

\bibitem{Hayashi:2009ge}
H.~Hayashi, T.~Kawano, R.~Tatar, and T.~Watari, {\it {Codimension-3
  Singularities and Yukawa Couplings in F- theory}},
  \href{http://xxx.lanl.gov/abs/0901.4941}{{\tt 0901.4941}}.

\bibitem{Bouchard:2009bu}
V.~Bouchard, J.~J. Heckman, J.~Seo, and C.~Vafa, {\it {F-theory and Neutrinos:
  Kaluza-Klein Dilution of Flavor Hierarchy}},
  \href{http://xxx.lanl.gov/abs/0904.1419}{{\tt 0904.1419}}.

\bibitem{Heckman:2009mn}
J.~J. Heckman, A.~Tavanfar, and C.~Vafa, {\it {The Point of $E_8$ in F-theory
  GUTs}},  \href{http://xxx.lanl.gov/abs/0906.0581}{{\tt 0906.0581}}.

\bibitem{Heckman:2009de}
J.~J. Heckman and C.~Vafa, {\it {CP Violation and F-theory GUTs}},
  \href{http://xxx.lanl.gov/abs/0904.3101}{{\tt 0904.3101}}.

\bibitem{Font:2009gq}
A.~Font and L.~E. Ibanez, {\it {Matter wave functions and Yukawa couplings in
  F-theory Grand Unification}},  {\em JHEP} {\bf 09} (2009) 036,
  [\href{http://xxx.lanl.gov/abs/0907.4895}{{\tt 0907.4895}}].

\bibitem{Conlon:2009qq}
J.~P. Conlon and E.~Palti, {\it {Aspects of Flavour and Supersymmetry in
  F-theory GUTs}},  \href{http://xxx.lanl.gov/abs/0910.2413}{{\tt 0910.2413}}.

\bibitem{Cecotti:2009zf}
S.~Cecotti, M.~C.~N. Cheng, J.~J. Heckman, and C.~Vafa, {\it {Yukawa Couplings
  in F-theory and Non-Commutative Geometry}},
  \href{http://xxx.lanl.gov/abs/0910.0477}{{\tt 0910.0477}}.

\bibitem{Hayashi:2009bt}
H.~Hayashi, T.~Kawano, Y.~Tsuchiya, and T.~Watari, {\it {Flavor Structure in
  F-theory Compactifications}},  \href{http://xxx.lanl.gov/abs/0910.2762}{{\tt
  0910.2762}}.

\bibitem{Marchesano:2009rz}
  F.~Marchesano and L.~Martucci,
  {\it Non-perturbative effects on seven-brane Yukawa couplings},
 \href{http://xxx.lanl.gov/abs/0910.5496}{{\tt
  0910.5496}}.

\bibitem{Tatar:2009jk}
R.~Tatar, Y.~Tsuchiya, and T.~Watari, {\it {Right-handed Neutrinos in F-theory
  Compactifications}},  \href{http://xxx.lanl.gov/abs/0905.2289}{{\tt
  0905.2289}}.
  
  \bibitem{Andreas:2009uf}
B.~Andreas and G.~Curio, {\it {From Local to Global in F-Theory Model
  Building}},  \href{http://xxx.lanl.gov/abs/0902.4143}{{\tt 0902.4143}}.

\bibitem{Donagi:2009ra}
R.~Donagi and M.~Wijnholt, {\it {Higgs Bundles and UV Completion in F-Theory}},
   \href{http://xxx.lanl.gov/abs/0904.1218}{{\tt 0904.1218}}.

\bibitem{Buican:2006sn}
M.~Buican, D.~Malyshev, D.~R. Morrison, H.~Verlinde, and M.~Wijnholt, {\it
  {D-branes at singularities, compactification, and hypercharge}},  {\em JHEP}
  {\bf 01} (2007) 107, [\href{http://xxx.lanl.gov/abs/hep-th/0610007}{{\tt
  hep-th/0610007}}].

\bibitem{Conlon:2009qa}
J.~P. Conlon and E.~Palti, {\it {On Gauge Threshold Corrections for Local
  IIB/F-theory GUTs}},  {\em Phys. Rev.} {\bf D80} (2009) 106004,
  [\href{http://xxx.lanl.gov/abs/0907.1362}{{\tt 0907.1362}}].

\bibitem{Tatar:2008zj}
R.~Tatar and T.~Watari, {\it {GUT Relations from String Theory
  Compactifications}},  {\em Nucl. Phys.} {\bf B810} (2009) 316--353,
  [\href{http://xxx.lanl.gov/abs/0806.0634}{{\tt 0806.0634}}].

\bibitem{Blumenhagen:2008aw}
R.~Blumenhagen, {\it {Gauge Coupling Unification in F-Theory Grand Unified
  Theories}},  {\em Phys. Rev. Lett.} {\bf 102} (2009) 071601,
  [\href{http://xxx.lanl.gov/abs/0812.0248}{{\tt 0812.0248}}].

\bibitem{Heckman:2008es}
J.~J. Heckman, J.~Marsano, N.~Saulina, S.~Schafer-Nameki, and C.~Vafa, {\it
  {Instantons and SUSY breaking in F-theory}},
  \href{http://xxx.lanl.gov/abs/0808.1286}{{\tt 0808.1286}}.

\bibitem{Marsano:2008py}
J.~Marsano, N.~Saulina, and S.~Schafer-Nameki, {\it {An Instanton Toolbox for
  F-Theory Model Building}},  \href{http://xxx.lanl.gov/abs/0808.2450}{{\tt
  0808.2450}}.

\bibitem{Donagi:2004ia}
R.~Donagi, Y.-H. He, B.~A. Ovrut, and R.~Reinbacher, {\it {The particle
  spectrum of heterotic compactifications}},  {\em JHEP} {\bf 12} (2004) 054,
  [\href{http://xxx.lanl.gov/abs/hep-th/0405014}{{\tt hep-th/0405014}}].

\bibitem{Blumenhagen:2006wj}
R.~Blumenhagen, S.~Moster, R.~Reinbacher, and T.~Weigand, {\it {Massless
  spectra of three generation U(N) heterotic string vacua}},  {\em JHEP} {\bf
  05} (2007) 041, [\href{http://xxx.lanl.gov/abs/hep-th/0612039}{{\tt
  hep-th/0612039}}].

\bibitem{Hayashi:2008ba}
H.~Hayashi, R.~Tatar, Y.~Toda, T.~Watari, and M.~Yamazaki, {\it {New Aspects of
  Heterotic--F Theory Duality}},  {\em Nucl. Phys.} {\bf B806} (2009) 224--299,
  [\href{http://xxx.lanl.gov/abs/0805.1057}{{\tt 0805.1057}}].


\bibitem{Blumenhagen:2009yv}
R.~Blumenhagen, T.~W. Grimm, B.~Jurke, and T.~Weigand, {\it {Global F-theory
  GUTs}},  \href{http://xxx.lanl.gov/abs/0908.1784}{{\tt 0908.1784}}.

\bibitem{Giudice:1997ni}
G.~F. Giudice and R.~Rattazzi, {\it {Extracting Supersymmetry-Breaking Effects
  from Wave- Function Renormalization}},  {\em Nucl. Phys.} {\bf B511} (1998)
  25--44, [\href{http://xxx.lanl.gov/abs/hep-ph/9706540}{{\tt
  hep-ph/9706540}}].


\end{thebibliography}

\providecommand{\href}[2]{#2}\begingroup\raggedright\endgroup


\end{document}